\documentclass[preprint]{aastex}

\def\kms{~km~s$^{-1}$\ }

\def\arcs{\char'175\ }
\def\arcsec{\char'175 }

\def\etal{et~al.\ }

\def\hub{\ifmmode H_\circ\else H$_\circ$\fi}

\shorttitle{Magellanic Cloud Clusters}
\shortauthors{Caldwell, Rose \& Concannon}

\begin{document}

\title{Magellanic Cloud Star Clusters as Simple Stellar Populations}
\author{Andrew J. Leonardi\altaffilmark{1}}
\affil{Department of Physics and Astronomy, CB \#3255, University of North 
Carolina, Chapel Hill, NC 27599 and Division of Natural Sciences \& 
Engineering, University of South Carolina - Spartanburg, Spartanburg, SC 29303}
\email{aleonardi@gw.uscs.edu}

\and

\author{James A. Rose}
\affil{Department of Physics and Astronomy, CB \#3255, University of North Carolina, Chapel Hill, NC 27599}
\email{jim@physics.unc.edu}

\altaffiltext{1}{Present address: Division of Natural Sciences \& 
Engineering, University of South Carolina - Spartanburg, Spartanburg, SC 29303}

\begin{abstract}
Integrated spectra have been obtained of 31 star clusters in the Magellanic
Clouds (MC) and of 4 Galactic globular clusters.  The spectra cover the
wavelength range $\lambda$$\lambda$3500 -- 4700 \AA \ at a resolution of
3.2 \AA \ FWHM.  The MC clusters primarily cover the age range $<$10$^8$ -- 3
Gyr, and hence are well-suited to an empirical study of aging post-starburst
stellar populations.  An age-dating method is presented that relies on two
spectral absorption feature indices, H$\delta$/Fe I $\lambda$4045 and Ca II, as
well as an index measuring the strength of the Balmer discontinuity.
We compare the behavior of the spectral indices in the observed integrated 
spectra of the MC clusters to that of indices
generated from theoretical evolutionary synthesis models of varying age and
metal-abundance. The synthesis models are based on those of Worthey (1994, ApJS,
95, 107), when coupled with
the combination of an empirical library of stellar spectra (Jones 1999, PhD
thesis, University of North Carolina) for the cooler stars and synthetic 
spectra, generated from Kurucz model atmospheres, for the hotters stars.
Overall, we find good agreement between the ages of the MC clusters derived from
our integrated spectra (and the evolutionary synthesis modelling of the spectral
indices) and ages derived from analyses of the cluster color-magnitude diagrams,
as found in the literature.  Hence the principal conclusion of this study is
that ages of young stellar populations can be reliably measured from modelling
of their integrated spectra.

\end{abstract}

 
\normalsize
\section{Introduction} 

While it is convenient to assume that most galaxies have experienced a
simple star formation history, for example, exponentially declining over time,
there is growing evidence that many galaxies have experienced multiple
major episodes of concentrated star formation (hereafter, starbursts) during 
their lifetimes, as opposed to a smoothly varying star formation rate.  Since 
the seminal study of Searle, Sargent, \& Bagnuolo (1973), in which starbursts
were advocated to explain the extreme UBV colors of the bluest galaxies,
starbursts have been implicated in a variety of situations in which galaxies 
have been found with blue colors and/or enhanced Balmer absorption line 
strengths.  Briefly, Dressler \& Gunn (1983) proposed starbursts to explain
the photometrically blue galaxies discovered by Butcher \& Oemler (1978a,b) in
galaxy clusters at z$\sim$0.3, starbursts have been advocated to explain the
excess of blue galaxies at faint apparent magnitudes (e.g., Ellis 1997), and
both numerical simulations of interacting and merging galaxies (Toomre \&
Toomre 1972; Mihos \& Hernquist 1994, 1996) and observations of such systems
(e.g., Larson \& Tinsley 1978; Kennicutt \etal 1987; Rifatto \etal 2001 and
references therein) indicate that starbursts should be common in merger events.
Clearly, there is much to be gained from obtaining information on the ages of
the starbursts in these situations.

Unfortunately, reliably extracting the ages of starburst episodes using 
population synthesis modelling has proven to be problematic, for several
reasons.  First, there is the difficulty in uniquely detecting the young
stellar population (YSP) produced in a starburst against the backdrop of an
underlying old stellar population (OSP).  This problem manifests itself as a
degeneracy in the age versus the strength of a burst.  Specifically, a 
younger, weaker burst is difficult to distinguish from an older, stronger burst
(e.g., Couch \& Sharples 1987; Bica, Alloin, \& Schmidt 1990; Charlot \& Silk 
1994; Leonardi \& Rose 1996).  The second problem is the degeneracy between
age and metal-abundance.  While this degeneracy is most notorious in the case
of modelling older stellar populations (Worthey 1994), it afflicts younger
populations as well (Nolan, Dunlop, \& Jimenez 2001).  Adding to these two
difficulties is the confusing effect of interstellar reddening on the spectral
energy distribution (SED) of a burst. 
Finally, there is the complicating factor that the behavior
of the SED and absorption line stengths of a post-starburst population is so
non-linearly dependent on age, with rapid evolution back to near-normal colors
taking place over the first few Gyr, followed by a much more gradual 
evolution thereafter (Searle, Sargent, \& Bagnuolo 1973; Bica, Alloin, \&
Schmidt 1990; Charlot \& Silk 1994).  In short, age dating of starburst 
populations is confronted by serious difficulties.

Despite the above-mentioned challenges, a variety of studies have been conducted
to extract the ages of starburst populations from population synthesis
modelling (Bruzual 1983; Bica, Alloin, 
\& Schmidt 1990; Schweizer \& Seitzer 1992;
Bernl\"{o}hr 1992; Leonardi \& Rose 1996; Kurth, Fritze-v. Alvensleben, \&
Fricke 1999; Bicker, Fritze-v. Alvensleben, \& Fricke 2002).  These 
investigations range in emphasis from broadband photometry to relatively high
resolution studies of specific spectral features.
 
The ideal testing ground for models of the integrated spectra of starburst 
populations are the observed integrated spectra of star clusters, and,
in particular, of star clusters in the Magellanic Clouds (MC).  Clusters are,
in essence, mini-starbursts, given that the stars in a cluster are coeval and 
of uniform chemical composition.  They are particualarly advantageous if their
stellar density is high enough to eliminate confusion from the underlying OSP
in the galaxy.  Clusters in the Clouds serve the purpose 
well, spanning a wide enough range in age and metallicity 
(e.g., Sagar \& Pandey 1989) to adequately 
assess the behavior of stellar population indicators.  They are particularly 
useful because there is no corresponding set of clusters in the Galaxy with 
sufficient diversity in both age and metallicity.  Milky Way globular clusters, 
while adequately encompassing metallicity space, are all old systems. 
Stellar populations of appropriate 
age are found in Galactic open clusters but they are not dense and populous 
enough to be readily observed in integrated light above the confusion of stars
in the Galactic disk.
In contrast, the young, populous clusters in the Clouds 
represent excellent analogs of a starburst YSP.  MC clusters are distant and
populous enough 
to be studied in integrated light, but are still sufficiently nearby so that the
results of integrated light techniques can be compared with the fundamental
age and metallicity information obtained from color-magnitude diagrams (e.g.,
Kerber \etal 2002) and
from spectroscopy of individual cluster members (e.g.,
Olszewski \etal 1991), including high dispersion (Hill \etal 2000; Spite \etal 
2001).

Given the attractiveness of MC clusters, there have been surprisingly few 
studies of their integrated spectra. 
The rationale behind early integrated light studies was to calibrate cluster age
under the photometric classification scheme of Searle, Wilkinson, \& Bagnuolo 
1980; hereafter SWB). In their landmark paper, SWB used two reddening-independent
parameters created from integrated Gunn photometry to show that MC clusters fall
along a one-dimensional sequence in this parameter space.  SWB divided the sequence
into seven photometric classification groups which, due to their usefulness,
are still referred to today.  A comparison to a similar sequence for Galactic clusters
led SWB to propose that the sequence is due to the combined effects of age and
chemical composition. Soon thereafter, Rabin (1982) demonstrated the usefulness 
of combining measurements of the equivalent widths of both Balmer and
metal-line absorption features, along with population synthesis modelling of
these features, to assign ages to the SWB classes.  In addition, Cohen, Rich,
\& Persson (1984) used the International Ultraviolet Explorer ($IUE$) to
study the integrated spectra of 17 MC clusters in the ultraviolet.
More recently, age and metallicity determinations based on spectroscopic 
indices defined by the Lick group (Faber \etal 1985) have been derived for a sample 
of 14 MC clusters (de Freitas Pacheco, Barbuy, \& Idiart 1998).  However, the
most comprehensive studies to date of the behavior of MC clusters from the near 
infrared to the near ultraviolet have been carried out by Bica, Alloin, and 
collaborators (Bica \& Alloin 1986a,b; Bica, Alloin, \& Santos 1990; Bica, 
Alloin, \& Schmidt 1994).  The purpose of their studies was to use the
integrated spectra of MC clusters as direct templates for galaxy evolution 
models.  Very recently, Beasley, Hoyle, \& Sharples (2002) have obtained optical
integrated spectra for 24 MC clusters with the FLAIR multi-fiber spectrograph
on the UK Schmidt telescope, covering a large range in cluster ages and
metallicities.

The goal of this study is to develop and test integrated light indicators which 
accurately detect and characterize the age and metallicity of a YSP.
Specifically, we characterize the behavior of diagnostic spectral 
absorption line indices generated from evolutionary synthesis models employing 
a combined empirical and synthetic spectral stellar library.  The age and 
metallicity dependence of the spectral indicators in the synthesis models are 
then compared to observed indices in the integrated spectra of a sample of MC 
clusters.  This paper represents an 
extension of the use of spectral indicators originally modeled in 
Leonardi \& Rose (1996, hereafter LR96) and further refined in Leonardi \&
Worthey (2000).  In LR96 we demonstrated how the degeneracy between age and
burst strength can be lifted, assuming solar composition populations, in age
dating post-starburst galaxies in the Coma cluster of galaxies. 
In Leonardi \& Worthey (2000) a preliminary approach was made in regard to the
age-metallicity degeneracy in detecting a young, metal-rich population in the 
candidate merger galaxy NGC~5018 .  
In this paper we present a detailed analysis of the
effects of age versus metallicity on the integrated indices of starburst models,
and compare them to the observed indices of MC clusters of known age and [Fe/H].
In \S2, we address the MC cluster selection criteria, and our sources for their
ages and [Fe/H].  In \S3 we summarize the spectral indicators used in this
study.  The details of the empirical and synthetic spectral libraries are
given in \S4, while the evolutionary synthesis models are described in \S5.
In \S6 we discuss the determination of ages and metallicities of MC clusters
from our integrated spectra and models, and in \S7 we briefly consider our
results in regard to the history of cluster formation in the LMC.
 
\section{The MC Cluster Sample} 
 
When LR96 investigated the degeneracy between young, weak starbursts and old, 
strong starbursts that exists when typical optical indicators are analyzed
(Couch \& Sharples 1987; Schweizer \& Seitzer 1992), the population synthesis
models employed were restricted to solar metallicity.  A limited metallicity 
analysis indicated that a difference of 0.4 dex in metallicity can result in a 
40\% difference 
in derived age for ages around 1 Gyr, even if the strength of the burst is held 
constant. Hence, the chemical composition of the YSP must be known before an 
accurate age can be determined.  To test diagnostic indices that can distinguish
age from metallicity effects, a grid of star clusters with a wide range in both 
age and metallicity is required.  The young, populous MC clusters span a 
significant range in metallicity within the effective age boundaries (0--3 Gyr) 
established in LR96.
 
\subsection{Selection Criteria} 
 
The indices used in LR96 and described in \S3 were demonstrated to 
be effective for YSPs younger than $\sim$2--3 Gyr (LR96). Targets were initially
selected from Sagar \& Pandey (1989), which
supplies a compilation of ages and metallicities derived from color-magnitude
diagrams in the literature.
Each candidate cluster was then viewed by eye on film copies of the
blue plate ESO/SERC Southern Sky Atlas and evaluated for
contamination from stars belonging either to the MC field or to the Galactic 
foreground.  
Compact populous clusters were primarily selected, since they present a true 
protrayal of an integrated light population; i.e., observations will suffer 
minimally from stochastic difficulties due to a few dominant stars and/or 
background subtraction problems.

The final sample consists of 28 LMC clusters and 3 SMC clusters.  Added to
the sample, for comparison purposes, are 4 Galactic globular clusters.
 
\subsection{Observations} 
 
All clusters were observed at the Cerro Tololo Inter-American  
Observatory (CTIO) over 5 nights in November 1995.  The data were obtained 
with a Cassegrain spectrograph attached to the 1.5m telescope 
at CTIO.  The detector was a Loral 1200x800 pixel CCD camera with 
a pixel size of 15 $\mu$m used with the 400 $l$/mm Grating \#58  in
2nd order.  This combination provides a reciprocal dispersion of 1.12 \AA/pixel 
and an average spectral resolution of 3.2 \AA\ (FWHM).  The 
effective spectral range is $\sim$ 3420--4760 \AA.  A slit of size 460\arcs by 
3\arcsec, with 
a fixed E-W orientation, was used for all the clusters. 

Typical exposure times were 45 minutes for each cluster.  Dome flats and 
HeAr lamps were observed for flat-field correction and wavelength 
calibration. Spectrophotometric standards LTT 377 and EG 21 were 
measured through a wide slit at the beginning and end of each night for flux calibration. 
Positions for the LMC clusters were taken from Olszewski \etal (1988) and 
Kontizas \etal (1990).  SMC cluster positions are from Welch (1991).
 
To obtain a true integrated light measurement, the slit was manually trailed  
across each cluster in the N-S direction during the exposure between specified
limits. The log of the observations, including the amount that
the slit was trailed, is given in Table~\ref{obs_tab}.

Image reductions were carried out using the IRAF package in the standard
\noindent manner: bias subtraction, flat-field correction, spectra extraction, 
wavelength
and flux calibration.  All cluster spectra are shown in Figure~\ref{repspec}.

\subsection{Adopted Ages and Metallicities}
The point of this paper is to test whether ages and metallicities derived 
directly from evolutionary synthesis models of the integrated spectral indices
of MC clusters are in accord with those determined from a ``fundamental''
analysis, i.e., based on cluster color-magnitude diagrams (CMDs) and on
individual star spectroscopy.
Thus to carry out such a comparison, we require a set of reliably known ages and
metallicities for the MC clusters.  To minimize the possible sources of 
discrepancies, we desire the
literature values that we adopt to be as homogeneous as possible.
Though there has been a long history of analyzing MC clusters,
few large studies of multiple clusters have been undertaken.  
Both observational or theoretical
difficulties in the techniques have given rise to disparate estimates of both
the ages and metallicities and so the field suffers from a lack of consistency.
Attempts to compile the various measurements
in single bibliographies (e.g., Sagar \& Pandey 1989; {Seggewiss \& Richtler
1989) have proven useful, but compilations in and of themselves
do not solve the discrepancies. Therefore, the selection of literature sources
must be undertaken carefully. The rationale for the adopted ages and
metallicities of the MC cluster sample is described below.

\subsubsection{Cluster Ages}

Ideally, an age comparison set would consist of a large sample of MC cluster 
ages derived from primary age indicators, which is defined here as an indicator 
developed from the analysis of star clusters in color-magnitude diagrams 
(CMDs).  However, the time-consuming effort needed to assemble a CMD,
as well as the difficulty for ground-based photometry to reach the
V$\sim$23 mag level needed to detect the MS turnoff in the older MC clusters, 
has prohibited any single study from including more than a few clusters.  
Instead, a major effort has gone into assembling
compilations of ages based on MS turnoffs (Hodge 1981, 1982; Sagar \&
Pandey 1989).
Though they are useful as a starting point, the lack of uniformity in MS turnoff
ages necessitates their careful use as a comparison set.
Since homogeneity is the fundamentally desired characteristic of our comparison 
set, we instead turn to secondary age
indicators, based on the integrated properties of the clusters, that have 
been calibrated against a primary age indicator.  
In particular, a secondary age indicator is desired 
that is easily measurable over a large sample of clusters,
is robust to stochastic sampling effects, and is applicable over a wide
range of ages.  Optical colors satisfy all these criteria, and the available 
data for MC clusters is extensive. Frenk \& Fall (1982) showed that
a parallel for the SWB diagram exists in UBV photometry.  It is equivalent to the
SWB sequence with the same class divisions when (U-B) color is plotted versus (B-V)
color. Elson \& Fall (1985) parameterized the sequence by drawing a smooth curve
through the clusters, dividing the curve into equal increments, and assigning a 
value to each cluster by projecting it onto the curve: their \emph{s} parameter.
The final calibration of \emph{s} with age is based solely on ages derived from MS
turnoff determinations (Elson \& Fall 1988).  Subsequently, the \emph{s}
parameter was refined further (Girardi \etal 1995; Girardi \& Bertelli 1998)
by more precisely assigning the \emph{s} value to a given cluster, accounting
for color dispersion effects cause by stochastic fluctuations in younger clusters
and metallicity differences in older clusters.  The modifications improved the
treatment of older clusters as well as outliers in the color-color diagram. With the
great increase in reliable integrated UBV photometry for LMC clusters 
(Bica, Clari\'{a}, \& Dottori 1992; Bica \etal 1996), the set of ages based on the
\emph{s} parameter has become the largest and most homogeneous sample of LMC cluster
ages in the literature.  We choose this set to be the basis of age comparison for 
this paper.  Ages of the clusters are taken preferentially from Girardi \etal (1995)
and Girardi \& Bertelli (1998), and then from Elson \& Fall (1988).
Adopted ages are listed in Table~\ref{litagemet_tab}.

\subsubsection{Cluster Metallicities} \label{sect:clustmet}
 
To this date, studies measuring homogeneous metallicities for a large number  
of MC clusters have been sparse.  In fact, the only one with a significant  
sample of clusters is that of Olszewski \etal (1991).  In that 
work,  equivalent widths of the infrared calcium triplet 
($\lambda \sim$ 8500~\AA) of one or two stars per cluster are used 
to derive metallicities for $\sim$80 clusters.  
In this paper, Olszewski \etal (1991) is the adopted reference for
MC cluster metallicities whenever possible.  For other clusters, metallicities 
are taken from the two major compilations of MC cluster data 
(Sagar \& Pandey 1989; Seggewiss \& Richtler 1989),
as noted.  
 
\section{Diagnostic Spectral Indices} \label{sect:indices}

\subsection{Strategy for Defining Spectral Indices}

It is now widely acknowledged that the cleanest way to decouple age from 
metallicity effects in the integrated spectra of stellar populations is to
combine the measurement of a Balmer line strength, which responds primarily
to the temperature (hence, age) of the MS turnoff, with a metal line strength,
which is primarily sensitive to the temperature (hence, metallicity) of the 
giant branch (e.g., Rabin 1982; Worthey 1994; Buzzoni, Montegazza, \& Gariboldi 1994).
In addition, there is advantage to using the higher order Balmer lines 
(H$\gamma$, H$\delta$, etc.) in order to maximize the contribution of the
MS turnoff stars to the integrated light (relative to the contribution of the
GB) and also to minimize any contamination from emission lines.  However, to
obtain clean measurements of the strengths of these spectral features is
challenging, due to line crowding, especially in the vicinity of the higher 
order Balmer lines.  Here we briefly consider different strategies for 
measuring spectral features before describing the indices to be used in this
study.

A particular difficulty for defining spectral indices in the case of 
post-starburst populations is the great range in spectral types that they cover.
The problem is that one is potentially looking at the superposition of two 
fundamentally different spectra, a young population, as illustrated by
NGC 1777 in Fig. 1(d), and an old population, as illustrated by 47 Tuc in
Fig. 1(h).  Naturally, it is important to use spectral indicators that can track
the behavior of the desired features over this great range in spectral type.
Examination of NGC 1777 reveals that to measure the total
equivalent width (EW) of a Balmer line in this spectrum requires using 
continuum sidebands that are widely separated from the line center, due to the
extensive Balmer line wings.  In contrast, in the spectrum of 47 Tuc the
Balmer line wings are lost amid the proliferation of metal lines that
dominate this cooler composite population.  An additional problem is that where
the Balmer lines dominate in the young populations, the metal lines are very
weak.  Hence, where the Balmer lines best lend themselves to definition through
wide bandpass EW indicators, the metal lines are so weak as to be difficult to
characterize in that manner.  In addition, defining the appropriate sidebands
in the case of the Ca II H and K lines, which we will see are highly useful as
age/metallicity indicators, is difficult due to the steep gradient in the
pseudocontinuum across these features.

In optimizing the definition of indices it is
important to keep in mind that an EW index will be independent of spectral
resolution (an important advantage) if the sidebands are located out beyond the
line wings and the line bandpass includes the entire line wings.  If, however,
this criterion is not met, then the EW index can be highly sensitive
to line broadening (either instrumental or Doppler), as the flux deficit in the line is
redistributed out of the line bandpass and into the sidebands with increasing
broadening. Thus the dilemma is that for the older populations one would like
to narrow the line definition as much as possible, to isolate the contribution
of the particular feature from the increasingly confusing contribution of other
features, a problem that afflicts not only the line bandpass but the sidebands
as well.  However, this strategy has the drawback of making the EW very 
sensitive to spectral resolution, and in addition leads to a serious
underestimate of the true Balmer line strengths in the young populations, where
much of the EW is in the extended line wings.  On the other hand, using broad
bandpasses introduces a great deal of contamination from other features in the
case of the older populations, and is not well-suited to the weak metal lines
in the younger populations.

An alternative strategy is to focus on obtaining a measure of the central line
depths, where the particular feature of interest achieves its greatest
contrast with other ``contaminating'' features.  An approach
formulated by Rose (1984, 1985, 1994) is to leverage a Balmer line center
against a neighboring metal feature.  Specifically, an index is defined by
taking the ratio of the counts in the bottoms of two neighboring spectral 
absorption features, without reference to the continuum levels.  The 
disadvantages of this approach are twofold.  First, the index loses sensitivity 
if one of the features becomes saturated in the line core, and also becomes more
difficult to measure, due to the lower S/N ratio in the line center with
increasing line depth.  Second, all indices represent relative
measurements of one feature against another, while a true EW measurement
of a single feature is ideal for resolving the degeneracy between
age and metallicity.  A key advantage of line ratio indices is in maximizing the
contribution of the particular spectral lines in the face of confusion from
other features, by working in the deepest part of the lines.  In addition, the
index is defined without requiring knowledge of the often problematic location
of the (pseudo)-continuum.  Because
neighboring absorption features are used, the measured indices are virtually
insensitive to reddening and to errors in spectrophotometry.  Moreover,
since both line centers tend to weaken equally with decreasing spectral 
resolution, the indices are only mildly sensitive to spectral broadening.
To summarize, while we make extensive use of line ratio
indices in this paper, there are indeed other means of characterizing the same 
key features that we utilize.

\subsection{Line Ratio Indices} \label{sect:hdcaind}

The age-dating technique presented in this paper is primarily based on two
of the spectral indices developed and calibrated in Rose (1984, 1985) and
explicitly modeled in LR96 and in Leonardi \& Worthey (2000).  The specific
features used, along with the Balmer discontinuity index passbands (cf. \S3.3),
are identified in the synthetic spectrum of a star (generated from the Kurucz 
(1993) SYNTHE code, as described in \S4.2) in Fig. \ref{features}.
The first spectral index, hereafter referred to as the H$\delta$/$\lambda$4045 
index, is calculated by taking the ratio of the residual central intensity in 
H$\delta$ relative to that of the neighboring Fe I $\lambda$4045 line.  The
absorption line indices are calculated by finding the minimum pixel value in a
small range around the line center for each of the two lines and then forming 
the appropriate ratio.  Due 
to the manner in which the index is defined, it \emph{decreases} in value as 
one proceeds from late type stars to earlier type stars, i.e., as H$\delta$ 
strengthens and Fe I $\lambda$4045 weakens.  It is essentially a measure of 
the spectral type of a star (or of the integrated spectral type of a stellar 
population).  The H$\delta$/$\lambda$4045 index decreases smoothly from spectral
type K through A0, where it reaches a minimum value.  For spectral types 
earlier than A0, H$\delta$/$\lambda$4045 increases again, due to the weakening 
of the Balmer lines in B and O stars and the virtual disappearance of Fe I 
$\lambda$4045.  The integrated spectral type as measured by the 
H$\delta$/$\lambda$4045 index will serve as the baseline for the age/metallicity
determination method. 
 
The second index, formed from the ratio of the Ca II H+H$\epsilon$ line relative
to Ca II K, and hereafter referred to as the Ca II index, behaves quite 
differently, due to the fact that Ca II H+H$\epsilon$ is dominated by Ca II H
for cooler stars and by H$\epsilon$ for hotter stars.  Due to saturation of the
Ca II H and K line cores, it has a constant value for stars with a spectral 
type later than about F5.  This 
constant value equals $\sim$1.2 and $\sim$1.4 for dwarf stars and giant stars, 
respectively.  The behavior of Ca II, when plotted against 
H$\delta$/$\lambda$4045 (as a surrogate for spectral type) is illustrated in 
Fig. \ref{CaII-stars} for a sample of 684 stars covering a wide region of
T$_{eff}$, log g , and [Fe/H].  There it can be seen that for H$\delta$/$\lambda$4045 
$\ge$ 0.7 (approximately spectral type F5), the Ca II indices are flat for all
later type dwarfs and giants. For earlier type stars, however, the index 
decreases dramatically as the Ca II lines weaken and H$\epsilon$ strengthens.  
As with H$\delta$/$\lambda$4045, it reaches
a minimum at spectral type A0 and then increases towards earlier spectral 
types as H$\epsilon$ fades at higher temperatures. 
The constant value of Ca II in late F-K stars and the sudden
drop at earlier types provides an unambiguous signature for stars hotter than F5
in the integrated light of a stellar population if the index value falls below 
the constant value for cool stars.  Since the MS turnoff is likely to be the 
major contributor to the integrated light in the visible part of the spectrum, 
Ca II can in principle be used as a YSP age indicator up to the MS lifetime of 
an F5 star, or approximately 4 Gyr. In addition, the T$_{eff}$ at which the
Ca II index begins its sharp drop is a function of [Fe/H].  This is illustrated
for dwarf stars in Fig.~\ref{caiisampdw}, where the Ca II index, as derived 
from a large sample of synthetic stellar spectra (described in \S\ref{sect:synthe}), is 
plotted versus T$_{eff}$.  For the low metallicity stars, the drop in Ca II
begins already at log(T$_{eff}$)$\sim$3.7 (T$_{eff}$$\sim$5000 K), while for 
metal-rich stars the drop begins close to log(T$_{eff}$)$\sim$3.85 
(T$_{eff}$$\sim$7000 K).  Consequently, the detailed behavior of the integrated
Ca II index has a metallicity-dependent component.  Finally, it should be
noted that
in active star-forming regions, the behavior described above is no longer valid 
because the H$\epsilon$ and H$\delta$ absorption lines can be contaminated by emission fill-in.

\subsection{Balmer Discontinuity Index} \label{sect:bdind}
 
Since it was determined in LR96 that chemical composition may play a significant
role in deriving the integrated age of a post-starburst stellar population, another
index is required to fully describe the age-strength-metallicity parameter space.
The Balmer discontinuity (BD) is an effective age indicator for young star 
clusters (e.g., Bica, Alloin, \& Schmidt 1994) and will serve as the third
index.  We define a BD index as the ratio of the average flux in 
the wavelength interval $\lambda\lambda$3700--3825 \AA\ to that in the interval
$\lambda\lambda$3525--3600 \AA, following Rose, Stetson, \& Tripicco (1987).
The BD index contains a great deal of dynamic range for hot stars, evolving 
rapidly as a function of spectral type.  As in the case of the two absorption 
line ratio indices, the BD index is based on Balmer transitions (in this case
the discontinuity at $\sim$3650 \AA \ produced by the bound-free transition), 
thus it has an extreme value at spectral type A0, where the Balmer lines are 
strongest, and falls off to both earlier and later types.
The behavior of the BD index with log T$_{eff}$ is shown in 
Figures~\ref{bdsampdw} and \ref{bdsampgi}. Of particular note is the strong
increase in the BD index with decreasing gravity, i.e.,  opposite to the trend 
in the Balmer lines.  With all three indices in this paper based on 
Balmer transitions, they are partially redundant in their ability to 
distinguish stellar populations.  However, there is enough difference in
behavior between them to act effectively in combination.

\subsection{Cluster Index Errors} \label{sect:errors}

Since most clusters were observed only once, the
errors were computed by using photon statistics for each spectral feature in
the absorption feature indices. The indices used in the age-dating
procedure are calculated from cluster spectra that have been gaussian-smoothed
from their original 3.2~\AA \ (FWHM) resolution, with a $\sigma$ of 2.0 \AA.
Hence, the number of counts used to formulate the photon statistics were
calculated by performing a weighted sum of the counts in the central pixel
and four pixels on either side for each of the index line feature components in
the raw cluster spectra.
A normalized gaussian with the same width as the smoothing gaussian was used
as the weighting function.  The number of counts resulting from the sum is
equivalent to the counts in the central pixel of the line feature in the
cluster spectrum after smoothing.
This value was combined in quadrature with the read noise to obtain
the total error in the measurement for the residual central line intensity
for each absorption feature in the smoothed spectra.
Errors for the BD index bandpasses were calculated by taking the square root of 
the average number of counts in each bandpass and dividing by the width of the
bandpass.  The effects of smoothing on the BD index errors and on the index
itself were deemed negligible.

The above errors are based solely on photon statistics.  Systematic errors may
be present as well.  In fact, as will be seen in Section~\ref{sect:results},
there is a discrepancy of $\sim$ 0.1 between the BD index values predicted by 
the models described in Section~\ref{sect:models} with the observed values of 
many of the clusters.  The possibility that the discrepancy could lie in the
evolutionary synthesis models and/or the synthetic spectra used as inputs for
those models will be considered later.  Here we evaluate the likelihood of
systematic errors in the observations, specifically in background subtraction
and in spectrophotometry.  

In regard to improper background subtraction affecting the observed BD indices,
the background measurements were taken from the
edges of the slit in the cluster observations, rather than from separate sky
exposures, and the average
signal-to-background ratio at 3600 \AA \ is $\sim$ 2.1, ranging from a minimum 
of 0.39 for NGC~2190 to a maximum of 4.9 for NGC~416.  These values are 
certainly low enough
to render errors in the background subtraction a real possibility.  However, the
likelihood that background errors play an important role is seriously undercut 
by noting the location of the BD index for 47 Tuc (see Figure~\ref{bdclustm07}).
The BD index for 47 Tuc is $\sim$ 0.06 directly below its proper age point for
its accepted metallicity of [Fe/H] = -0.7, yet the cluster-to-background ratio 
is 9.5.  If background subtraction problems were the principle issue, one
would expect to see an inverse correlation between the deviation of models from data
and the the cluster-to-background ratio.  However, 
47 Tuc presents a striking counterexample to such a hypothesis.

Errors in spectrophotometry could also account for the outlying clusters.  The main
emphasis of the cluster observations was on obtaining the absorption feature 
indices, hence only two flux calibration stars were observed at the 
beginning and end of each night.  With the outlying clusters departing from
the models by only $\sim$ 0.1 in the BD index, an error of $\sim$10\% in the
relative spectrophotometry across the wavelength region $\lambda$3500 -- 
$\lambda$3800 \AA \ can account for the discrepancy.  Achieving 10\% accuracy
in spectrophotometry at $\lambda$3500 \AA \ is ordinarily quite challenging,
requiring more than a couple of spectrophotometric measurements.  On the other
hand, the principal problem for accurate spectrophotometry in the blue, 
differential atmospheric refraction, is probably not an issue for our
observations.  While the slit was not rotated to the parallactic angle, in the
case of the spectrophotometric standards we used a wide slit, and in the case
of the clusters they are all extended objects, {\em and the spectrograph slit
was trailed at a uniform rate over the core diameter of the cluster},
so that the atmospheric 
dispersion effect is avoided.  Another potential problem for our
spectrophotometry is that we used the standard CTIO extinction curve, thus
leaving us vulnerable to possible systematic differences in extinction during 
our run, when compared with the standard curve.  We note that atmospheric 
extinction tends to be particularly variable in the near-UV.

We do have a limited number of stars observed
which have reasonably well-defined atmospheric parameters.  These stars, 
however, were observed through the narrow slit and not at the parallactic angle,
thus are subject to all of the uncertainties regarding atmospheric dispersion.
Overall, due to the steep dependence of the BD index on both T$_{eff}$ and
log g, coupled with the uncertainties in the fundamental 
determinations of these parameters for the limited number of observed stars,
and with the possibility of spectrophotometric errors introduced by
atmospheric dispersion, we cannot determine at the 10\% level whether the 
observed stars agree with the synthetic spectrum values or not.  One additional
handle on the situation is that if spectrophotometric errors are involved, one
might expect that the errors would be particularly severe on one night.  
However, the clusters with the discrepancy in BD index between observations and
models were observed over several nights.   

In conclusion, we cannot rule out systematic errors in the BD indices of the
clusters at the 10\% level.  However, the above discussion leads us to suspect 
that discrepancies between data and models at that level are more
likely due to the models themselves rather than to observational errors in the
cluster indices.

\section{Spectral Library} \label{sect:lib}

Modeling stellar populations requires building up the integrated properties of
the population from the individual, well-determined properties of its 
components.  To do so, an
extensive spectral library (for integrated spectroscopic properties) is needed
as building blocks for the population models.  The ideal library is
composed of high signal-to-noise (S/N) ratio, high-resolution stellar spectra,
obtained with a single stable telescope/spectrograph/detector combination, 
which 
fully sample all possible stellar atmospheric parameters and stages of stellar
evolution. In reality, due to the limitations imposed by the particular star
formation and chemical enrichment history of our Galaxy, certain areas of the
parameter space are not represented in the Solar Neighborhood.  In particular,
for the young systems we are interested in, the hot stars in the spectral 
library must be well represented across a large metallicity range.  Given the
lack of hot, metal-poor, high-gravity stars in the Galaxy, we cannot hope to
generate adequate coverage of YSPs using a purely empirical spectral library.
On the other hand, as is further discussed in \S\ref{sect:interface}, correctly
reproducing the detailed absorption spectra in cool stars is challenging, and
so the use of empirical spectra is favored.  Thus, in modelling YSPs of all
chemical compositions, we found it is necesssary to assemble a spectral library
containing both empirical stellar spectra and synthetic spectra.  

\subsection{Empirical Spectra} \label{sect:empspec}

The W94 models utilize polynomial functions derived from fits to stellar
spectral features in the spectral library.  The fitting functions express how the
stellar features vary as a function of effective temperature, gravity, and
metallicity. Consequently, they
require spectra of stars with well-determined atmospheric parameters covering a 
wide range of the parameter space.  For our 
purposes, we also require high resolution spectra to accurately measure the 
absorption line indices.  Such a library was compiled by 
Jones (1999; see also Leitherer \etal 1996) at the KPNO Coud\'{e} Feed 
telescope.
The spectra are available from the NOAO FTP archive at 
ftp://ftp.noao.edu/catalogs/coudelib/, and this homogeneous library of
stellar spectra is hereafter referred to as the Coud\'{e} Feed Spectral
Library (CFSL). The CFSL consists of two spectra, a red one and a blue one, 
for each of 684 stars.  We limit our discussion here to only
the blue spectra.  These spectra cover the wavelength range $\lambda\lambda$ 
3820--4509 \AA\ and have a pixel sampling of 0.622807 \AA/pix with a spectral 
resolution of 1.8 \AA\ FWHM.  The library is well populated at low effective
temperatures in a large metallicity range, and covering a suitable range in
surface gravity, but at higher temperatures the coverage is limited, primarily
due to the lack of hot, metal-poor stars. To insure adequate
sampling throughout the metallicity range, we restrict the effective temperature
range for the empirical library to be 4350--6300 K.  Of the stars in this range,
some were eliminated either for having chromospheric emission, which 
contaminates the absorption line indices (Rose 1984), or for having uncertain 
atmospheric parameters, which includes both undetermined parameters and 
parameters which clearly do not agree with their
measured indices.  In all, 523 CFSL stars are included in the empirical spectral
library used for the models. 

To produce the fitting functions, we employ a fitting program kindly provided by
G. Worthey.  It uses a series of least-squares regressions to fit a third order
polynomial in $\Theta$ (= 5040 / T$_{eff}$), [Fe/H], and log g. In an iterative 
procedure, the terms in the polynomial are either included or excluded until no
systematic trends appear in the residuals of the atmospheric parameters versus 
the polynomial fit. Each absorption line
used in the spectral indices (e.g., H$\delta$) is fitted with one polynomial 
across the entire temperature range. The coefficients
for the adopted polynomials are given in Table~\ref{coeffs_tab}.  Before
the fitting procedure, the stellar spectra were normalized to unity at 4000 \AA\ and
rebinned to 1 \AA\ per pixel dispersion.  Unfortunately, the BD index could not be 
fit since the empirical spectra do not go far enough into the blue.  Hence, all
results and conclusions derived from the BD index are based \emph{solely} on the
synthetic spectral library described below.

\subsection{Synthetic Spectra} \label{sect:synthe}

The CFSL has complete metallicity coverage only for temperatures
cooler than 7000 K, due to the aforementioned lack of young, metal-poor stars
in the Galaxy.  To remedy this shortcoming, as well as to supply wavelength
coverage down to $\lambda$3500 \AA \ for the BD index, the spectral library was 
augmented with 2095 synthetic spectra.  These spectra were generated from ATLAS 
model atmospheres (Kurucz, 1994) and the SYNTHE synthetic spectrum codes 
(Kurucz 1993) that were kindly 
supplied by R. Kurucz (1995, private communication).

The SYNTHE program generates a synthetic spectrum
from a given model stellar atmosphere and a spectral line list.  The model 
atmospheres are generated from Kurucz's (1994) ATLAS program, and a packaged
set of 2095 of those atmospheres were taken from the available CD-ROM provided
by R. Kurucz.  The SYNTHE program is actually a
series of VMS Fortran programs that were converted to run on a SUN/UNIX workstation.
A stellar spectrum was computed for every available stellar atmosphere contained
on the CD-ROM of Kurucz (1994).  
For reasons of available time, ATLAS was not itself used to increase the number 
of stellar atmospheres.

The spectral line list is the same for each computed spectrum.  Rather than
individually assessing which species are important for a particular set of 
stellar atmospheric parameters, every species is included for every star.
These include Fe, CN, C$_2$, CO, SiO, and H$_2$.  The SYNTHE CD-ROM has a series
of data files that contain the line list information for the atomic and
molecular species.  Once the line list was generated, the basic parameters for 
the spectra were selected.  Specifically, the wavelength range was chosen to be
$\lambda\lambda$ 3500--5500 \AA\ and the resolution is 
$\lambda$/$\Delta$$\lambda$ = 72700.  The default microturbulent velocity of 
2.0 \kms was used for each spectrum.

The final atmospheric parameter space covered by
the grid is given in Table~\ref{synthegrid_tab}. 
The calculated spectral indices of the SYNTHE spectra vary systematically
as a function of the atmospheric parameters, and contain no observational
error. Hence, instead of applying fitting functions to the variation in spectral
indices with atmospheric parameters, we linearly
interpolate between the line strengths of the synthetic spectra to find line
strengths for a given set of atmospheric parameters between the grid points.

Although we have greatly extended our ability to model YSPs with the addition of the
synthetic spectra, the theoretical HR diagram is still not covered completely.
Specifically, the sampling suffers for very hot giant stars, which make
significant contributions to the integrated light of very young ($<$0.1 Gyr)
stellar populations, especially in the blue part 
of the spectrum.  With few library spectra, synthetic or empirical, near these 
points in atmospheric parameter space, interpolation within the grid at these
points, or extrapolation beyond it, needs to be handled with care.  

As seen in Table~\ref{synthegrid_tab}, there are no spectra with log g $<$ 2.0 for all
T$_{eff}$, log g $<$ 3.0 for T$_{eff} >$ 10500 K, and log g $<$ 4.0 for T$_{eff} >$
26000 K.  We consider temperature extrapolation of the spectral indices beyond
either of the synthetic grid absolute endpoints, T$_{eff} <$ 4000 K or T$_{eff} >$
35000 K, to be highly uncertain,  and we ignore all contributions from such 
stars.  In other cases, e.g., for hot giants, we use the following scheme.
We do not have enough spectra to bracket the isochrone points
in temperature, so we drop down in temperature to the next nearest spectrum to 
maintain the ability to
interpolate.  There is still the need to have spectra on each side of the isochrone
point in all three atmospheric parameters.  See Table~\ref{interp_tab} for an
example of this interpolation scheme.  For the given isochrone point, the first
three columns are the library spectra that would be used for the interpolation if
the synthetic grid were fully populated. The fourth column shows which spectra
are not present in the library and the last three columns show the library specta
that are actually used in the model calculations.  Since T$_{eff}$ = 10500 K is the
hottest available spectrum with log g = 2.0, it serves as the lower interpolation
in multiple instances. Its large distance from the
isochrone point in temperature space, however, causes its actual weight to be small.
It should be noted that this interpolation scheme only applies when T$_{eff}$ is
hotter than 10500 K, hence they only significantly affect the results for ages 
less than $\sim$
0.1 Gyr.  The net effect is to produce some
discontinuity in the index trajectories at the youngest ages, since 
some isochrone points
remain excluded, but the probability of observing a star cluster or galaxy at such
a young age is low.

\subsection{Empirical/SYNTHE Interface} \label{sect:interface}

The reasons for combining the synthetic and empirical libraries to serve as the
input library for the evolutionary synthesis models are compelling.  The 
advantages of the synthetic spectra are threefold.  First, this study is 
primarily concerned with
YSPs and the necessary observed stellar spectra in a wide metallicity range
do not exist.  Second, to take advantage of the age discrimination power of the
BD index, we require regions of the spectrum that the empirical CFSL spectra do not
reach.  Finally, the increase in the HR diagram sampling is quite significant.
On the other hand, keeping empirical spectra in the mix is not only satisfying 
since they represent reality, but necessary, since the synthetic spectra diverge
systematically from the CFSL spectra at low temperatures.
In Figure~\ref{synemp}, we plot the index values for both the 
synthetic and empirical stellar spectra at solar metallicity.  The curves 
represent the synthetic spectra with [Fe/H] = 0.0, while the symbols 
represent the empirical stars in a small metallicity range around solar metallicity.
In each set, the spectra are grouped into gravity subsets.  Overall, the agreement
between the two libraries is quite good.  There are two overt differences, however.
First, the Ca II index for the dwarf empirical stars (square symbols)
has a flatter slope than the synthetic curves for lower temperatures 
(H$\delta$/$\lambda$4045 $>$ 1.0).  Second, the empirical giant stars (triangle
symbols) have a wider range of Ca II index values for a given 
H$\delta$/$\lambda$4045 index at cooler temperatures (H$\delta$/$\lambda$4045 $\sim$
1.2) than the synthetic giants.  At higher temperatures (H$\delta$/$\lambda$4045 $<$
1.0), the libraries track very well though there is a lack of hot giant stars. 
Similar behavior is seen at other metallicities
as well, but there are significantly fewer empirical stars with which to make
comparisons.  Divergence at lower temperatures is not entirely unexpected since the
sheer abundance of metallic lines in real stars makes their spectra very
complex.  Including, with accuracy, every spectral feature of every species is 
extremely difficult.

To interface the two libraries, we chose a simple effective temperature cutoff.
If the temperature prescribed by an isochrone is higher than 6000 K, synthetic spectra
are used in the calculations.  If lower, the empirical fits are used.  The cutoff
temperature applies for all metallicities and gravities.  Recall, however, that for
the BD index, synthetic spectra are used at all effective temperatures.
The interface is shown in Figure~\ref{cutoff}.  This is the same plot
as Figure~\ref{synemp}, but the synthetic curves have been truncated at 6000 K
to illustrate where the empirical library takes over.  The cutoff temperature was
chosen to be well within the temperature range of the empirical fits to avoid any
edge effects at the boundaries of the fits.  A cutoff temperature of 5000 K was
investigated, to evaluate whether the increased continuity of the synthetic
spectra would improve the results in the low temperature region.  The differences
were not significant, hence a cutoff of 6000 K was retained to maximize the use of
empirical stars.  At this temperature, the metallicity sampling of the empirical
library is good througout the entire metallicity range.  No steps were taken to
induce continuity between the libraries at the interface.

\section{Evolutionary Synthesis Models} \label{sect:models}

LR96 employed the evolutionary synthesis models of 
Bruzual \& Charlot 1993; hereafter BC93).
Unfortunately, both the original BC93 spectral library and the higher resolution
library of Jacoby, Hunter, \& Christian (1984) which replaced it in LR96, 
consist of solar
metallicity stars only.  Because chemical composition may have a 
significant impact on the derived age of a YSP, the present study utilizes the
models of W94.  In the W94 models, the isochrones of
Bertelli \etal (1994; hereafter referred to as the Padova isochrones),
which cover a range in metallicity, form the basis of the
evolutionary synthesis approach.

The Worthey models employ empirical polynomial functions derived from fits to stellar
index values for a library of stellar spectra.  The fitting functions express how
the stellar index strength varies as a function of effective temperature, gravity,
and metallicity.  From a Padova isochrone of a given age and metallicity and an 
assumed IMF, the models calculate the number of stars at each given point on the
isochrone.  Weighting the index value given by the fitting functions at that 
isochrone point by the luminosity and the number of stars for each point and then
summing along the entire isochrone gives the index value for the total stellar
population (Worthey \& Ottaviani 1997).  See Charlot, Worthey, \& Bressan (1996)
for how this methodology differs from other evolutionary
synthesis models and the sources of discrepancies between different model sets.
 
\subsection{Resolution Matching} \label{sect:resol}

Spectral broadening, due to both instrumental resolution and to stellar velocity
broadening, redistribute the flux deficit out of the line center into the wings.
Although the line ratio indices used here are fairly insensitive to
spectral broadening, since
both features which make up an index should be equally affected by broadening,
it is still necessary to match the spectral resolutions of the
model inputs and the observational data as closely as possible.

The broadening of the MC cluster spectra relative to the CFSL spectra used in
the evolutionary synthesis models is determined as follows.  First, since all
MC cluster spectra are from a single observing run and all clusters have a 
similar low internal velocity dispersion, we use the very high S/N ratio 
integrated spectrum of the cluster 47 Tuc as our reference, and compare it
with the CFSL G2 stellar template star HD10307.  This latter template star is
broadened with a variety of gaussian dispersions, and then cross-correlated
against itself using the IRAF task {\em fxcor}, and a fit made to the width of
the cross-correlation peak versus the $\sigma$ of the smoothing.  Then from a
cross-correlation of the 47 Tuc spectrum with that of HD10307, it is determined
that a broadening of $\sigma$=1.17 \AA \ is required to match the CFSL spectra
to that of the MC clusters.  Similarly, it is determined that the spectra in
the synthetic library need to be broadened by $\sigma$ = 0.78 \AA \ to match
the MC cluster spectra.  Recalling that the MC cluster spectra are broadened
with a $\sigma$ of 2.0 \AA \ before the spectral indices are measured, we 
apply the additional 2 \AA \ broadening to both the CFSL and SYNTHE spectra
as well. The indices measured from clusters and stars can now be intercompared.

\subsection{Calculation of Integrated Line Features} \label{sect:intind}

Since the diagnostic spectral indices involve quotients of
the central intensities of pairs of absorption lines, taking linear combinations
of the line ratios found in individual stars in building a model integrated stellar population 
will not produce correct results for the integrated 
indices.  Instead, to model the integrated spectral 
indices it is necessary to determine the 
individual residual central line intensities in the integrated model
population before the final indices are
formed.  The W94 models use the input spectral libraries in the form
of the fitting functions for the empirical data and an interpolation grid
for the synthetic spectra.  In each case, the data used are the residual 
central line intensities of the library spectra, where the library spectra have
been normalized to unity at 4000 \AA.
The central line intensities are measured by finding the minimum pixel value in
a small range centered on the absorption line of interest.  Spectral
features H$\delta$, Ca II H + H$\epsilon$, Ca II K, and Fe I $\lambda$4045
are measured for each star in the smoothed versions of the empirical and 
synthetic libraries.  Polynomials are fit to the individual empirical features,
as discussed in Section~\ref{sect:empspec}.  Additionally, for each synthetic
spectrum, the average flux in the two wavebands for the BD index is
calculated.   To produce the integrated indices for a stellar population of a
given age and metallicity, the W94 models calculate the integrated line features
from the Padova isochrone of the desired age and [Fe/H].  The isochrones specify
the number of stars with a given set of atmospheric parameters present in the
theoretical HR diagram.  Weighting by the number and luminosity of stars of each
type, an integrated continuum for the population is developed.  By also adding 
up the normalized line features weighted by the number of stars and applying
the continuum, the total integrated line fluxes are calculated (see W94 for
details).

The MC clusters are already integrated
stellar systems, hence their spectral indices are measured directly.  The 
cluster indices, and their errors, are given in Table~\ref{clustind_tab}.

\section{Application of Age Determination Method}\label{sect:results}

\subsection{Integrated Spectral Indices for Simple Stellar Populations}\label{sect:traj}

Utilizing the above modelling procedures we
form the integrated spectral indices for stellar populations covering a variety
of age and metallicity.  In this manner we can track the time evolution of the indices for
different chemical compositions. The observed cluster indices can be compared 
with the model indices to determine the ages and metallicities of the MC
star clusters from their integrated spectra.   Those ages and metallicities can
then be checked for consistency with the more primary ages and metallicities
derived for the clusters from their CMDs and from spectroscopy of individual
cluster stars.  In this Section we first present the evolution of 
spectral indices as a function of age and chemical composition in two diagnostic
diagrams, then discuss how we determine the cluster ages from the diagrams,
and finally compare the derived ages to those found in the literature.

To begin with, we plot the Ca II index against the
H$\delta$/$\lambda$4045 index as a function of age to form a trajectory in
this two-index space.  A trajectory is computed for each desired metallicity. By
overplotting the observed MC cluster indices, the metallicity of a cluster can be determined
from the relative location of the cluster point with respect to the
trajectories, and its age determined by where it falls along a particular track. The process is
repeated for the BD index resulting in two separate determinations.
The trajectories are illustrated in Figure~\ref{cabig}, where the Ca II index 
is plotted against the 
H$\delta$/$\lambda$4045 index for integrated stellar populations at six 
different metallicities. The indices were computed at the same spectral 
resolution as the observed MC cluster spectra following the prescription in
Section \ref{sect:resol}.  A similar plot for the
BD index is given in Figure~\ref{bdbig}. The upper left panel in each figure
shows the evolution for [Fe/H] = -0.4, and this trajectory is plotted in all
the succeeding panels for comparison purposes.
The symbols represent the index values for
an integrated stellar population of the given metallicity at different
ages, marked in Gyr. The trajectories show the path that a cluster travels 
in the index space as it grows older. The need to model metallicity, which 
prompted the change to the W94 models from the BC93 models used in LR96,
is apparent in the figures. Without adequate
account for metallicity, an old, metal-poor cluster could be mistaken for a 
significantly younger cluster at solar metallicity. 
Although the general behavior of the indices 
is very similar for different metallicities, the absorption feature index 
evolution is compressed
as metallicity decreases. Since the BD index is relatively insensitive to
metallicity, the compression is less drastic in the BD index
tracks, with most of the effect arising from the H$\delta$/$\lambda$4045 
index.

Some significant features of the figures should
be noted. For the most metal-poor models, there
is a degeneracy in the Ca II index values for different ages.
It is particularly evident for the [Fe/H] = -1.7 trajectory in 
Figure~\ref{cabig}. The youngest clusters (age = 0--0.4 Gyr) have the same
Ca II and H$\delta$/$\lambda$4045 indices as those of the intermediate 
clusters (age = 0.5--5.0 Gyr). It is primarily this degeneracy which 
necessitates the inclusion of the BD index.  As can be seen in 
Figure~\ref{bdbig}, the BD index significantly separates the youngest
clusters from the intermediate age clusters and enables simultaneous age and
metallicity determination.  

There is another degeneracy in the figures which affects the youngest clusters. 
The sections of the trajectories from an age of 0.004 Gyr to the ``hook'' at 
$\sim$ 0.4 Gyr (for [Fe/H]=-0.4) are superimposed for all metallicities.  This behavior occurs for
both the Ca II index and the BD index, thereby rendering the determination of
metallicities for these youngest populations essentially impossible.  The
derived age is also affected since the timescale from 0 Gyr to the ``hook'' 
depends on the metallicity of the cluster. 

The observed index values for the MC clusters listed in Table~\ref{clustind_tab}
are plotted on the model 
trajectories in Figures~\ref{caclustm17}--\ref{bdclustp00b}
(note the expanded scales in
Figures~\ref{caclustm17}, \ref{caclustm13}, and \ref{bdclustm17}).  
The clusters are
assigned to the figures based on their literature metallicities given in
Table~\ref{litagemet_tab}.  Since no clusters have a derived [Fe/H] $\sim$ +0.4,
that plot was omitted.  The error bars were calculated following the 
procedure described in \S\ref{sect:errors}. It is apparent in the figures 
that the clusters generally fall along the index tracks, especially in the Ca II
versus H$\delta$/$\lambda$4045 diagrams.  This agreement provides encouraging
indication that the model trajectories are in accordance with reality. 
However, there are some clusters in the BD index plots which do not
coincide with the model predictions.  The most glaring examples
are in Figures~\ref{bdclustp00a} and \ref{bdclustp00b}.  As described in 
\S\ref{sect:errors}, spectrophotometric and/or background subtraction errors could be
responsible for the $\sim$0.1 discrepancy between modelled and observed BD
indices.  However, we note that the discrepancy appears to be correlated with 
metallicity, such that the more metal-rich clusters in general are the ones 
with the largest discrepancies between models and observations.  Thus we 
speculate that the problem is due to deficiencies in either the isochrones or
the synthetic spectra for the more metal-rich systems.  

\subsection{Age and Metallicity Determination}

Ages and metallicities for the MC clusters have been derived in the following 
manner.  As is evident from the Figures, model trajectories have been computed
at six metallicities, in both the Ca II versus H$\delta$/$\lambda$4045 and the
BD versus H$\delta$/$\lambda$4045 diagrams.  For a given metallicity trajectory
in the Ca II diagram, at each computed model age point a test is made to
see whether the observed indices for a particular cluster, when projected 
perpendicularly onto the model trajectory defined by the given age point and 
the next higher age point, lies between the two model ages.  If so, a distance
is computed for the line connecting the cluster indices and the projected 
position on the metallicity trajectory.  As well, an age is interpolated based 
on the relative position of the projected cluster between the two model ages.
For each metallicity trajectory, candidate ages and distances are calculated in
this way.  Then the minimum distance, and associated cluster age, is determined
for each metallicity trajectory, thereby producing six candidate ages with their
associated metallicity.  A final age and metallicity for the cluster is computed
by using a weighted average of three values.  These are the values associated
with the minimum distance among the candidates, and also the values associated
with the next higher and lower metallicities.  The weights used are simply the
inverse of the three distances.  If no candidate values have been found for a
particular metallicity, then the next available metallicity is used.  

The procedure used for computing ages and metallicities from the BD diagram is
slightly different.  Since the BD index provides
virtually no metallicity information (see Figure~\ref{bdbig}), we use the Ca II
diagram results to constrain the final metallicity.  For the BD results we 
produce candidate ages and metallicities only at the two metallicity 
trajectories that bracket the final metallicity found from the Ca II diagram.
We then take a weighted average (by inverse distance) of the minimum distance 
ages and metallicities determined from the two bracketing metallicities.  Thus
we emphasize that the BD metallicities are not independent of the Ca II
metallicities.

As mentioned in the previous section, there can be a degeneracy in the age
inferred from the Ca II diagram if a cluster is either very young or in the 
``hook'' region of the diagram.  If such a degeneracy is evident, then we use 
the location of the cluster in the BD diagram to determine whether to go with 
the younger, as opposed to the older, age possibility from the Ca II diagram.
Likewise, if the 
cluster is either very young or in the ``hook'' region of the diagram, then no
reliable metallicity information can be extracted.  
For instance, NGC~1818 in Figure~\ref{caclustm07} lies on the young
age arm of the trajectory for all six metallicities.  The BD index in
Figure~\ref{bdclustm07} provides no additional insight, since NGC~1818 lies in
a similar position for all six BD trajectories.
Fortunately, in these cases, the ages given by all the trajectories are quite
close, with a slight trend to younger ages at higher metallicity, so that
the derived age is not sensitive to the weighting of different metallicity
trajectories in the final result.

For some clusters, even though a well-determined Ca II age can be obtained,
deriving an age from the
BD index is inconclusive.  The measured value places the cluster in a region of
the diagram that is not close to \emph{any} of the model trajectories.  As is
discussed in \ref{sect:errors} and \ref{sect:traj}, it is unclear whether the
discrepancy between models and data is due to errors in the observations or in
the models, or in both.

Errors have been determined in the ages derived from the Ca II diagram and the
BD diagram, as well as in the derived metallicities, in the following manner.
We computed revised Ca II, BD, and H$\delta$/$\lambda$4045 indices by adding
the 1$\sigma$ index errors to the original indices, and computed another index
set by subtracting the 1$\sigma$ index errors.  For these revised indices we
utilized the same age and metallicity determination procedure as described
above to derive ages and metallicities.  We then used the mean absolute 
difference between the original age and these revised ages as a measure of the
$\pm$1~$\sigma$ error in the age, and similarly for the metallicities.  {\it Because
the uncertainties in age determinations are more symmetric when considered in
log age, we present all our age results in log format.}

The derived ages from the Ca II indices, and their $\pm$1 $\sigma$ errors, are 
given in columns 2 and 3 of Table~\ref{agefin_tab}, while the derived ages and
errors from the BD indices are given in columns 4 and 5.
Final ages are assigned to the cluster by averaging the
individual determinations of the Ca II and BD indices.  For the clusters whose 
BD index is inconclusive, the Ca II age is assigned as the final age.  Though
the Ca II ages are more reliable due to the metallicity discrimination inherent
in the diagrams, no specific weightings are given to one index or the other.  
It will be seen that this is a moot issue for a majority of the clusters because
the two determinations are very close in most cases.  The final derived ages are
given in column~6 of Table~\ref{agefin_tab} and the associated 1 $\sigma$ 
errors are in column~7.  Similarly, the [Fe/H] values derived from the Ca II
diagram, and associated $\pm$1 $\sigma$ errors, are listed in columns 8 and 9,
while those determined from the BD diagram are given in columns 10 and 11.
Final values for [Fe/H], taken from a straight average of those determined from
the Ca II and BD methods, are listed in column 12.  Note that the errors given
in column 13 are identical to those from the Ca II index determinations.  The
[Fe/H] values extracted from the BD diagram are sufficiently coupled to those
obtaind from thr Ca II diagram that we do not consider the errors to be
lowered by averaging the two results.
In the next section various clusters are discussed on an individual basis.

\subsection{Notes on Individual Clusters} \label{sect:clustnote}

The following clusters have an observed BD index that does not allow a 
conclusive age determination:  \emph{NGC 1651}, \emph{NGC 1751}, \emph{NGC 1777},
\emph{NGC 1795}, \emph{NGC 1978}, \emph{NGC 2155}, \emph{NGC 2193}, \emph{NGC
2203}, \emph{NGC 411}, and \emph{NGC 416}

The following clusters are in 
the ``hook'' region for the Ca II index but the
BD index clearly selects a young age so the young solution for Ca II is chosen. No
metallicity determination is possible: \emph{NGC 2010}, \emph{NGC 2133}, 
\emph{NGC 2164}, \emph{NGC 2214}

\emph{NGC 1754} and \emph{NGC 2210}- According to the literature, these two
clusters are of Galactic globular cluster
age and very metal-poor.  The $\sim$4-6 Gyr ages derived from our models are most 
likely due to the contamination of the indices by a blue horizontal branch. 
The W94
models only treat the horizontal branch as a red clump (Leonardi \& Worthey 2000). A blue
horizontal branch will decrease the value of the Ca II index, causing the
clusters to appear to have a younger age.  As can be seen in 
Figure~\ref{caclustm17}, these two clusters reside in the same region
of the diagram as , a very old, metal-poor Galactic globular 
clusters with a blue horizontal branch.

\emph{NGC 1818} - The cluster is very young so no metallicity determinations
can be made.

\emph{NGC 1831} - This cluster potentially has another set of determinations.  It
is situated in the ``hook'' region of both the Ca II and the BD diagrams so it 
has both a younger solution (age = 0.19 Gyr and [Fe/H] = -0.50) and an older solution.
The older solution is 
selected because it has more self-consistency between the Ca II and BD age
derivations.

\emph{NGC 1846} - Determinations from the BD index are uncertain because the index
point is significantly above all the metallicity trajectories.

\emph{NGC 1866} - The Ca II (0.09 Gyr) and BD (1.0 Gyr) ages are in serious 
disagreement.

\emph{NGC 2134} - The Ca II index places the cluster in the ``hook'' region but
the BD index clearly selects an older solution.

\emph{NGC 2136} - The Ca II index has multiple solutions but the BD index indicates
a young age so the younger solution is selected. No metallicity determination
is possible.

\emph{NGC 2190} - The BD index determinations are somewhat uncertain because the
index point falls in between all the trajectories, making interpolation difficult.

\emph{NGC 2249} - The age determinations for the BD index (0.08 Gyr) and the 
Ca II index (0.8 Gyr) are
in serious disagreement.  The Ca II age is closer to the literature value of 
0.5 Gyr, as given in Girardi \etal (1995).

\subsection{Reliability of Age and Metallicity Determinations} \label{sect:analy} 

To check the self-consistency of the age determinations, a least-squares fit
was calculated for the derived Ca II ages and the BD ages.  The best fit line
is displayed in Figure~\ref{regress} for the clusters with age determinations
for both indices.  The agreement is excellent, with only NGC~1846 
and NGC~2249 showing serious deviations.  The 
calculated slope of the least-squares fit to the data is 0.82$\pm$0.10, with
an $r$ value of 0.89.  If we make a least-squares fit in \emph{linear} age, 
rather than in the \emph{log} of the age, the resulting slope is 1.01$\pm$0.07,
with an $r$ value of 0.96.
Thus the two spectral indices produce nearly the same derived age for
most of the clusters.

If we similarly compare the final age determinations with literature values, the
agreement is still encouraging.  In comparing our ages with literature data,
we exclude the two oldest clusters, NGC~1754 and NGC~2210, which have the 
aforementioned problem that our derived ages are too young, due to the lack of
a blue HB in our models.  The fit is shown in
Figure~\ref{figlogages}.  The correlation coefficient for the fit is $r$ = 
0.90, while the slope is 0.85$\pm$0.08, in the sense that our modelled ages 
show a larger age range than those from the literature.  If we make the
comparison in age, rather than in the $log$, we find a slope of 0.86, with an
$r$ value of 0.83. We have also plotted  
(as asterisks) age data for the clusters which we have in common with the
Beasley \etal (2002) integrated light study.  The Beasley \etal (2002) ages
are the average of the two values for each cluster given in their Table B1.
While there are five clusters in common between our two studies, we exclude the 
results for NGC~1754, as mentioned above.

While the agreement between our age determinations and those from the literature
is encouraging, the comparison of our [Fe/H] determinations with those from the
literature, plotted in Fig.~\ref{figmets}, is less satisfactory.  For the sake
of homogeneity, we have only taken literature [Fe/H] values from the work of
Olszewski \etal (1991), and also plotted (as asterisks) results for the five
clusters in common between us and Beasley \etal (2002).  While overall
there is some trend between our values and those in Olszewski \etal (1991), there is a
large scatter present as well as a systematic offset at higher [Fe/H].   The
large scatter ($\sim$$\pm$0.3 rms) seen at [Fe/H]$>$-0.7 (according to the
Olszewski \etal 1991 values) is actually quite consistent with the fact that
our methods for extracting age and metallicity from the integrated spectra is
more sensitive to age than to metallicity.  However, the offset between our
[Fe/H] values and those of Olszewski \etal (1991) at high [Fe/H] is more
troublesome.  Restricting the data to only those clusters with 
Olszewski \etal (1991) [Fe/H] values higher than -0.7, we find a mean offset 
of $\Delta$[Fe/H] = -0.42, in the sense that our values are systematically
lower than those of Olszewski \etal (1991).  As can be seen in 
Fig.~\ref{figmets}, there is a suggestion tha the [Fe/H] offset between
our results and those of Olszewski \etal (1991) increases with decreasing
cluster age, in that the metallicity offset for the younger clusters (designated
as open circles) appears to be less than for the older clusters (filled
triangles).  However, the result is not definitive.  No offset appears to be 
present between our metallicities and those from Beasley \etal (2002), which
are plotted as asterisks.

Given the more fundamental nature of the [Fe/H] determinations carried out
by Olszewski \etal (1991), viz., Ca II triplet spectroscopy of individual
giants, it is most likely that our abundances are systematically in error.
On the other hand, Bica \etal (1998)
noted a discrepancy of $\Delta$[Fe/H]$\sim$-0.3 between their MC cluster
metallicities (derived from the GB in their CMDs) with the work of
Olszewski \etal (1991), i.e., their discrepancy is of the same magnitude and 
direction as ours.  Moreover, Geisler \etal (2003) find a metallicity offset
with respect to Olszewski \etal (1991) in the same direction as ours, but with
only about half the size.  The Geisler \etal (2003) studies, as in the case of
Bica \etal (1998), come from characteristics of the CMDs (in the Washington
photometric system).  In addition, while the number statistics are small, our 
results show no systematic offset, with respect to Beasley \etal (2002).
Thus, while the likelihood remains that our abundances need
revision, a zeropoint problem may be present as well in the Olszewski 
\etal (1991) abundance scale.

Returning to the comparison between our ages and literature ages in 
Fig.~\ref{figlogages}, and to the fact that the slope of the relation between 
literature
ages versus our ages is significantly less than unity (0.85), it is
evident that our ages show a greater spread than the literature ages.  The most
likely explanation is that observational errors in the cluster spectral indices
give rise to correlated errors in age and metallicity.  Specifically, upon 
examining the location of NGC 1783 in Fig.~\ref{caclustm07}, it is evident that
if we assume a fixed metallicity for the cluster, e.g., [Fe/H]=-0.4, then the
$\pm$1-$\sigma$ observational errors will tend to affect the derived age by 
$\sim$$\pm$0.1 Gyr, while for an assumed fixed age, the $\pm$1-$\sigma$ errors
will affect the derived [Fe/H] by $\sim$$\pm$0.4.  The problem is that an error
in the [Fe/H] determination of 0.4 will lead to a correlated error in the age
determination of $\sim$0.6 Gyr, due to the slower evolution of the spectral
indices for a more metal-poor population.  Naturally, the specifics of the
systematic age error associated with an error in [Fe/H] is dependent
on both the age and [Fe/H] of a cluster, but the effect always works in the
direction that an error in [Fe/H], in the sense that [Fe/H] is underestimated,
leads to an associated overestimate of the age.

The previous discussion should clarify why our age determinations for the MC
clusters spread to older ages than the literature values.  The reasons are
twofold.  First, assuming that our [Fe/H] determinations are systematically
too metal-poor by -0.4 in [Fe/H], the implication is that our ages are 
accordingly overestimated, which will draw out the higher end of the age
distribution.  Second, random errors in [Fe/H] will create a corresponding
scatter in age that will broaden out the age distribution and create an
artificial age-metallicity relation (AMR) for the MC clusters.  This latter 
problem has played a
role in the long-standing controversy over the reality of an AMR in the 
Galactic Disk (McClure \& Tinsley 1976; Feltzing, Holmberg, \& Hurley 2001 and
references therein).

Based on the above discussion, we arrive at two principal conclusions:

\noindent 1) The ages determined from direct modelling of the spectral indices 
of young stellar populations in integrated light are overall in good accord 
with those determined from the more fundamental analyses of CMDs.

\noindent 2) The main impediment to obtaining high accuracy in age 
determinations is the difficulty in independently determining the metallicity
from the integrated spectrum.  Thus the age-metallicity degeneracy that
afflicts the reliable extraction of ages for older stellar populations is a
significant factor for young populations as well.

\section{History of Cluster Formation in the Magellanic Clouds}

The integrated-light age determination method described here can be a powerful tool
for examining the cluster formation history of the LMC, since the ages are all
derived from a single method and from integrated spectra of fairly uniform
quality.  By simultaneously determining
cluster ages and metallicities, the distribution of clusters in time
and chemical composition can be determined, thus revealing the complete
history of star formation and chemical enrichment.  Having verified the basic
reliability of the age/metallicity determinations from our integrated spectra,
we now look at the distribution in age and metallicity of these clusters, with
a particular view towards assessing the time dependence of the cluster formation
rate.

It is known from a variety of studies that the star formation rate 
in the LMC has not been constant over its lifetime.  From age data based 
on compilations of cluster CMDs it is now generally agreed 
that a pronounced gap appears in the LMC cluster age distribution from $\sim$3 Gyr
to $\sim$12 Gyr (Jensen, Mould, \& Reid 1988; Da Costa 1991; Van den Bergh 1991;
Geisler \etal 1997; Rich, Shara, \& Zurek 2001; Piatti et al 2002).  A second,
smaller minimum in the cluster formation rate appears to be present from 0.2 -- 
0.7 Gyr, or log \emph{t} = 8.3 -- 8.8 (Barbero \etal 
1990; van den Bergh 1991).  While the
cluster age distribution determined from integrated UBV colors initially did
not reveal age gaps or peaks, a subsequent reevaluation, incorporating
both stochastic effects and metallicity effects on the integrated colors,
does indeed produce similar age gaps as for the CMD-based age distributions
(Girardi \etal 1995).  In addition, the derived age
gap between 3 and 12 Gyr appears to be accompanied by a metallicity gap 
between [Fe/H] = -0.7 and [Fe/H] = -1.5 (Olszewski \etal 1991).  In contrast,
the cluster formation history of the SMC appears to be quite different from,
and in fact complementary to, that of the LMC, in that very few truly old SMC
clusters are found, while several are found within the large LMC age gap (Da Costa 
1991; Piatti \etal 2002).
Studies of the star formation history in the LMC field stars are generally in
accord with the cluster age distribution, in that the star formation rate
appears to have gone through a major upswing starting about 3 Gyr ago.
However, the deep lull between 3 and 12 Gyr found in the cluster formation
rate is not as evident in the field star formation rate (Butcher 1977; Geha 
\etal 1999; Olsen 1999; Holtzman \etal 1999; Harris \& Zaritsky 2001).
Thus on the whole 
the latter studies find some decoupling between field and cluster star formation
and chemical enrichment histories.

Unfortunately, the age determination method used
in this paper is not well suited to explore the intermediate age gap.  The 
spectral indices employed here are optimized to distinguish the younger
ages and lose their discriminating power for ages greater than $\sim$4
Gyr. Moreover, the MC clusters studied were selected for their younger ages and
thus do not represent an unbiased sample of MC clusters.  However, we can 
examine the age distribution of our sample of clusters for nonuniformity in
the cluster formation rate over time, and compare that to the literature data.
Hence in Figure~\ref{histnoold} we plot a histogram of the LMC cluster
ages derived in this study as well as one derived from literature values.
The width of the bins is 0.2 in log(\emph{t}) in Gyr, and again we have excluded
the two oldest clusters (NGC 1754 and NGC 2210) from the analysis for which we 
derive discrepant ages, 
due to the lack of blue HB stars in the population models.  
The peak in the age distribution derived
from the spectral indices is offset by about one age bin from the literature
distribution. 
As mentioned above, this shift probably arises from the systematically lower
metallicities that we derive for the clusters, which then cause correspondingly
older ages to be determined. Otherwise, the two
distributions agree well.  In particular, the sharp cutoff in the histogram
of the literature-derived cluster ages at $\sim$3 Gyr (log \emph{t} = 9.5)
is also seen in the age distribution
derived from our integrated light methods.  In addition, given the logarithmic
stretching of the age scale at younger ages in the histogram, it appears
that cluster formation has been more concentrated at log \emph{t} $\sim$ 8.0 
than at log \emph{t} $\sim$ 8.5.  Specifically, a pronounced depletion in 
clusters between 0.2 Gyr and 0.7 Gyr in age is readily seen when the clusters
are plotted in linear age. In fact, the two clusters with ages near 
log \emph{t} = 8.5 represent cases for which the Ca II ages and BD ages are in
disagreement with each other, and averaging the two values places the clusters 
in the intermediate log \emph{t} = 8.5 age zone.  
Thus the main features seen in the CMD based age distribution
are also seen in the age distribution determined from the integrated spectra.

We can examine the chemical evolution of the LMC by looking at the AMR
for the star clusters.  Given the short formation timescale for an indvidual
cluster compared to the LMC lifetime, the metallicity of each
cluster is a snapshot of the local chemical conditions at the time of formation.
In Figure~\ref{agemeto91} we show the AMR for all clusters which have primary
chemical abundance measurements from Olszewski \etal (1991).  The open squares
represent the literature values (with [Fe/H] coming from Olszewski \etal (1991)
and the ages coming from CMD based determinations), while the filled squares 
represent our determinations of age and [Fe/H] for the same sample of clusters.
The AMR derived from literature values is basically flat, i.e, no trend in
[Fe/H] with age, and with considerable scatter in [Fe/H] at all ages.  The
lack of a trend in the AMR is further supported by the recent Geisler \etal 
(2003).  On the 
other hand, our determinations appear to support a significant AMR slope in
the sense of steady enrichment over the past few Gyr from [Fe/H] = -1.0 up to
[Fe/H] = -0.3 at the present day.  While such an enrichment pattern is perhaps
not out of the question, we propose that it is likely to be an artifact of the
correlated errors in age and metallicity that arise from errors in the observed
spectral indices of the clusters.  Moreover, as previously discussed in \S6.4,
in Fig~\ref{figmets} we find
some indication that the discrepancy between our ages and the literature ages
increases with decreasing metallicity, which has the effect of producing a
spurious AMR.  Clearly it would be interesting to obtain
definitive [Fe/H] values for the five clusters (NGC 1651, NGC 1846, NGC 2162,
NGC 2173, and NGC 2213) which occupy the older-age, lower-[Fe/H] part of the
plot for our determinations.

\section{Conclusions}

The principal result of this paper is that ages of young ($<$4 Gyr) star 
clusters can be determined through evolutionary synthesis modelling 
of the integrated light of the clusters.  An effective discrimination of age 
from metallicity effects in most cases can be accomplished using a pair of
spectral indices that leverage the different behavior of two Balmer lines
(H$\delta$ and H$\epsilon$) against neighboring metal line features.  To
adequately distinguish young ages in populations with [Fe/H]$<$-1, an index
measuring the strength of the Balmer Discontinuity needs to be incorporated as
well.  Age-metallicity degeneracy effects remain for young ($\leq$0.4 Gyr),
metal-poor ([Fe/H]$<$-1) clusters.  However, this degenerate region represents a
relatively small portion of the entire age-metallicity parameter space.  We
base our conclusions regarding the reliability of age-dating young stellar
populations from modelling of their integrated spectra on a comparison
between synthetic spectral indices generated from evolutionary synthesis
models and observed spectral indices in MC clusters.  We find that the agreement
between models and observations is particularly good in the case of the two
line ratio indices, while the agreement is less convincing in the case of the
Balmer Discontinuity index.  There is still uncertainty as to whether the 
disagreement is due to spectrophotometric errors or to deficiencies in the
models.  However, with the emphasis shifting to working as far into the 
rest-frame blue as possible, so as to perform such analyses at appreciable
redshifts, the Balmer Discontinuity index holds considerable promise, since
observations of rest frame 3500 \AA \ are simplified at moderate redshift.

While this study has been largely successful at disentangling the effects of
age and metallicity on the integrated spectra of young star clusters, the
ultimate goal is to carry out such an analysis in the more general case where
a young post-starburst population is superposed on an underlying older 
population.  Such an analysis has a variety of applications towards 
disentangling starburst scenarios in galaxies, which is becoming an increasingly
important task for early epochs of galaxy evolution.  In this case we are faced
with the additional degeneracy of the ratio of burst light to underlying old
light versus age of the burst.  Specifically, a younger weaker (relative to the
old population) burst is difficult to distinguish from an older, stronger
burst.  Thus in the more general case of a post-starburst population superposed
on an older population there is the threefold degeneracy to consider of burst
age, burst strength (relative to the old population), and chemical
composition.  The burst age versus burst strength degeneracy can be 
readily distinguished if one can assume solar composition for the population
(LR96).  Moreover, in the particular case of NGC 5018 the threefold degeneracy
was effectively resolved (Leonardi \& Worthey 2000), but largely due to the 
fact that in this case the starburst population completely dominated the 
integrated light within the spectrograph aperture.  In short, additional work
remains to be carried out if an entirely robust method of determining 
post-starburst ages is to be achieved.

\acknowledgements
We are greatly indebted to Dr. Robert Kurucz, for making available his stellar 
atmosphere models and SYNTHE code, and to Dr. Guy Worthey, for supplying his 
population synthesis code.  We also thank the anonymous referee for careful
reviews which led to considerable improvements to
the manuscript.  This research was partially supported by NSF grants
AST-9320723 and AST-9900720 to the University of North Carolina.

\clearpage

\begin{figure}
\plotone{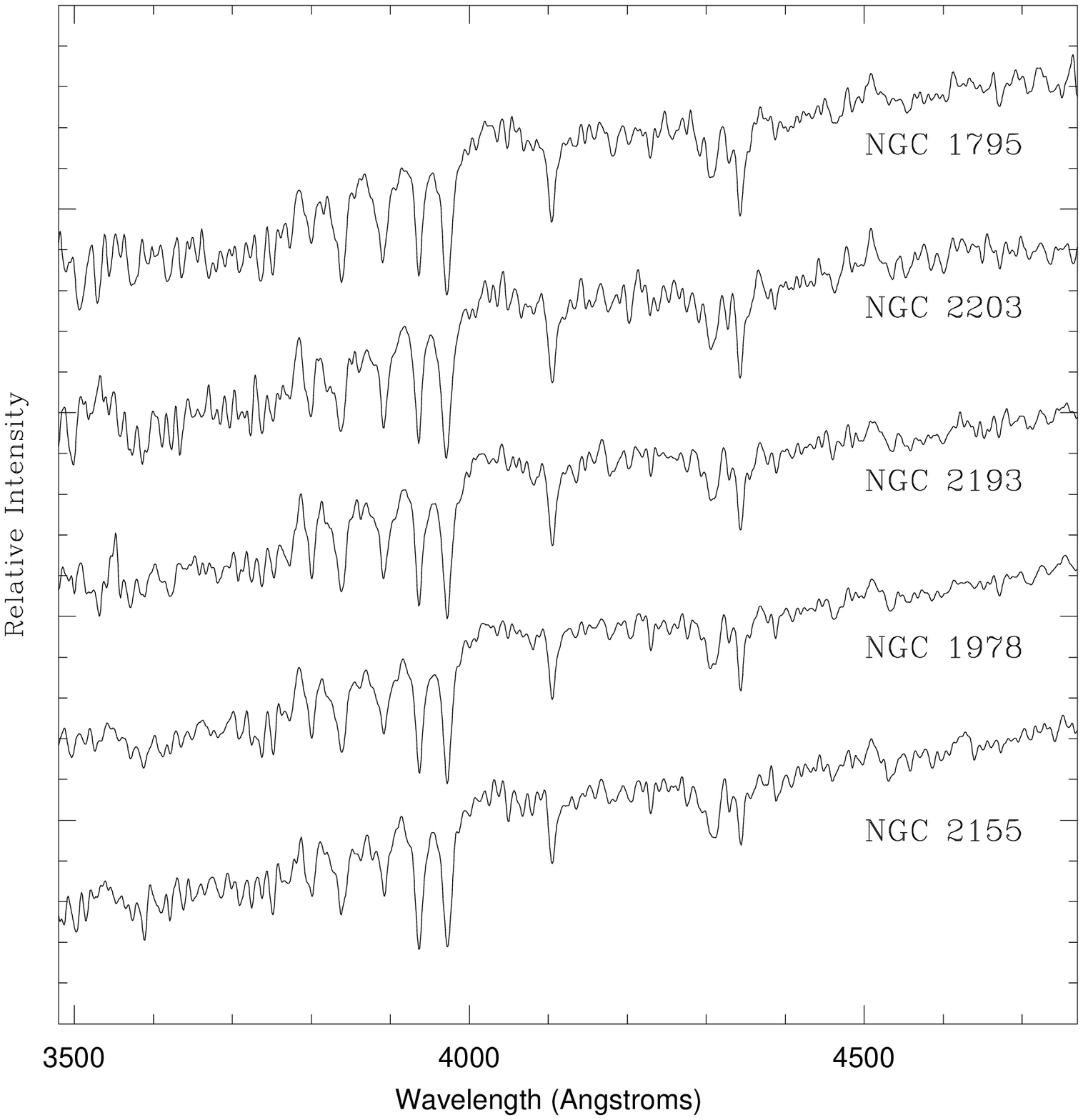}
\caption{The observed MC cluster spectra smoothed with a gaussian of width 2 \AA.
The spectra have been put in the order of increasing Ca II index (See Section
\S3.1) from the bottom to the top of each panel.
(a)--(f) LMC clusters.  (g) SMC clusters.  (h) Galactic globular clusters.}
\label{repspec}
\end{figure}
\setcounter{figure}{0}
\begin{figure}
\plotone{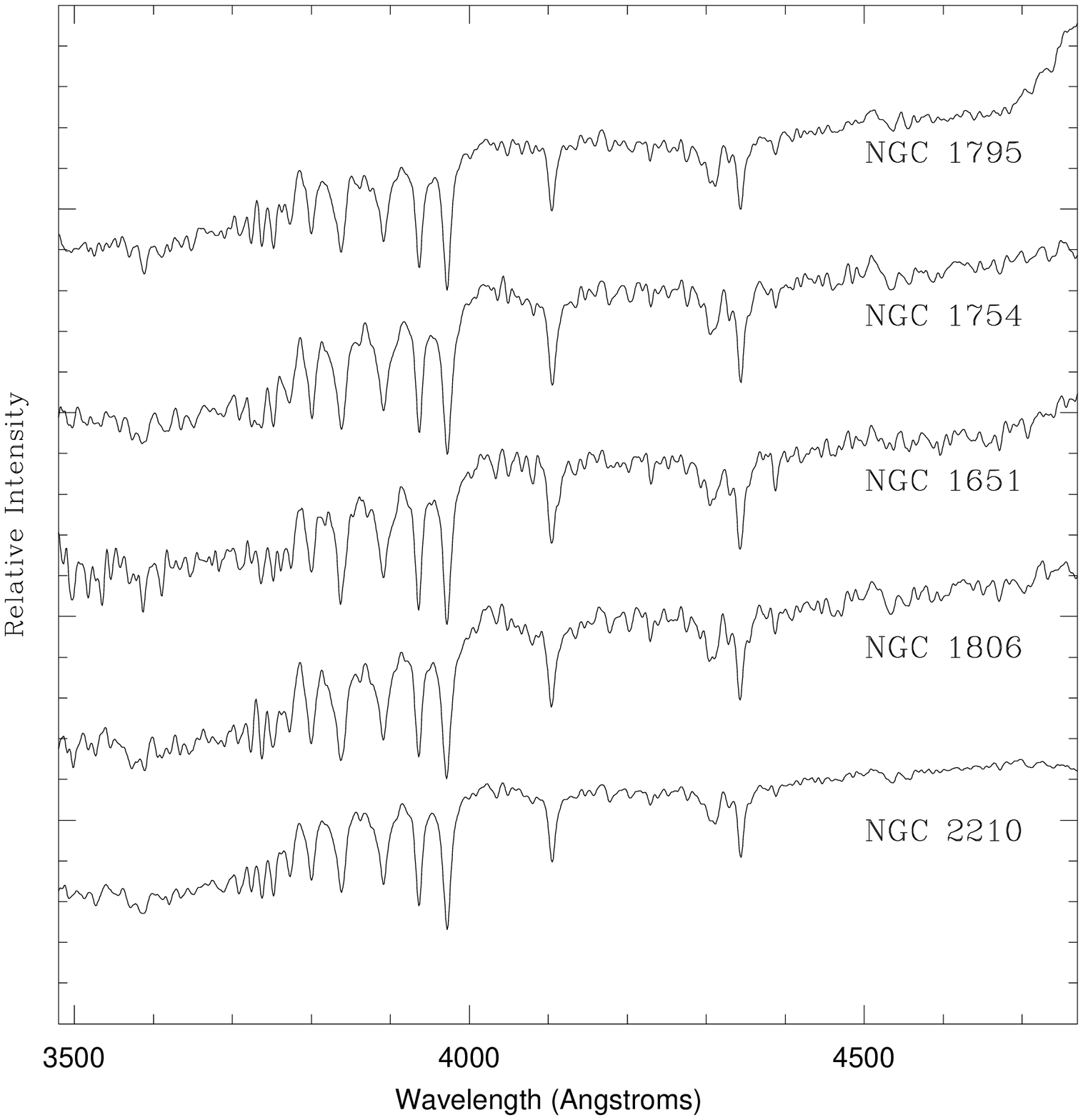}
\caption{b. LMC clusters (cont.)}
\end{figure}
\setcounter{figure}{0}
\begin{figure}
\plotone{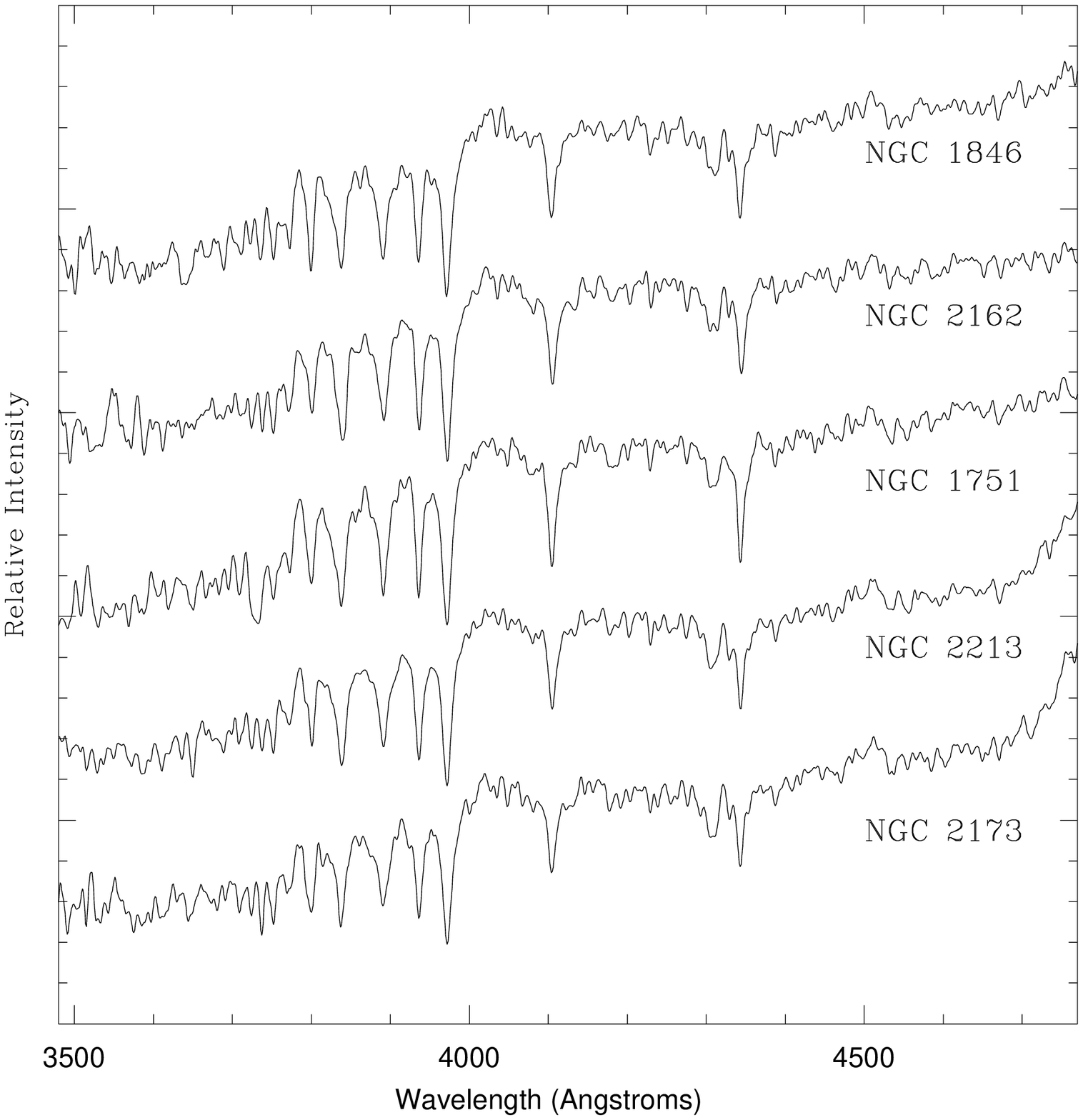}
\caption{c. LMC clusters(cont.)}
\end{figure}
\setcounter{figure}{0}
\begin{figure}
\plotone{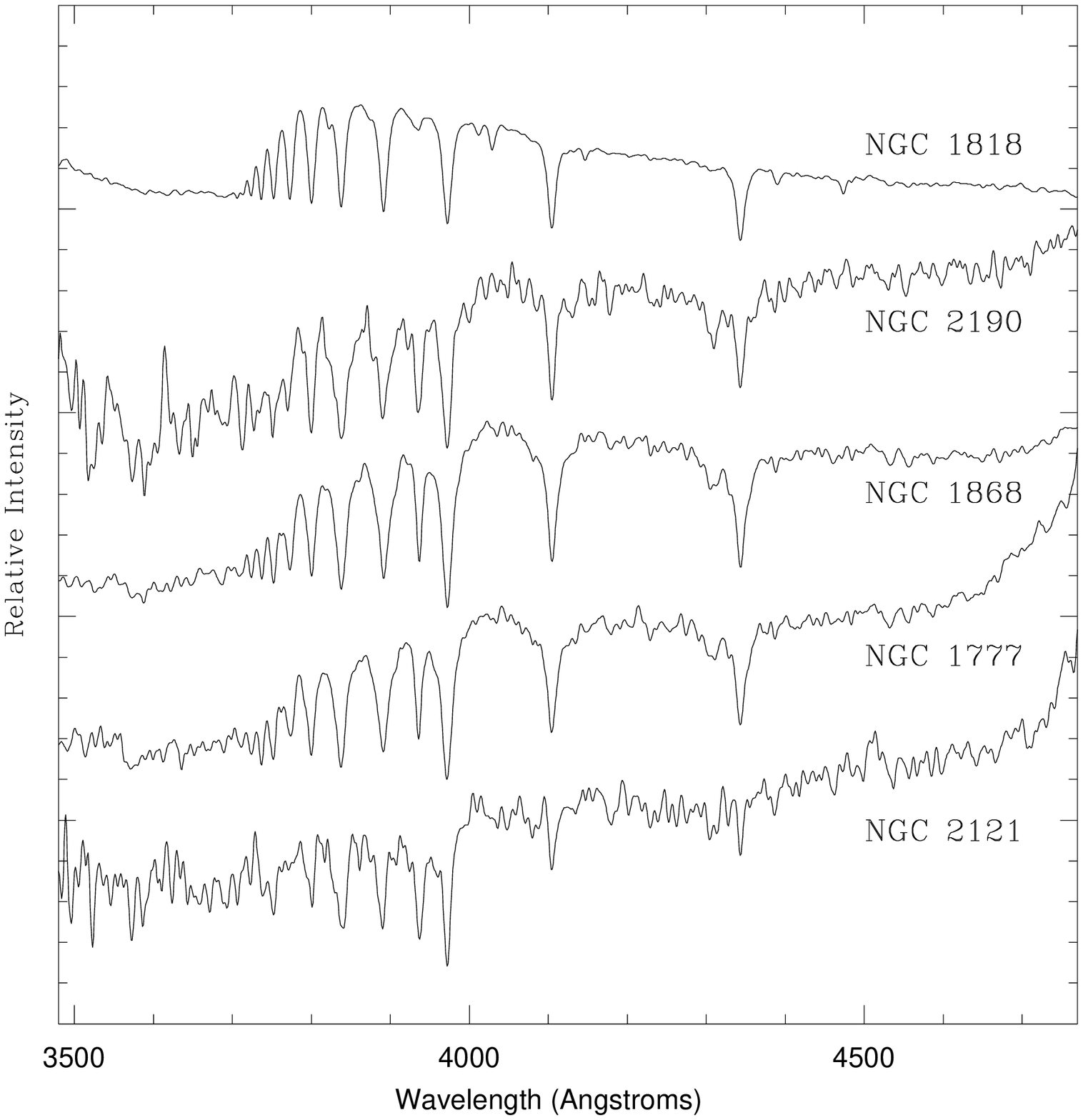}
\caption{d. LMC clusters(cont.)}
\end{figure}
\setcounter{figure}{0}
\begin{figure}
\plotone{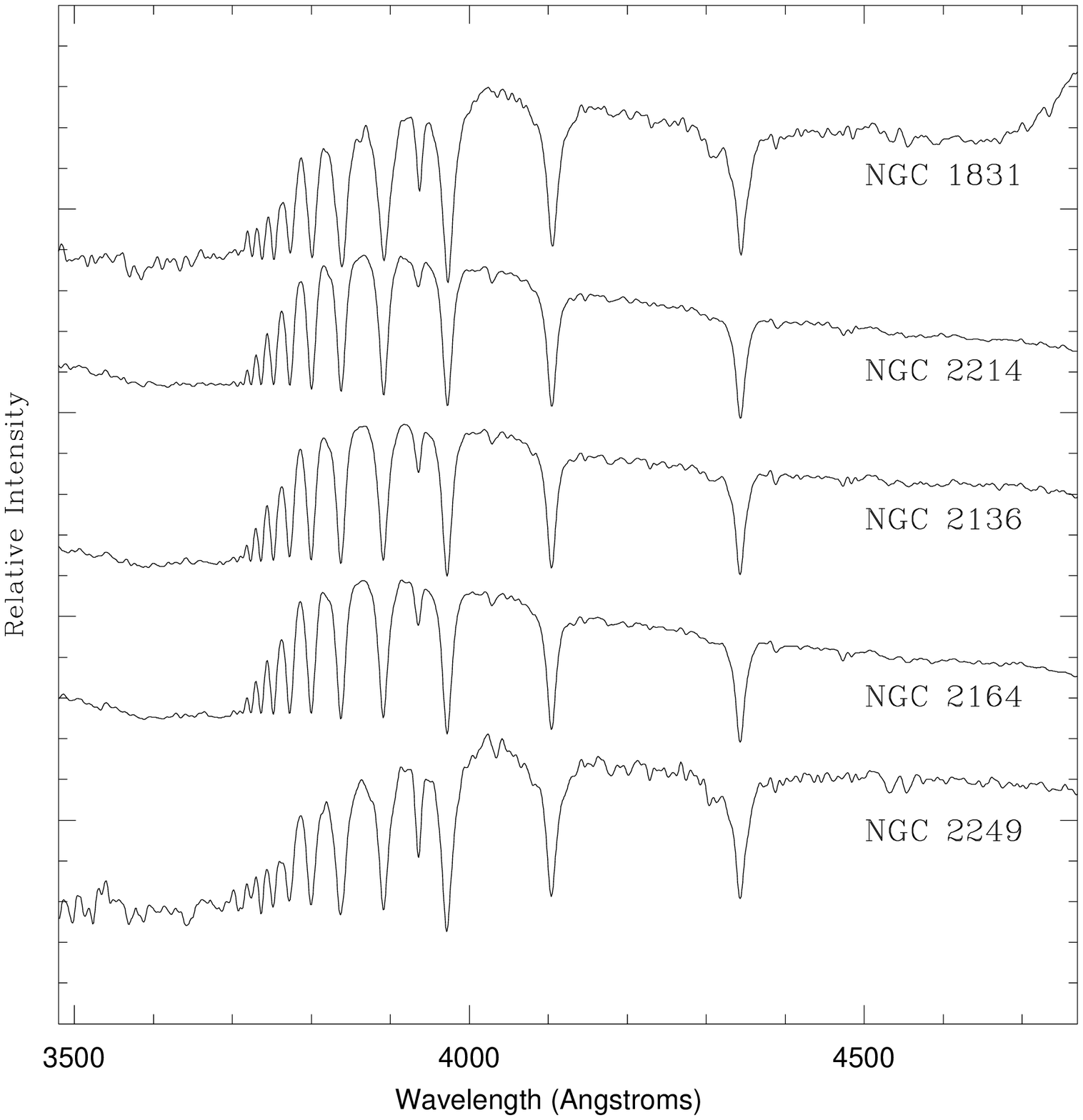}
\caption{e. LMC clusters(cont.)}
\end{figure}
\setcounter{figure}{0}
\begin{figure}
\plotone{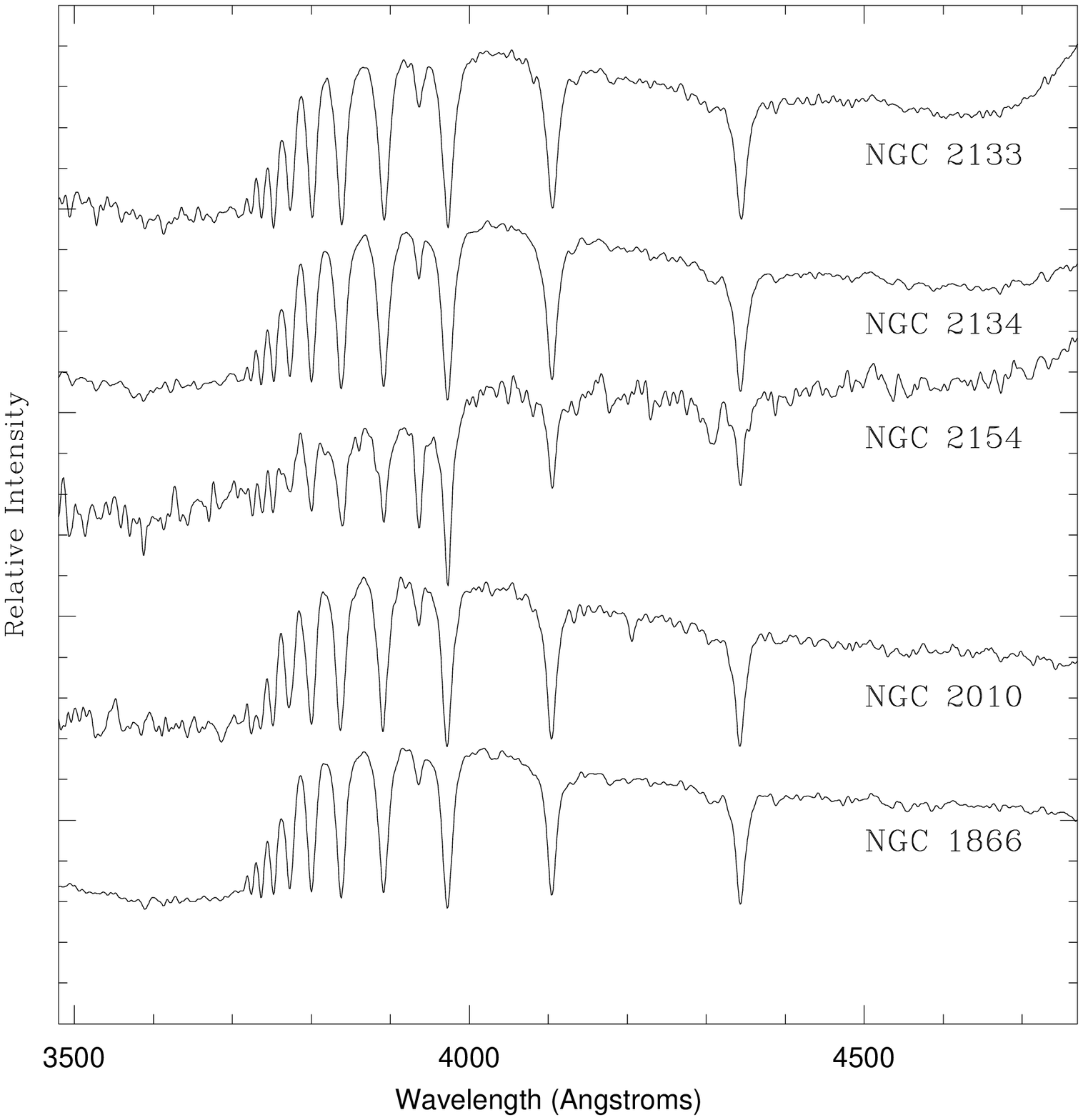}
\caption{f. LMC clusters(cont.)}
\end{figure}
\setcounter{figure}{0}
\begin{figure}
\plotone{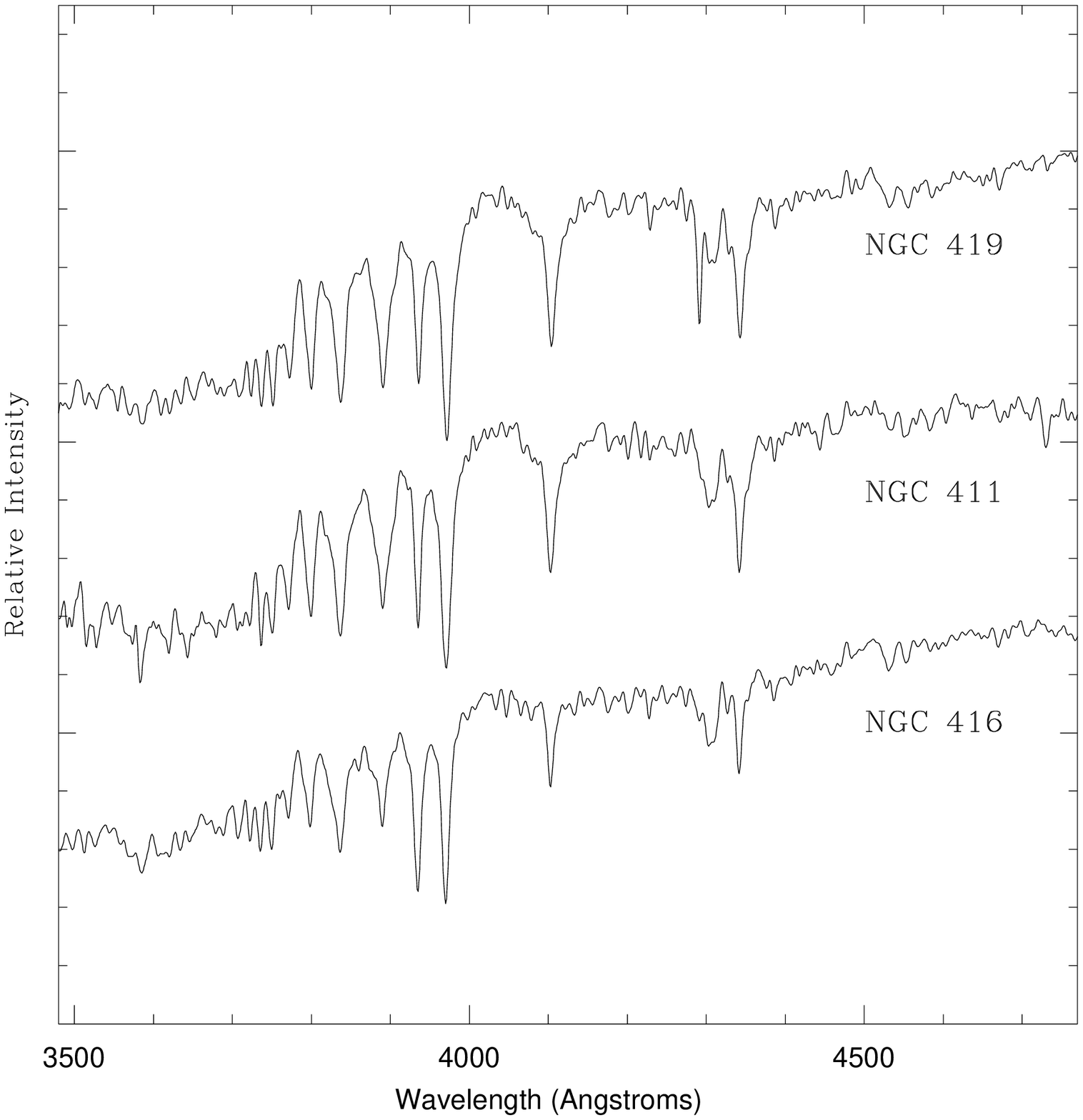}
\caption{g. SMC Clusters}
\end{figure}
\setcounter{figure}{0}
\begin{figure}
\plotone{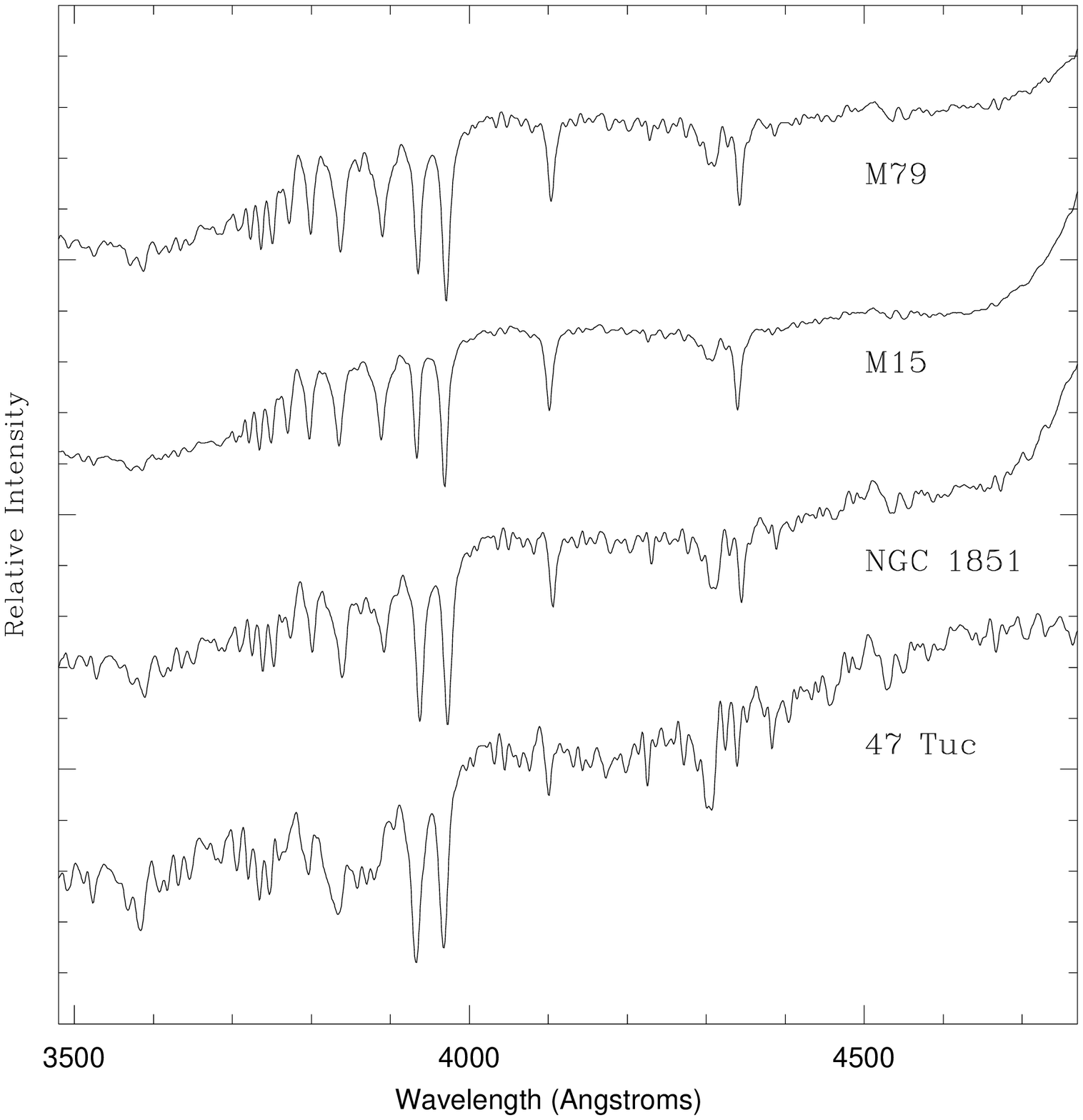}
\caption{h. Galactic Globular Clusters}
\end{figure}

\begin{figure}
\plotone{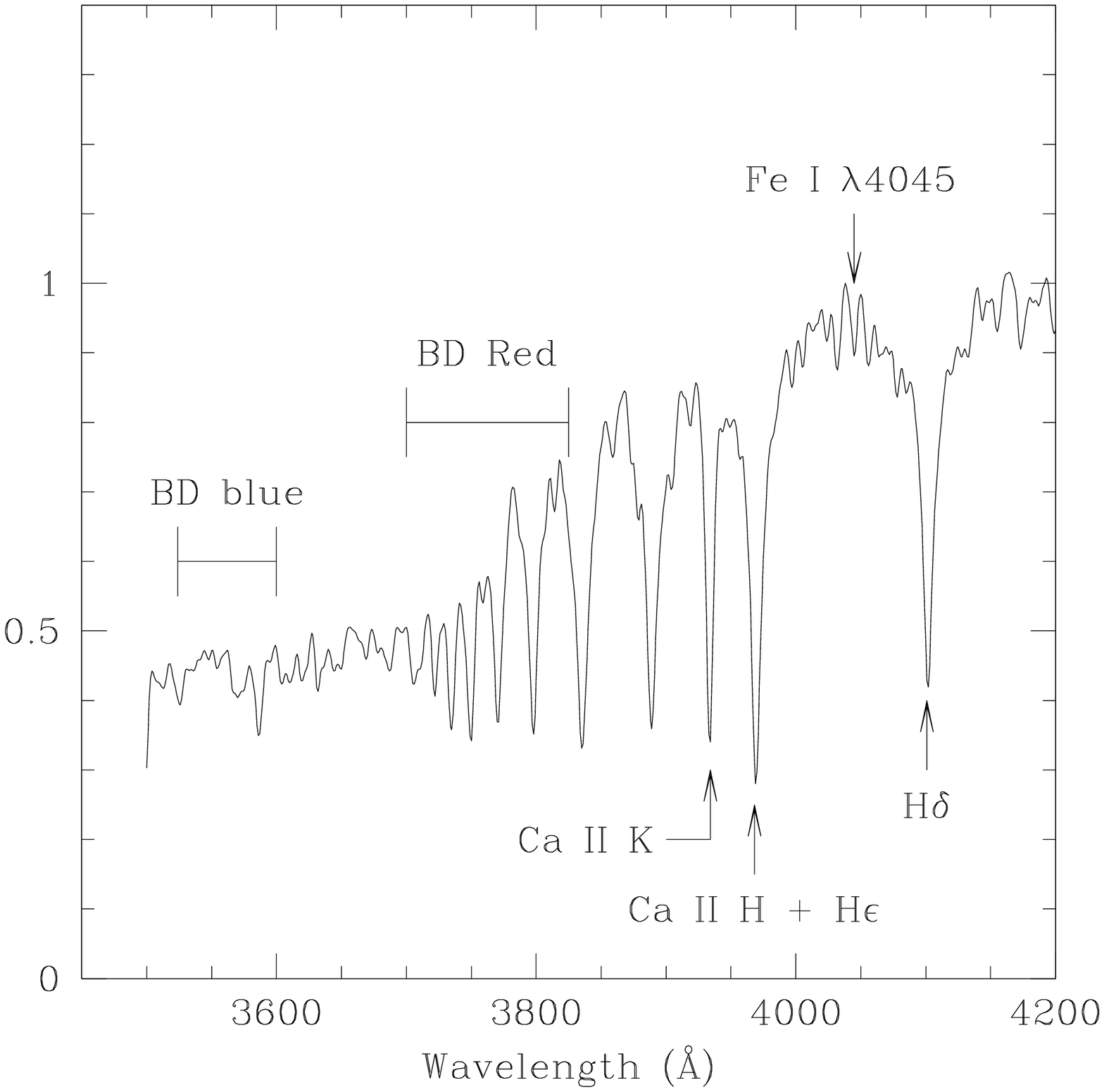}
\caption{The synthetic spectrum of a star with T$_{eff}$=7500 K, log g = 5.0, and
solar chemical composition is plotted, with all spectral features identified
that are used in this paper.}
\label{features}
\end{figure}

\begin{figure}
\plotone{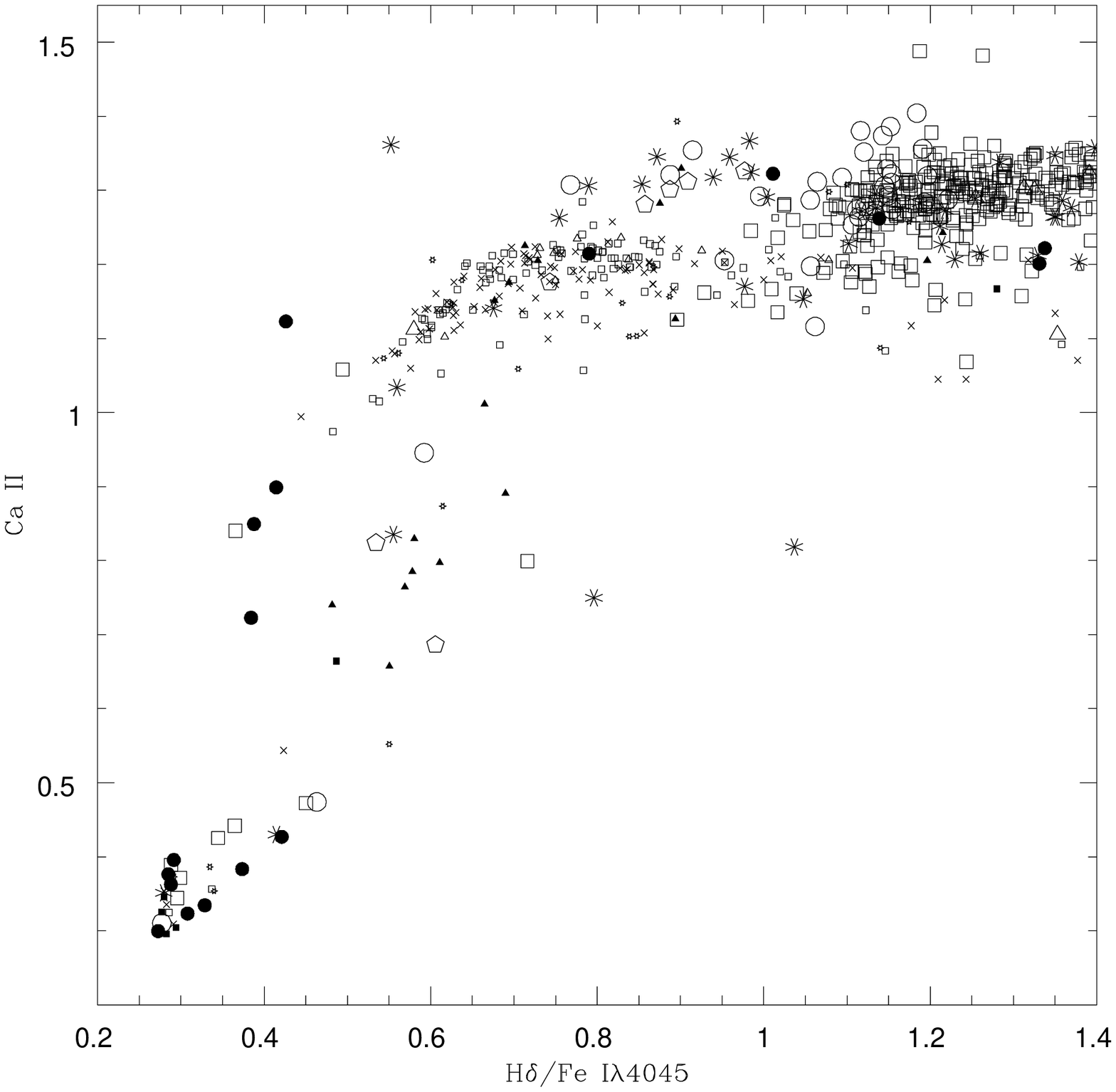}
\caption{The Ca II index is plotted versus the H$\delta$/Fe I $\lambda$4045 
index for stars in the CFSL.  Different symbols denote stars in various
gravity and metallicity bins.  Large filled circles are for stars with
intermediate gravity of 3.3 $<$log g $\le$ 3.7.  Large octagonal asterisms are
for low-gravity (supergiant) stars with 1.8 $\ge$log g. Giant stars (1.8$<$log g
$\le$3.3) are plotted as large open symbols, with triangles, squares, circles,
and pentagons referring respectively to the following metallicity intervals:
[Fe/H]$>$+0.3; +0.3$\ge$[Fe/H]$>$-0.4; -0.4$\ge$[Fe/H]$>$-1.5; -1.5$\ge$[Fe/H].
Dwarfs are plotted as small symbols, with the following scheme: filled squares
for [Fe/H]$>$+0.3; x's for +0.3$\ge$[Fe/H]$>$-0.2; unfilled squares for
-0.2$\ge$[Fe/H]$>$-0.75; unfilled triangles for -0.75$\ge$[Fe/H]$>$-1.25; filled
triangles for -1.25$\ge$[Fe/H].
 }
\label{CaII-stars}
\end{figure}

\begin{figure}
\plotone{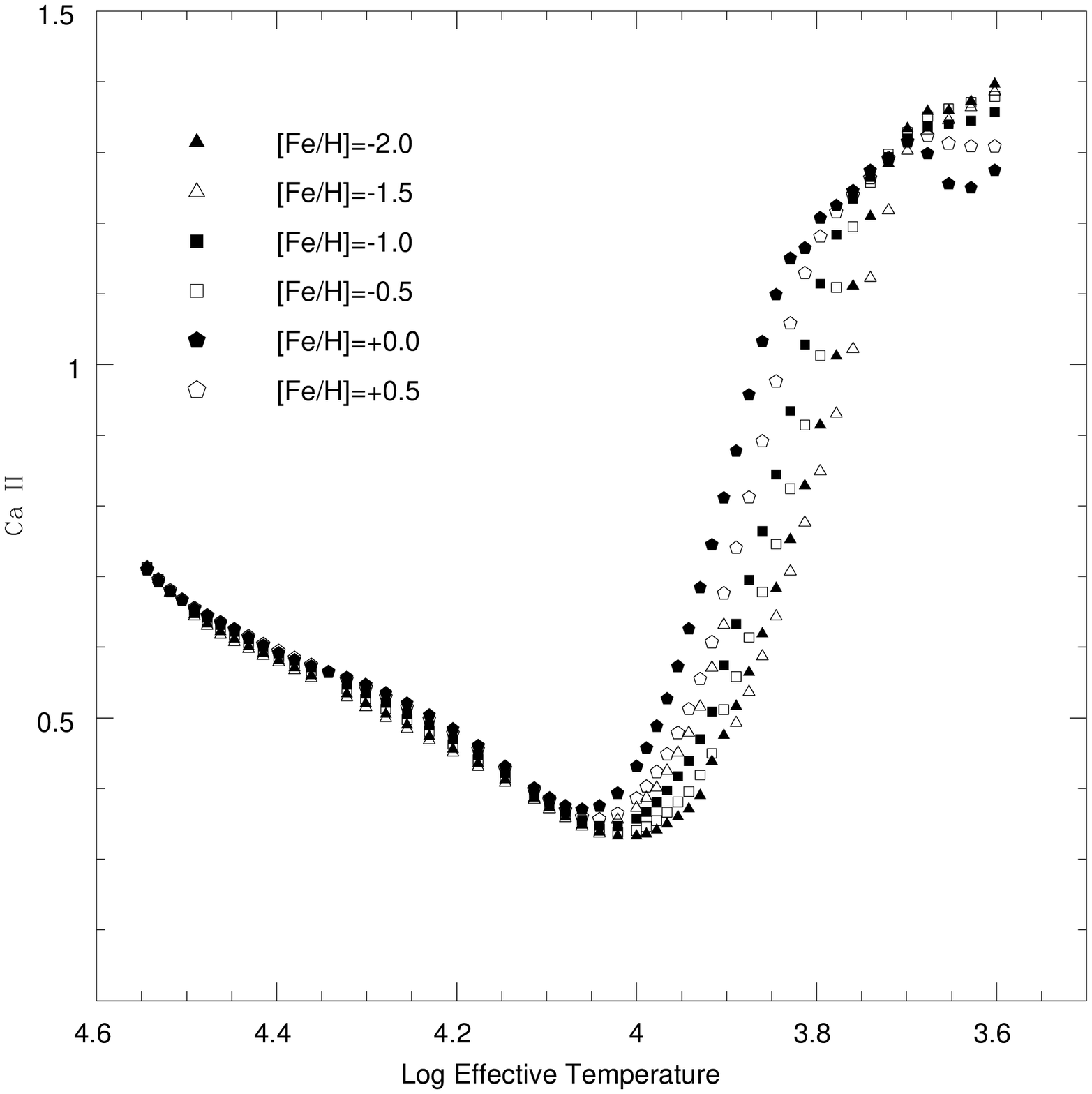}
\caption{The Ca II index plotted vs. log(T$_{eff}$) for synthetic dwarf stars.  
Note the separation in metallicity for intermediate T$_{eff}$.}
\label{caiisampdw}
\end{figure}

\begin{figure}
\plotone{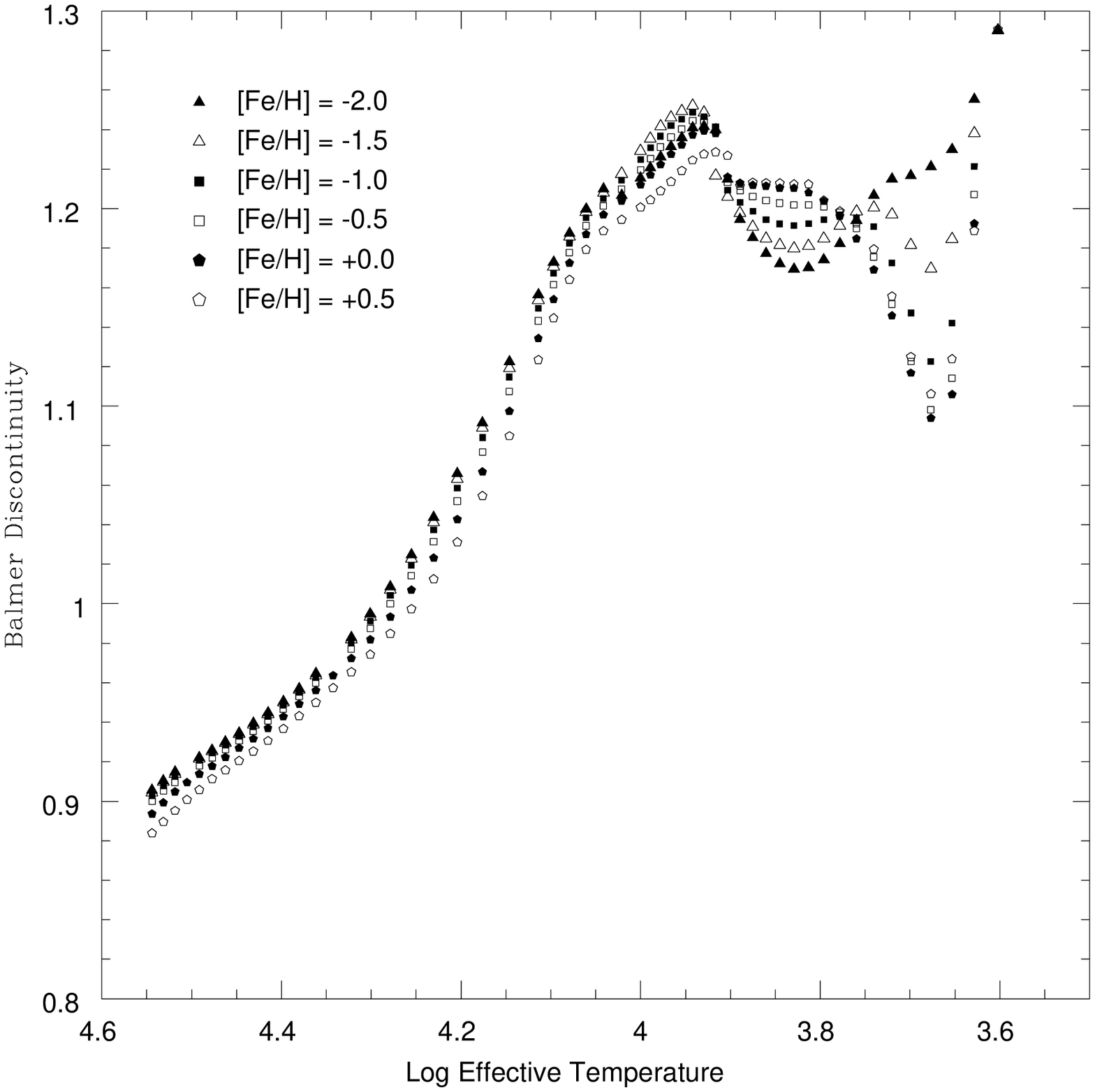}
\caption{The Balmer Discontinuity index plotted vs. log(T$_{eff}$)
for synthetic dwarf stars.
The steep dependence on T$_{eff}$ makes the BD index a very good age indicator
for young populations.}
\label{bdsampdw}
\end{figure}

\begin{figure}
\plotone{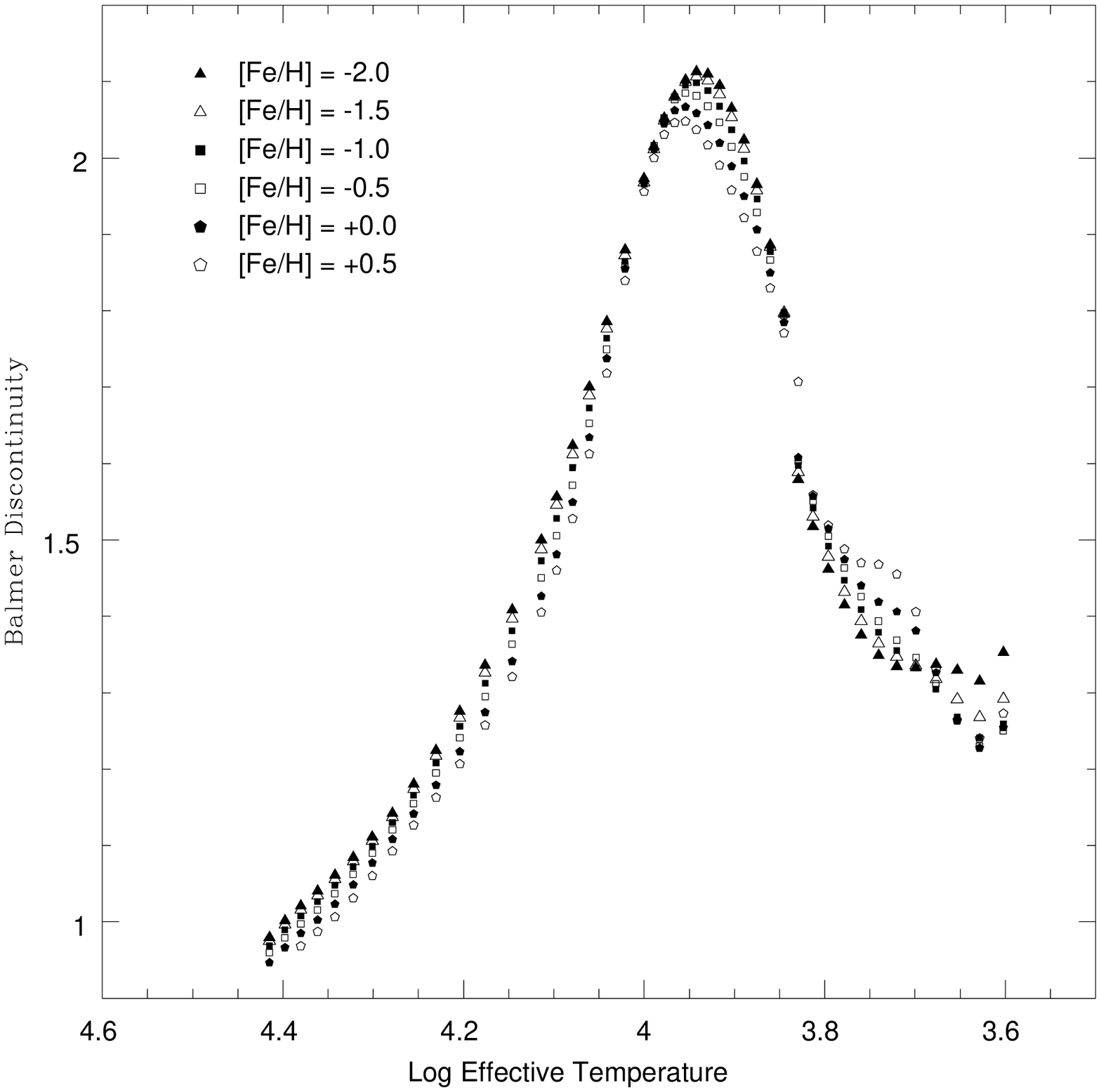}
\caption{The Balmer Discontinuity index plotted vs. log(T$_{eff}$)
for synthetic giant stars.}
\label{bdsampgi}
\end{figure}

\begin{figure}
\plotone{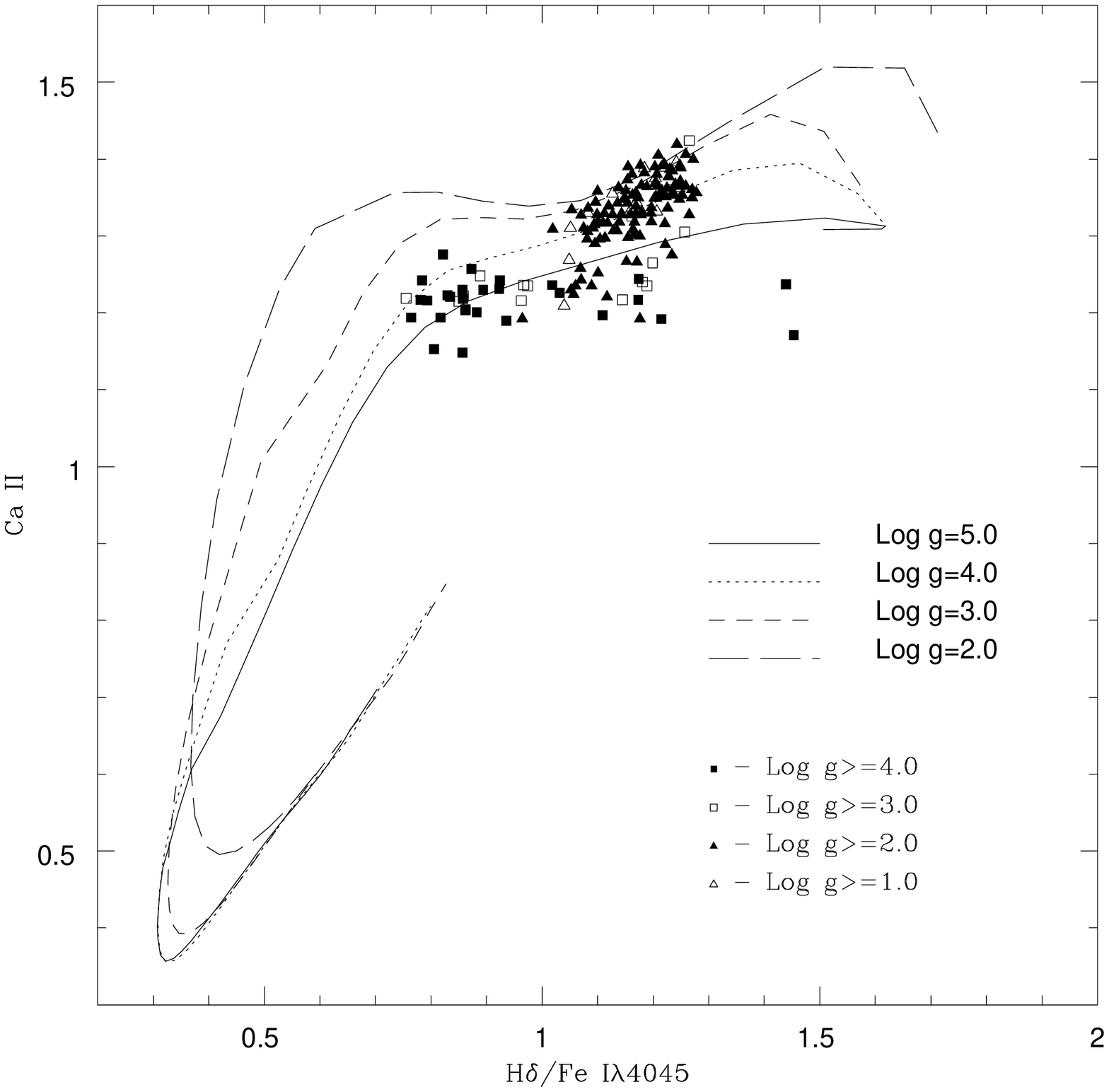}
\caption{H$\delta$/Fe I $\lambda$4045 vs. Ca II (at solar metallicity) for
both the empirical and synthetic libraries.  Empirical stars are plotted
with symbols while the indices for the synthetic stars are plotted as lines.
Both libraries have been further separated by gravity as shown in the figure
legend.  Notice that the empirical dwarfs have a flatter slope than the
synthetic dwarfs as H$\delta$/Fe I $\lambda$4045 increases past 1.0.}
\label{synemp}
\end{figure}

\clearpage

\begin{figure}
\plotone{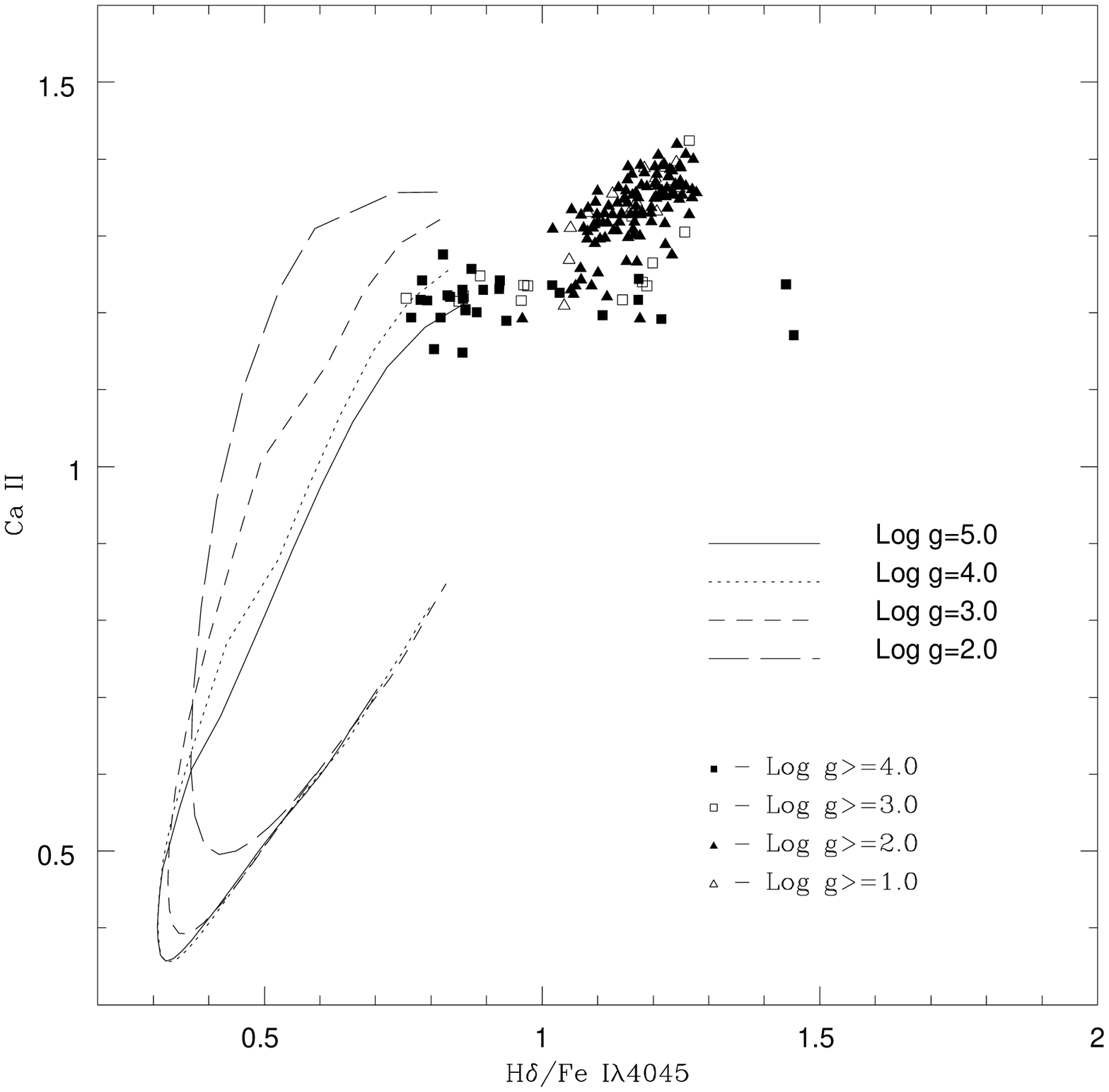}
\caption{Same as Figure \ref{synemp} but with the synthetic curves
truncated at an effective temperature of 6000 K.}
\label{cutoff}
\end{figure}

\begin{figure}
\plotone{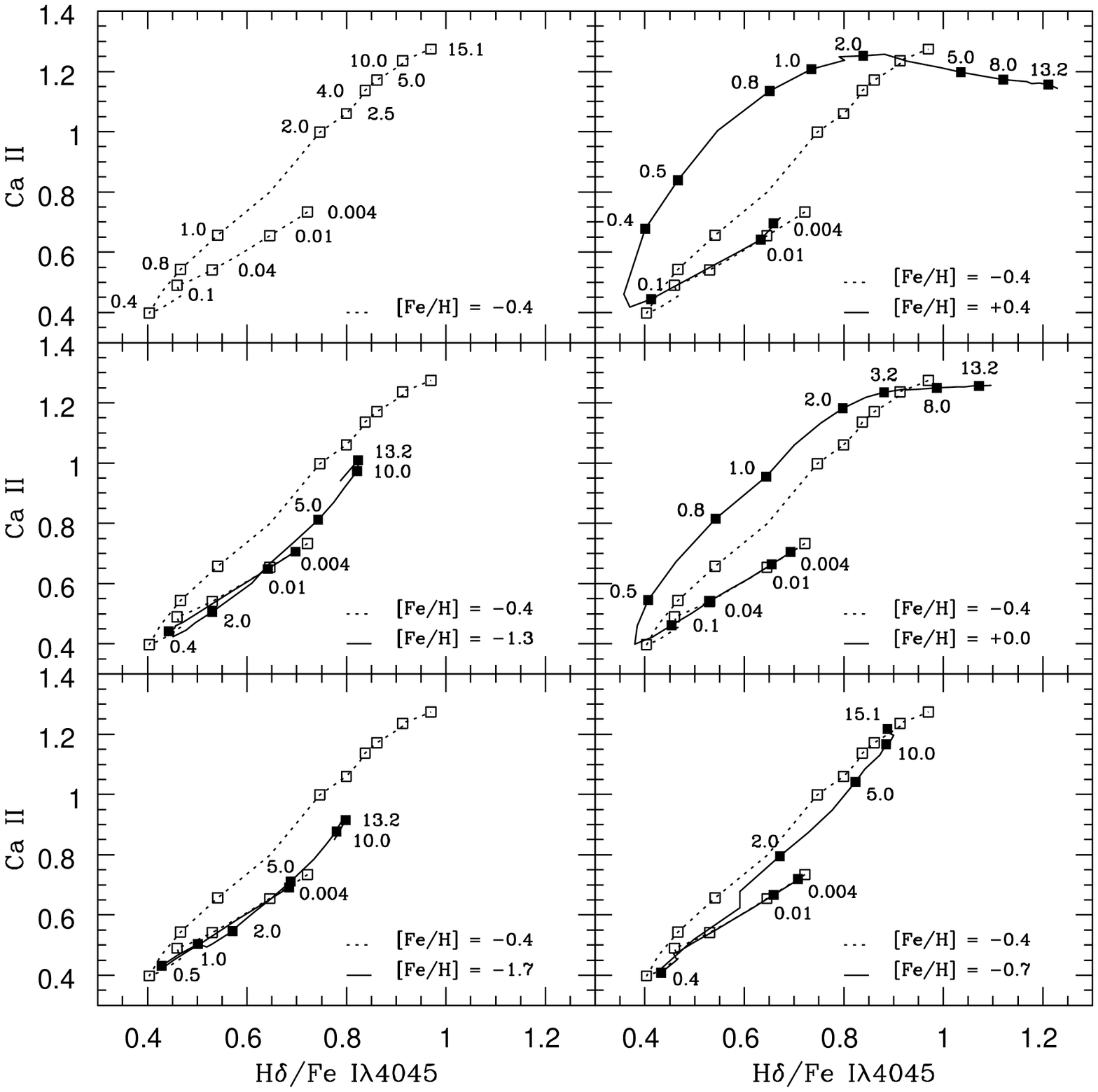}
\caption{H$\delta$/Fe I $\lambda$4045 vs. Ca II as a function of age and
metallicity.  The upper left panel shows the index evolution for [Fe/H] =
-0.4.  The open squares are the index values for the stellar population
at a given age labelled in Gyr.  This curve is reproduced as a dotted line in
the other panels for comparison purposes.  The other panels display,
as solid lines, the index trajectories for other metallicities as labelled
in the figure.}
\label{cabig}
\end{figure}

\begin{figure}
\plotone{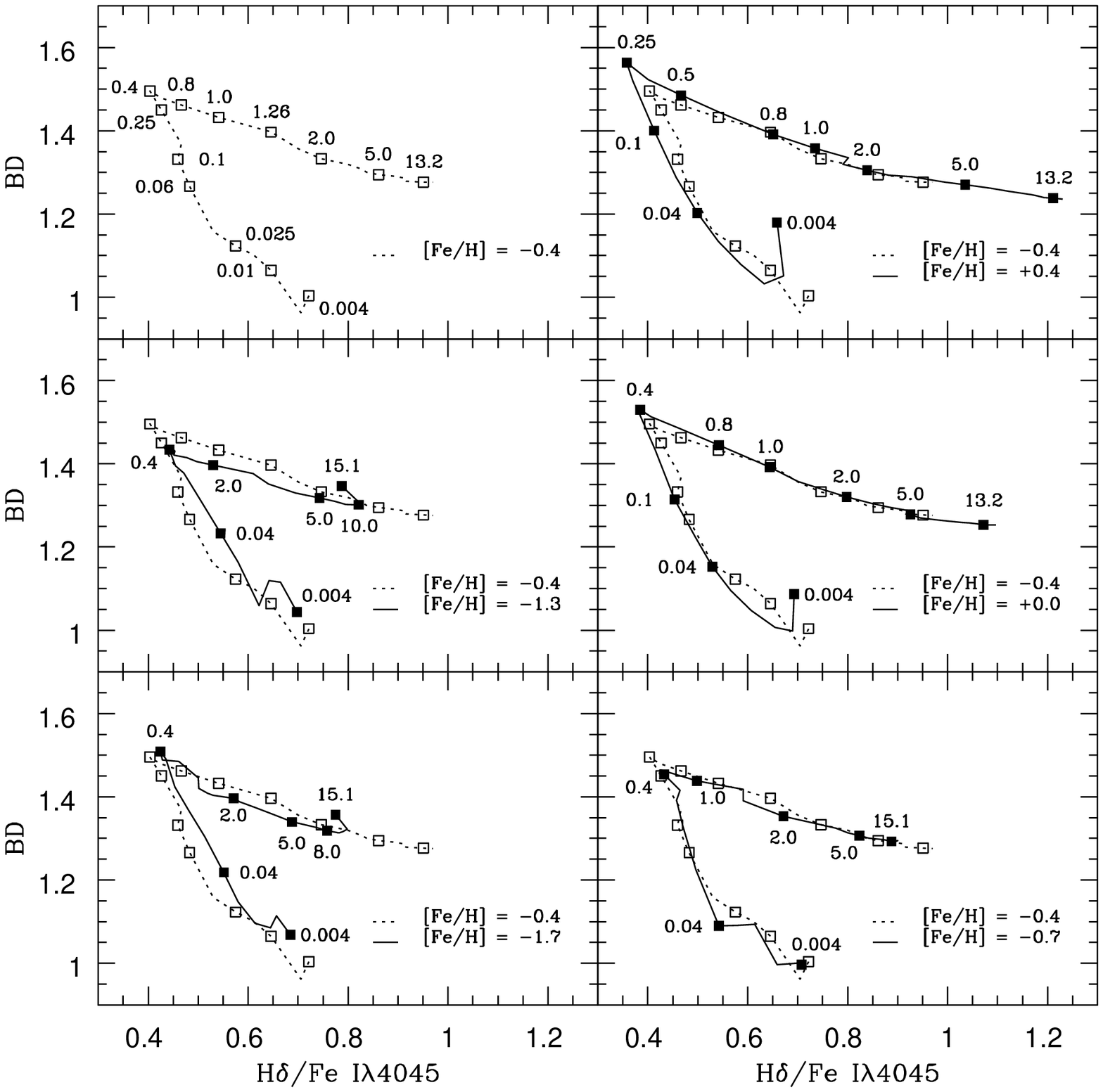}
\caption{Same as Figure \ref{cabig} but with the Balmer Discontinuity
index plotted vs. H$\delta$/Fe I $\lambda$4045.}
\label{bdbig}
\end{figure}

\begin{figure}
\plotone{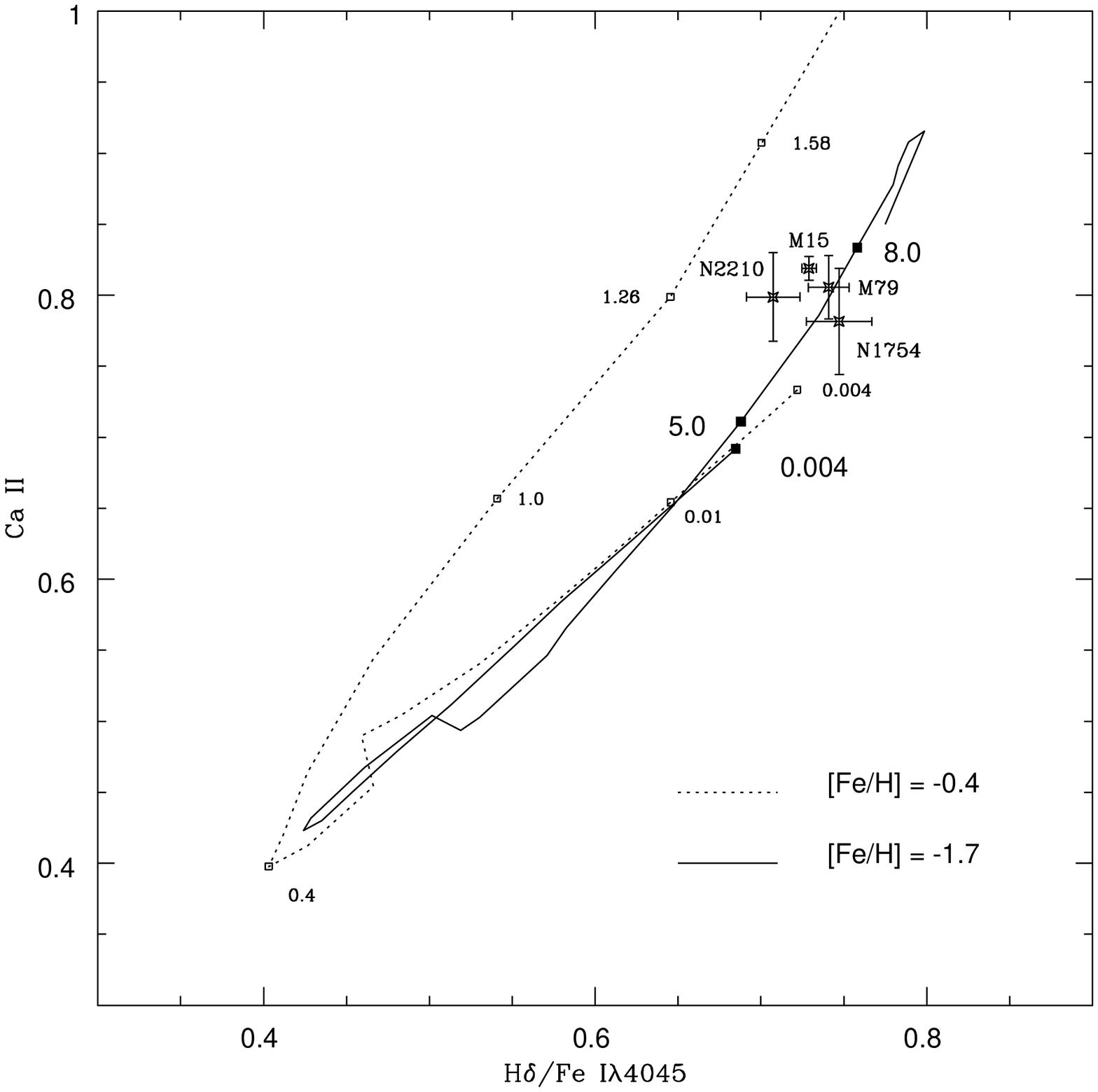}
\caption{H$\delta$/Fe I $\lambda$4045 vs. Ca II for [Fe/H] = -1.7.  The
trajectories are the same as in Figure~\ref{cabig} but now the
MC cluster index data are included. Clusters are placed on this and the
subsequent Ca II figures based on their literature metallicity values given in
Table~\ref{litagemet_tab}. The dotted line is the trajectory
for [Fe/H] = -0.4 with ages displayed as small labels, while the solid line
is the track for [Fe/H] = -1.7 with large labels to mark the ages.  All
labelled ages are in Gyr.  The two LMC clusters, NGC 2210 and NGC 1754,
occupy the same region of the diagram as the Galactic globular 
M79.  These two clusters may have blue horizontal branch stars contaminating
their spectral indices (see text).}
\label{caclustm17}
\end{figure}

\begin{figure}
\plotone{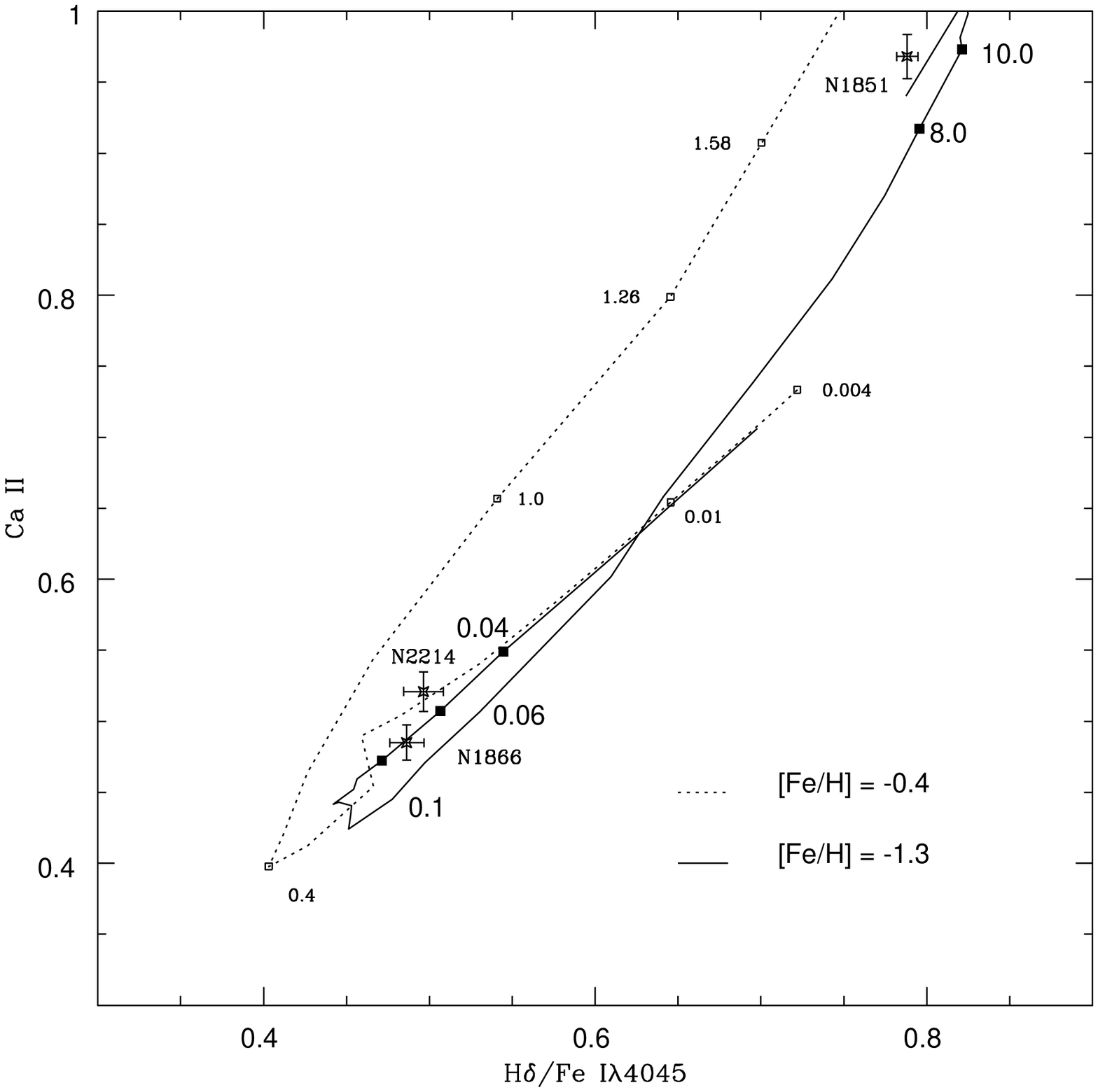}
\caption{Same as Figure \ref{caclustm17} but with [Fe/H] = -1.3.}
\label{caclustm13}
\end{figure}

\begin{figure}
\plotone{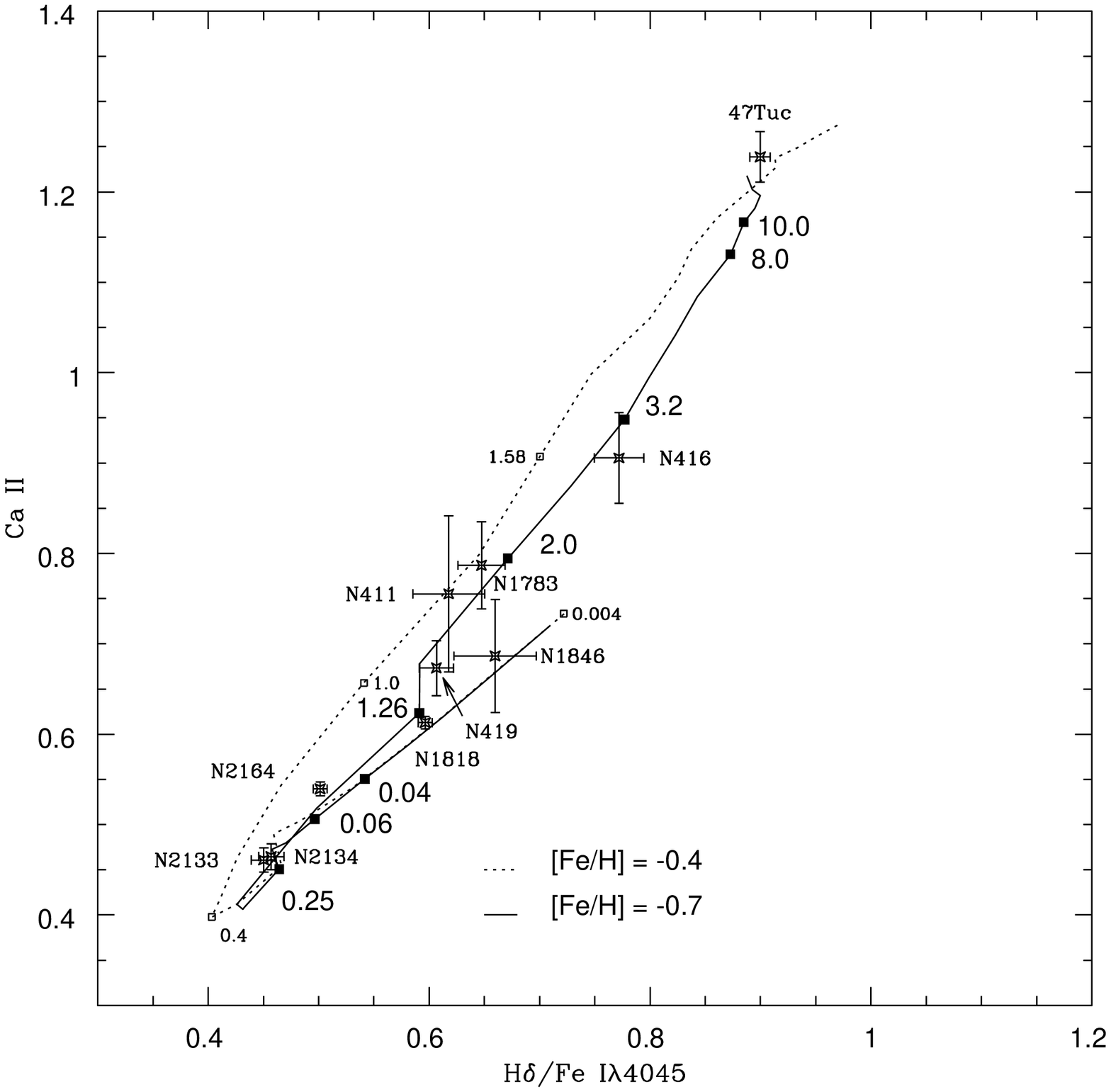}
\caption{Same as Figure \ref{caclustm17} but with [Fe/H] = -0.7.}
\label{caclustm07}
\end{figure}

\begin{figure}
\plotone{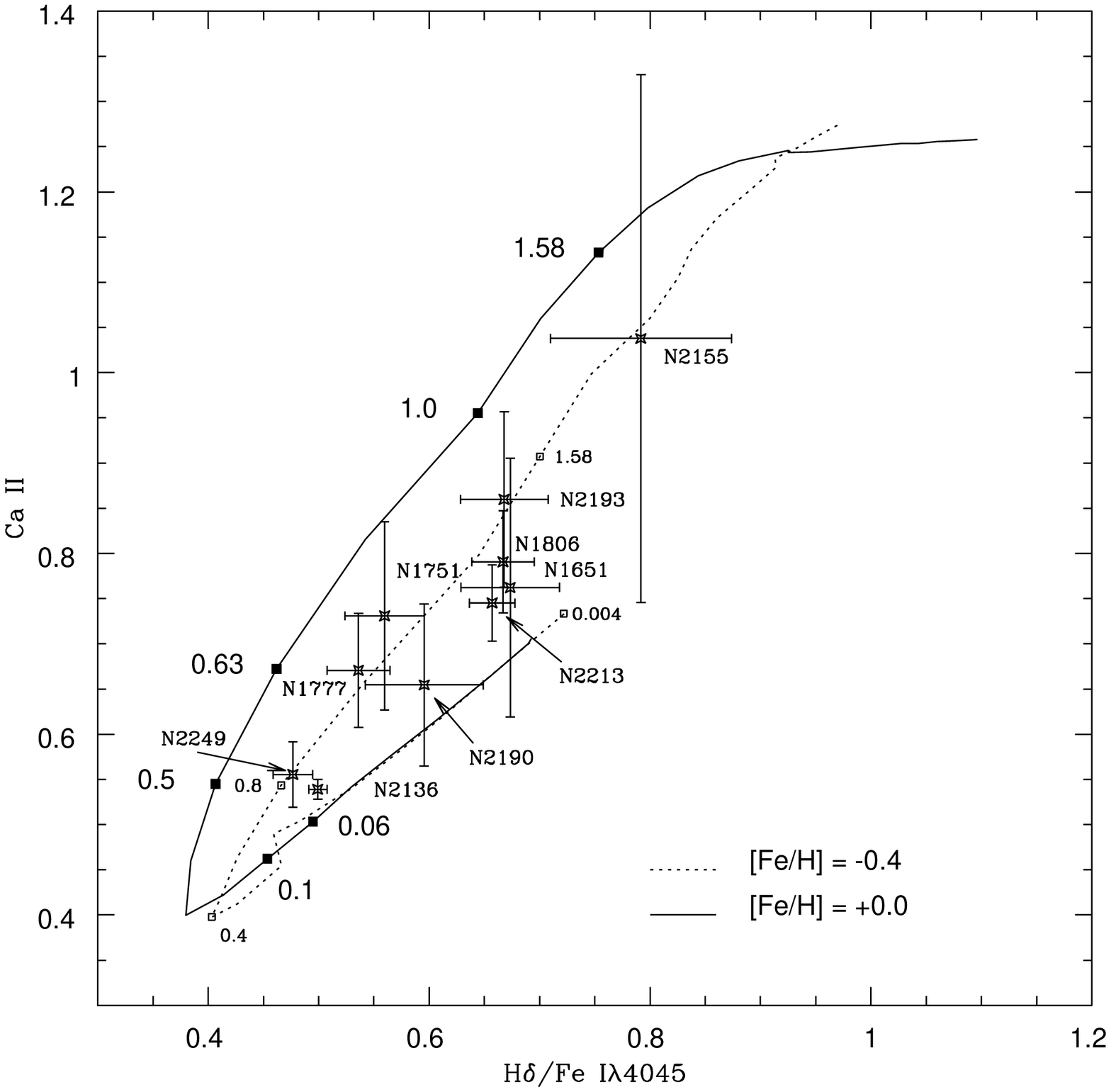}
\caption{Same as Figure \ref{caclustm17} but with [Fe/H] = +0.0.}
\label{caclustp00a}
\end{figure}

\begin{figure}
\plotone{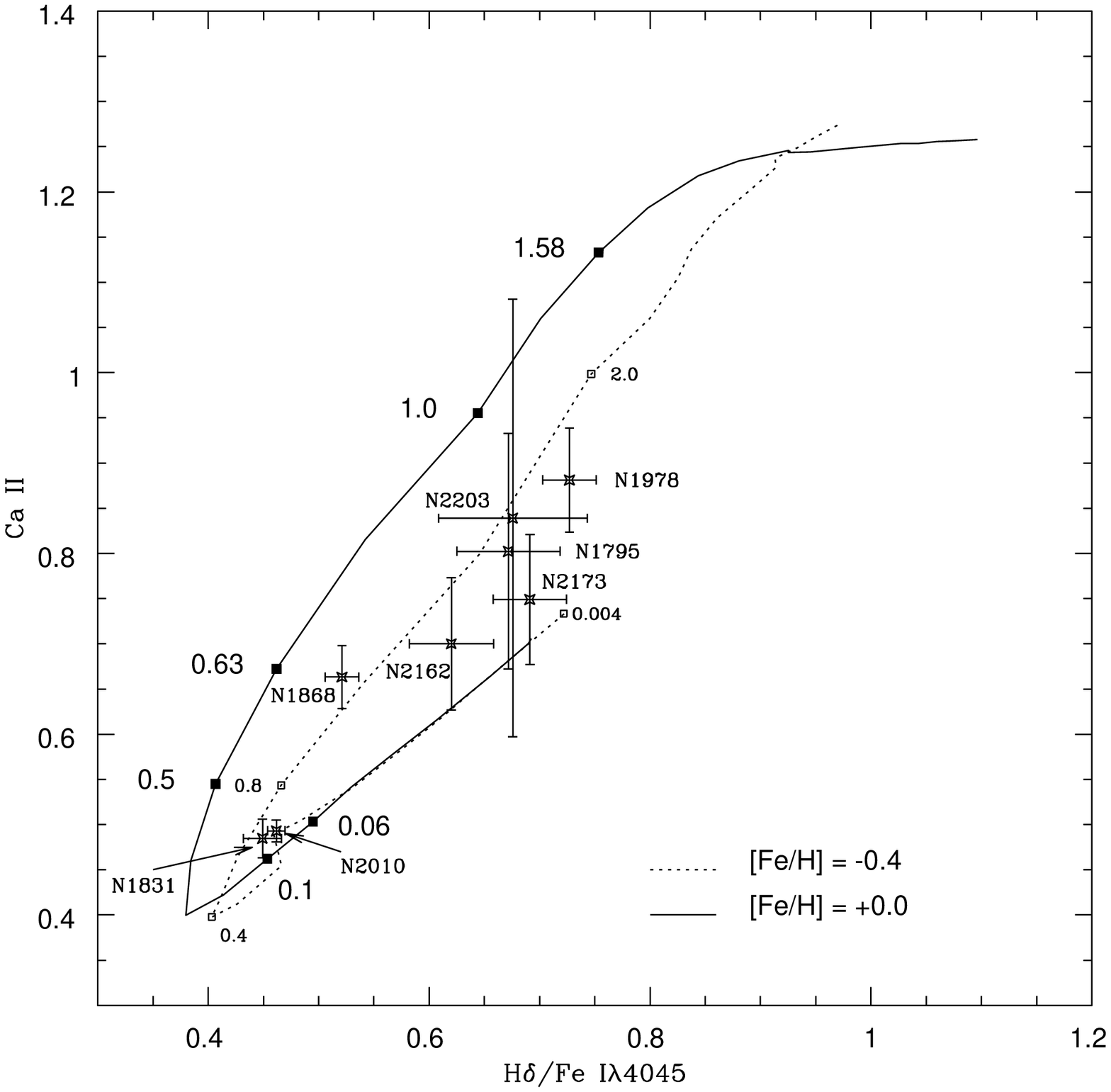}
\caption{Same as Figure \ref{caclustp00a}}
\label{caclustp00b}
\end{figure}

\clearpage

\begin{figure}
\plotone{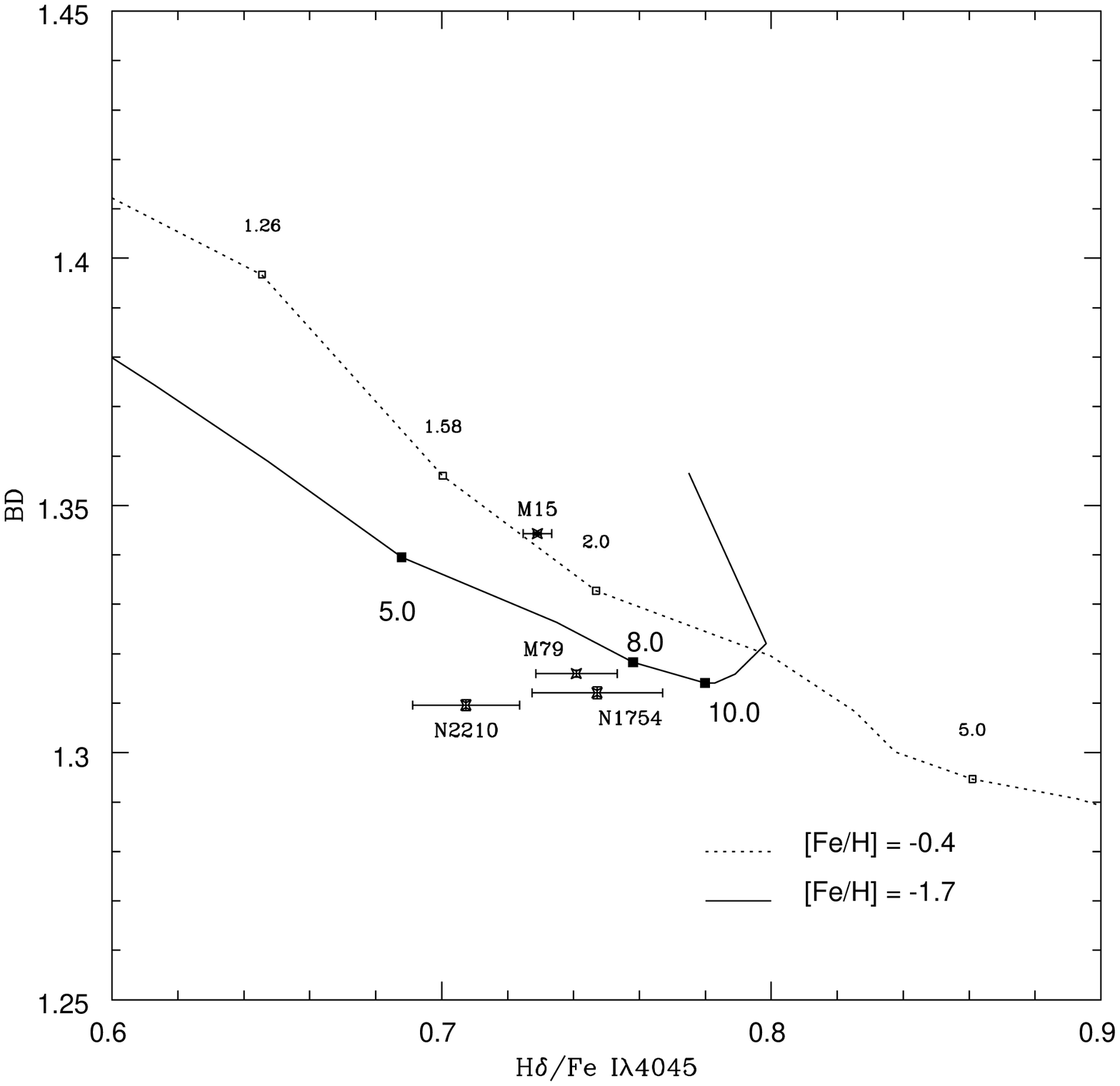}
\caption{H$\delta$/Fe I $\lambda$4045 vs. BD for [Fe/H] = -1.7.  The
trajectories are the same as in Figure~\ref{bdbig} but now the
MC cluster index data are included. Clusters are placed on this and the
subsequent BD figures based on their literature metallicity values given in
Table~\ref{litagemet_tab}. The dotted line is the trajectory
for [Fe/H] = -0.4 with ages displayed as small labels, while the solid line
is the track for [Fe/H] = -1.7 with large labels to mark the ages.  All
labelled ages are in Gyr. Again, NGC 2210 and NGC 1754
occupy similar regions of the diagram as M15 and M79.}
\label{bdclustm17}
\end{figure}

\begin{figure}
\plotone{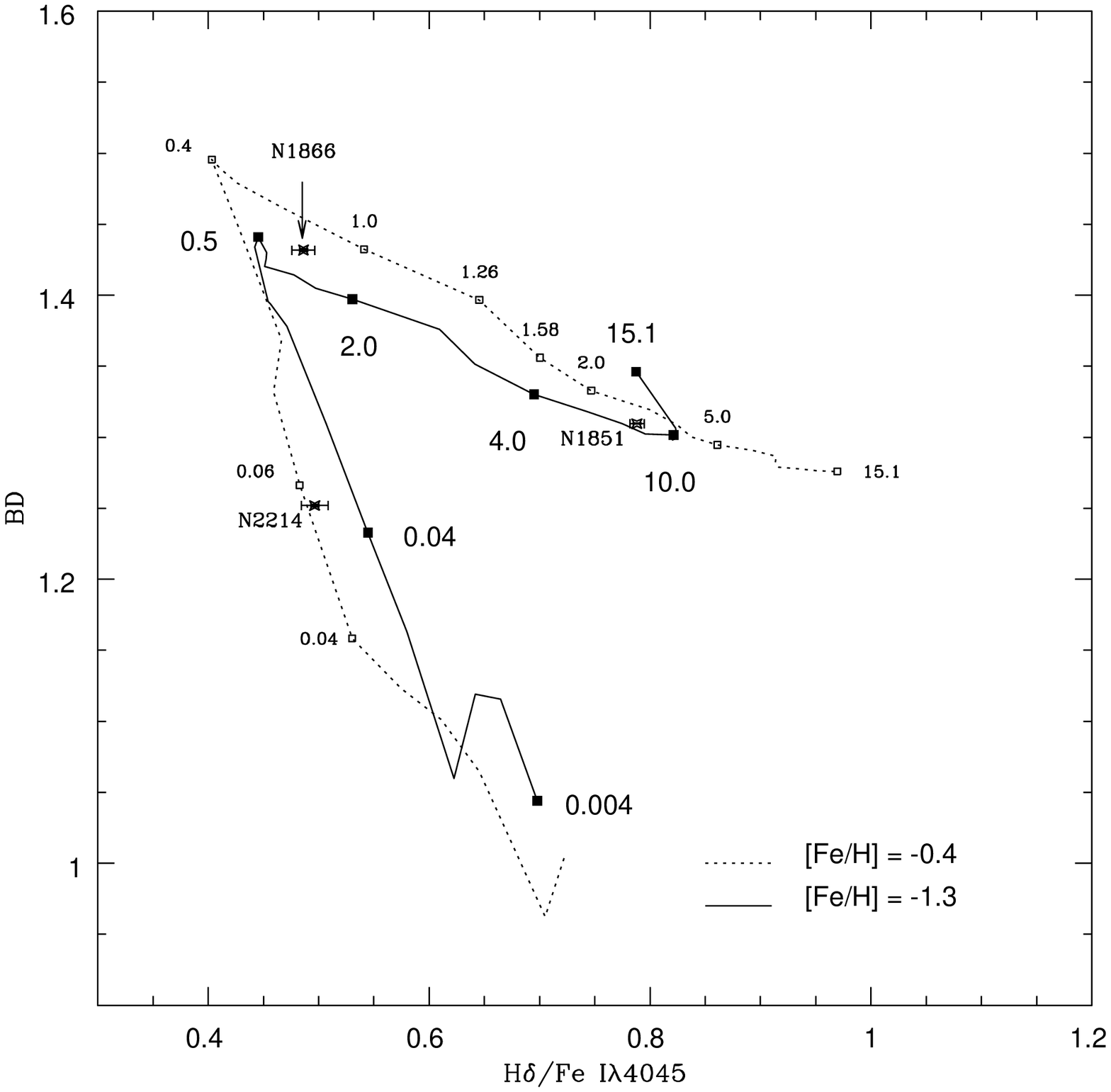}
\caption{Same as Figure \ref{bdclustm17} but with [Fe/H] = -1.3.}
\label{bdclustm13}
\end{figure}

\begin{figure}
\plotone{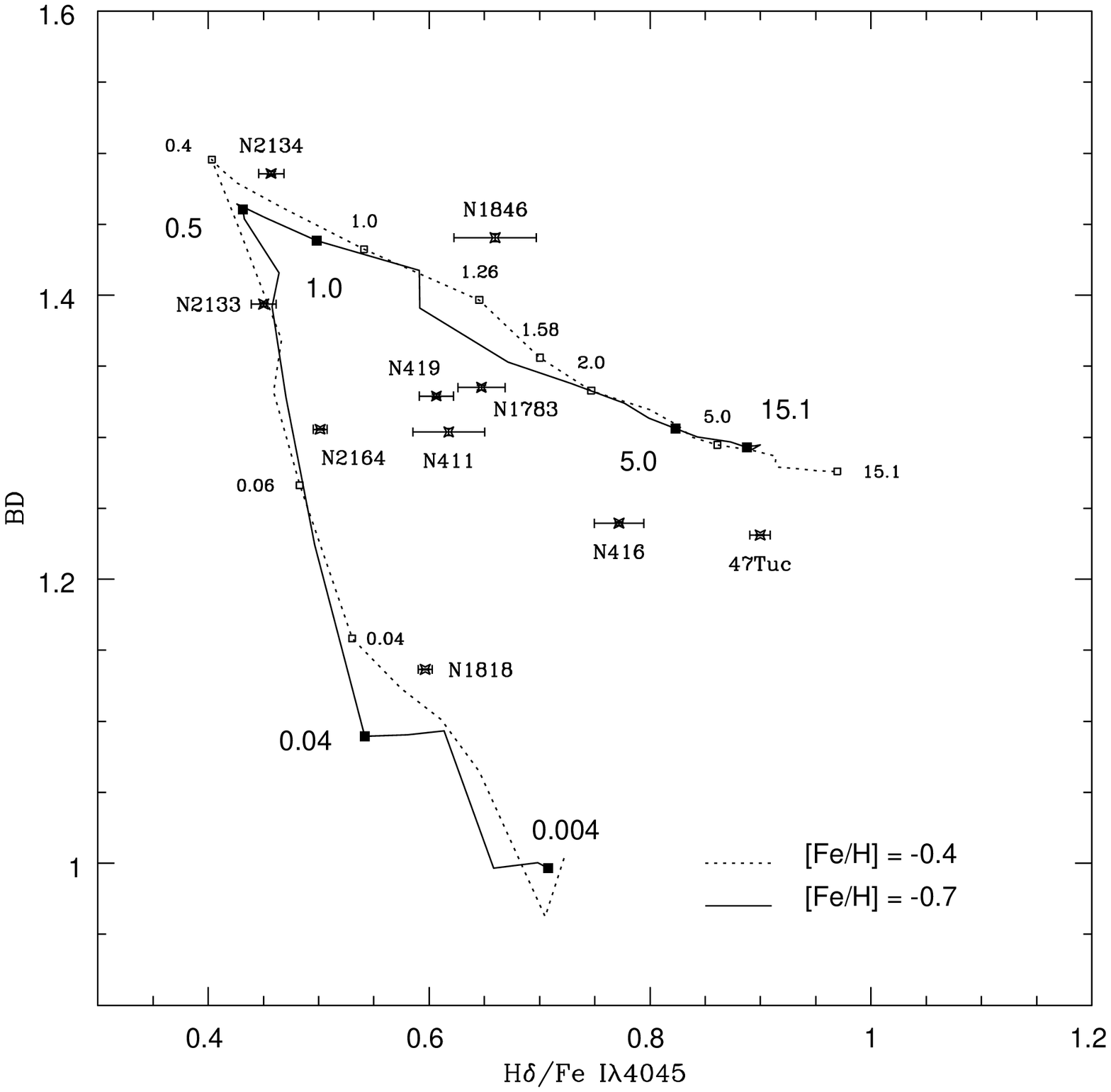}
\caption{Same as Figure \ref{bdclustm17} but with [Fe/H] = -0.7.  The
SMC clusters, NGC~411 and NGC~416, both lie far enough from all the BD index
tracks to make age determinations impossible for these clusters.}
\label{bdclustm07}
\end{figure}

\begin{figure}
\plotone{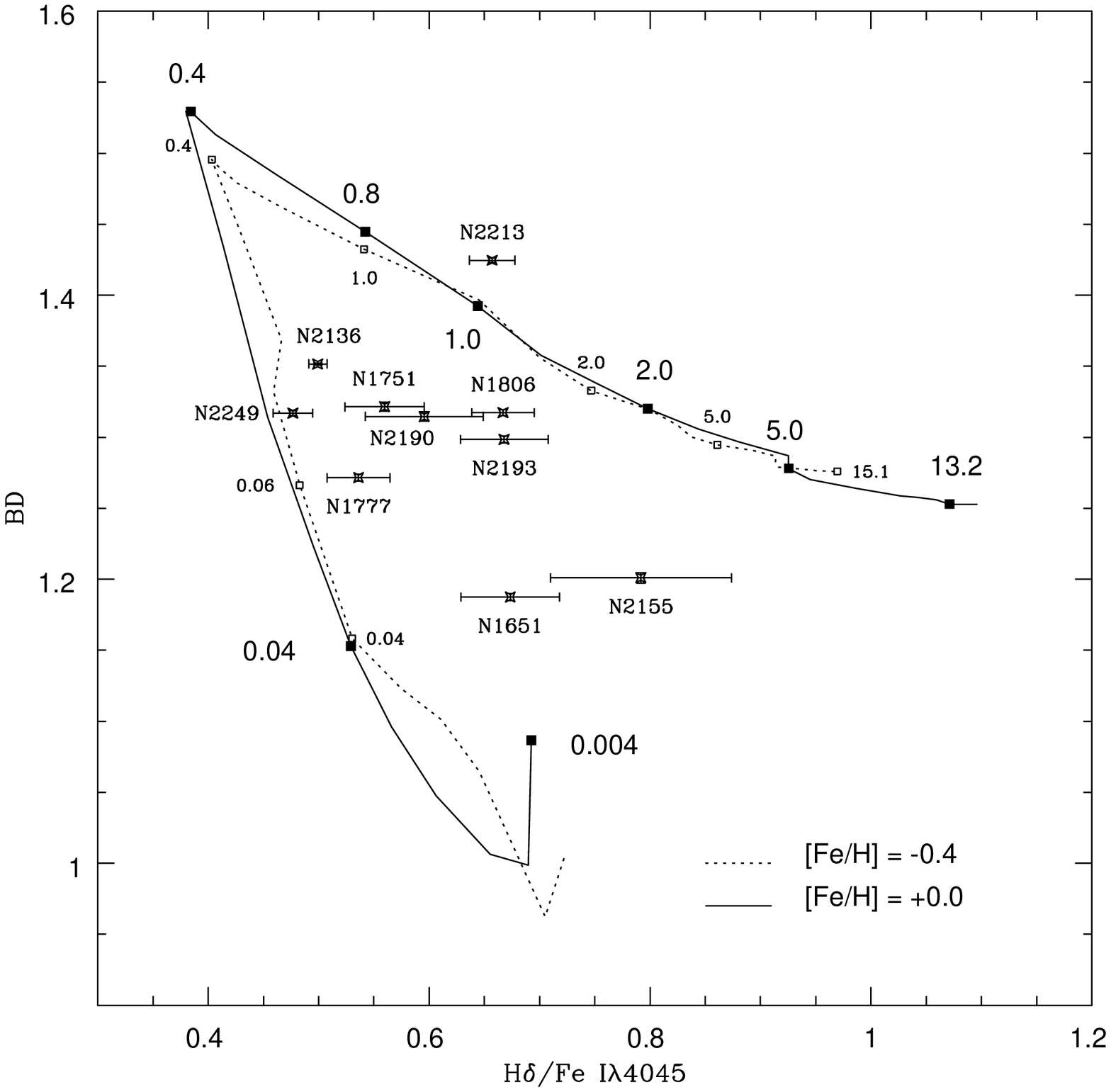}
\caption{Same as Figure \ref{bdclustm17} but with [Fe/H] = +0.0.  Again,
note here and in Figure~\ref{bdclustp00b} the group of clusters whose
BD index values prevent any age determinations due to their locations in
the index space.}
\label{bdclustp00a}
\end{figure}

\begin{figure}
\plotone{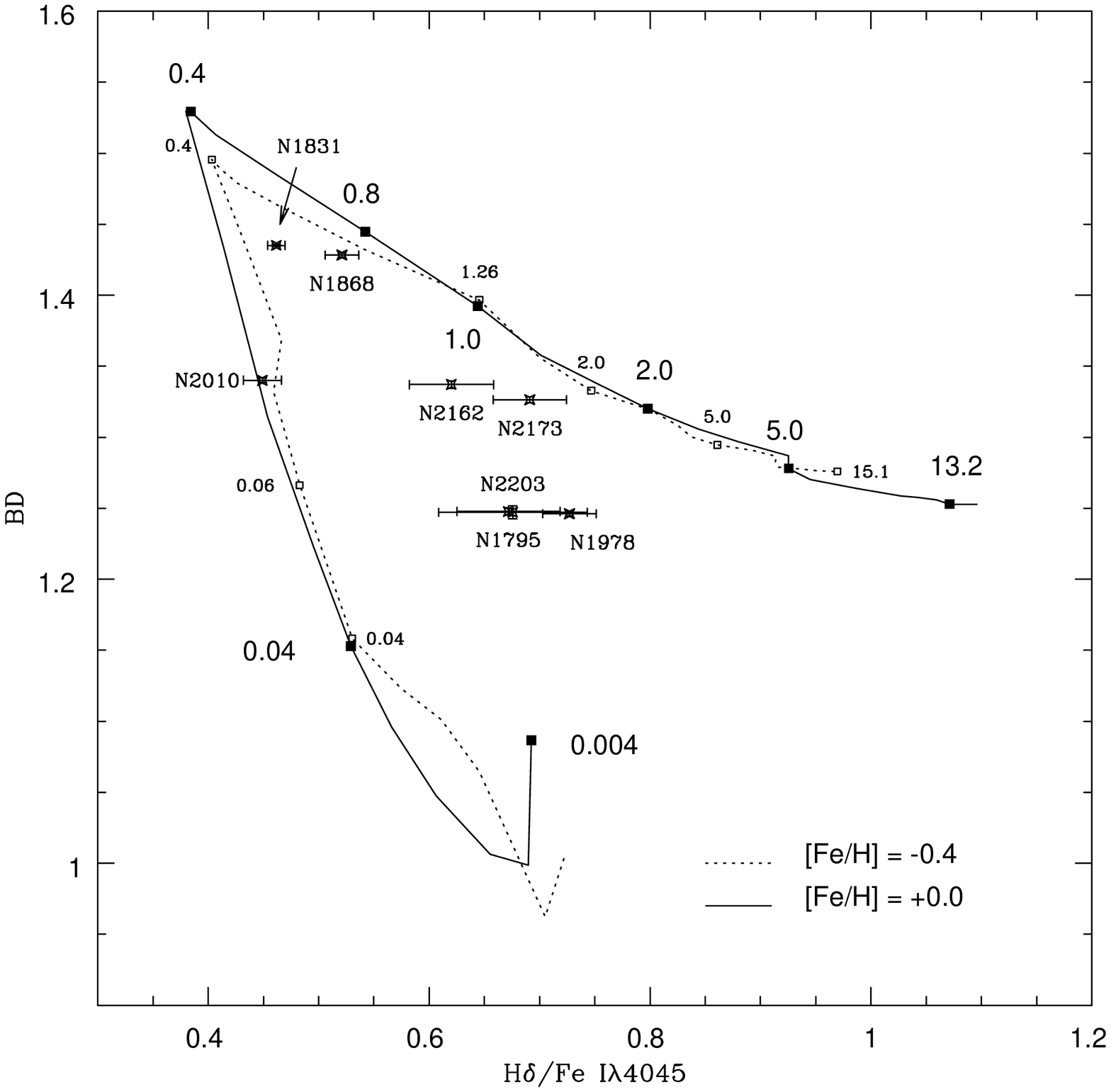}
\caption{Same as Figure \ref{bdclustp00a}.}
\label{bdclustp00b}
\end{figure}

\begin{figure}
\plotone{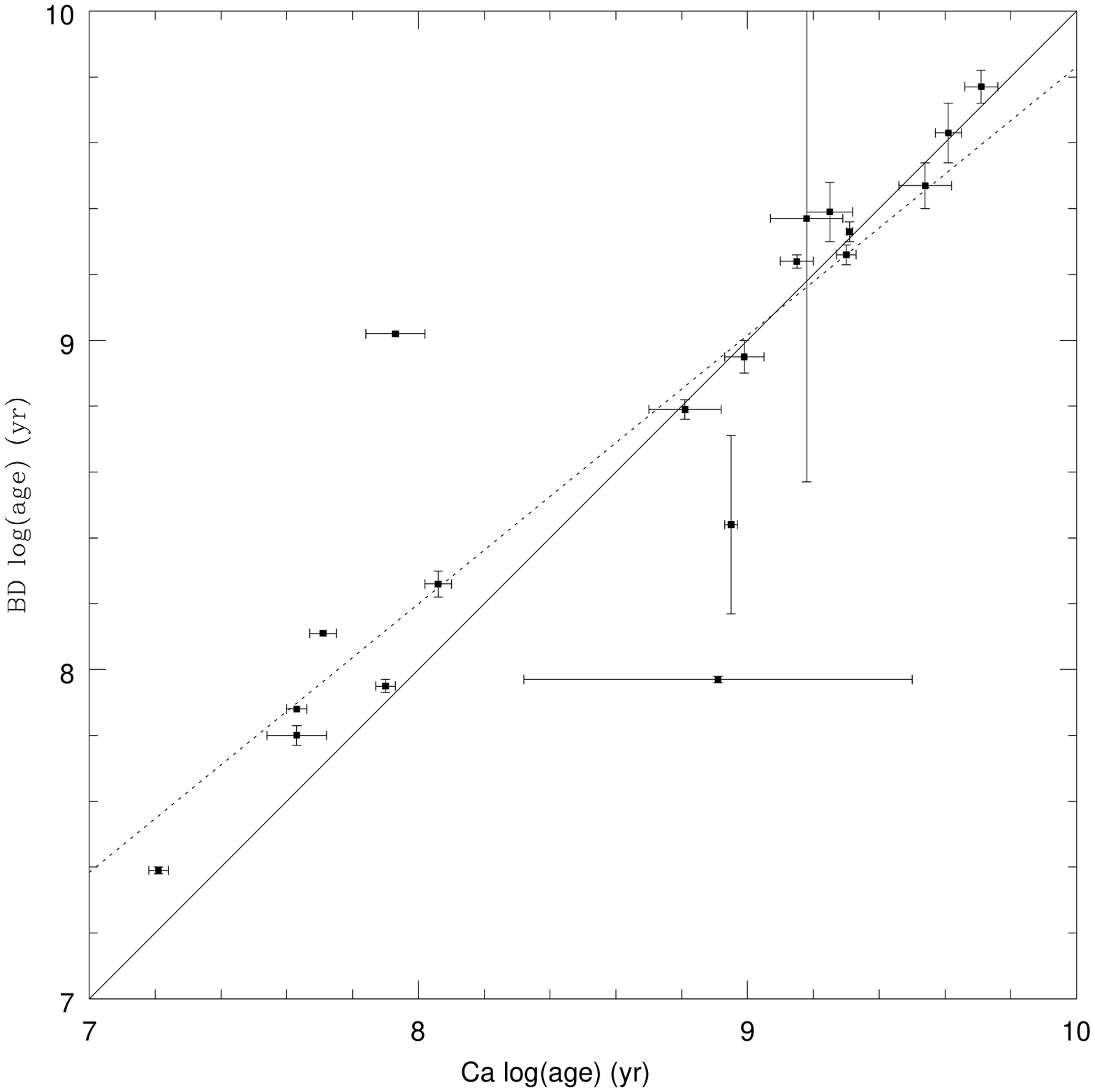}
\caption{The MC cluster ages derived from the Ca II index and the BD index
are plotted against each other (solid squares) for the clusters which have
determinations from
both indices.  The dotted line is a least-squares fit to the data while the
solid line denotes the identity relation.}
\label{regress}
\end{figure}

\begin{figure}
\plotone{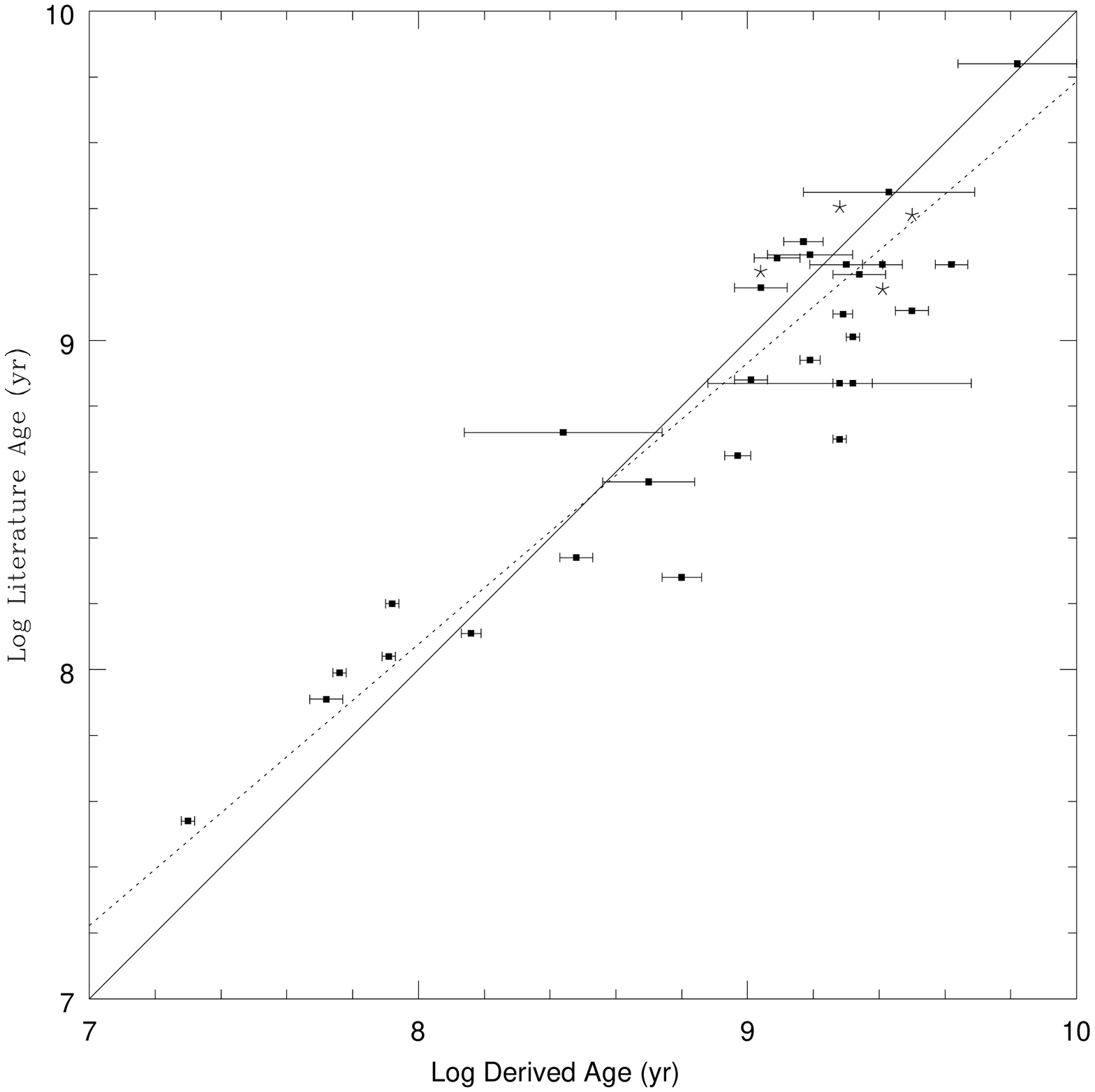}
\caption{The final derived ages for the MC clusters are plotted vs. the
literature values from Table~\ref{litagemet_tab} (solid squares). The dashed
line is the best-fit line to the data, whle the solid line denotes the identity
relation. The asterisks show age determinations in common with the Beasley \etal
(2002) results. }
\label{figlogages}
\end{figure}

\clearpage

\begin{figure}
\plotone{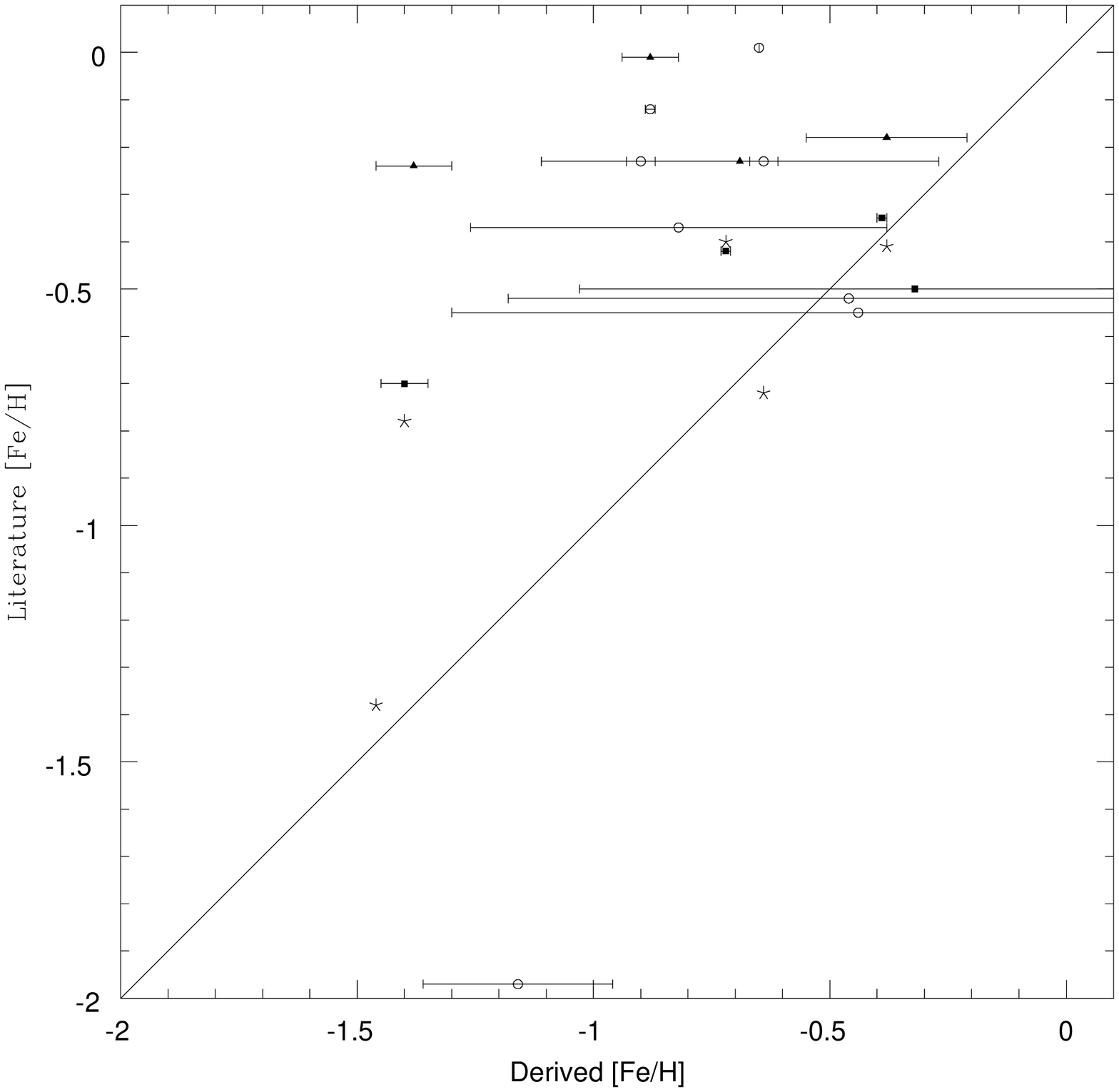}
\caption{The final derived metallicities for the MC clusters are plotted vs.
the Olszewski et al. (1991) values from Table~\ref{litagemet_tab}. The data are
divided into three age bins.  Specifically, the open circles represent the
youngest clusters with ages less than 9.1 in the log.  The filled squares
represent clusters with ages between 9.1 and 9.3 in the log, and the filled
triangles are for the oldest clusters with ages (in Gyr) greater than 9.3 in the
log.  The solid line shows the identity line.  The derived metallicities are
systematically more metal-poor than the Olszewski et al. (1991) metallicities.
Asterisks denote metallicity determinations in common with the Beasley \etal
(2002) results.}
\label{figmets}
\end{figure}

\begin{figure}
\plotone{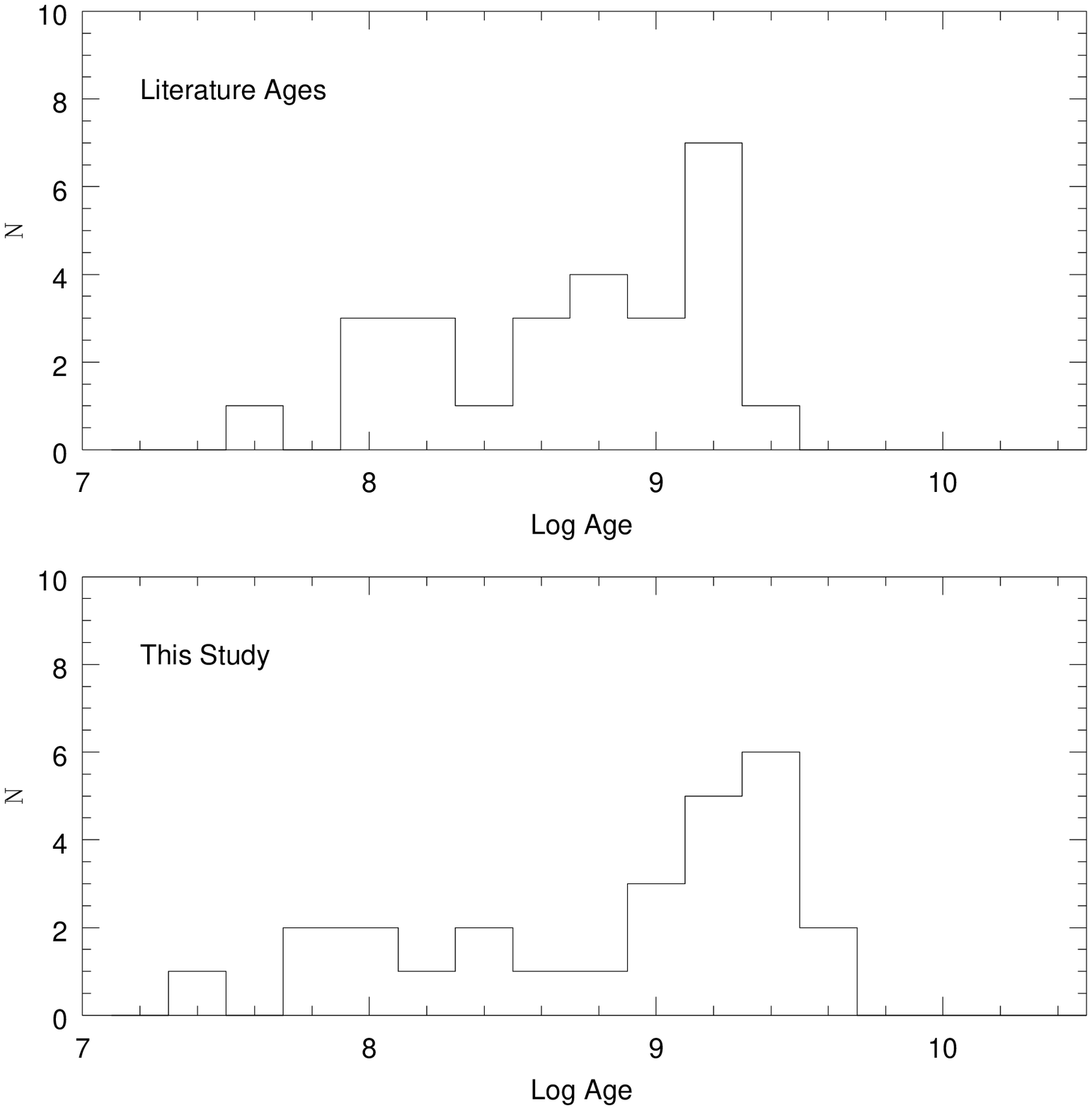}
\caption{Histogram comparison of the derived LMC cluster age distribution 
(lower panel) vs. 
their literature ages (upper panel).  We have excluded the two clusters (NGC 
1754 and NGC 2210) with a literature
age $\sim$ 15 Gyr.}
\label{histnoold}
\end{figure}

\begin{figure}
\plotone{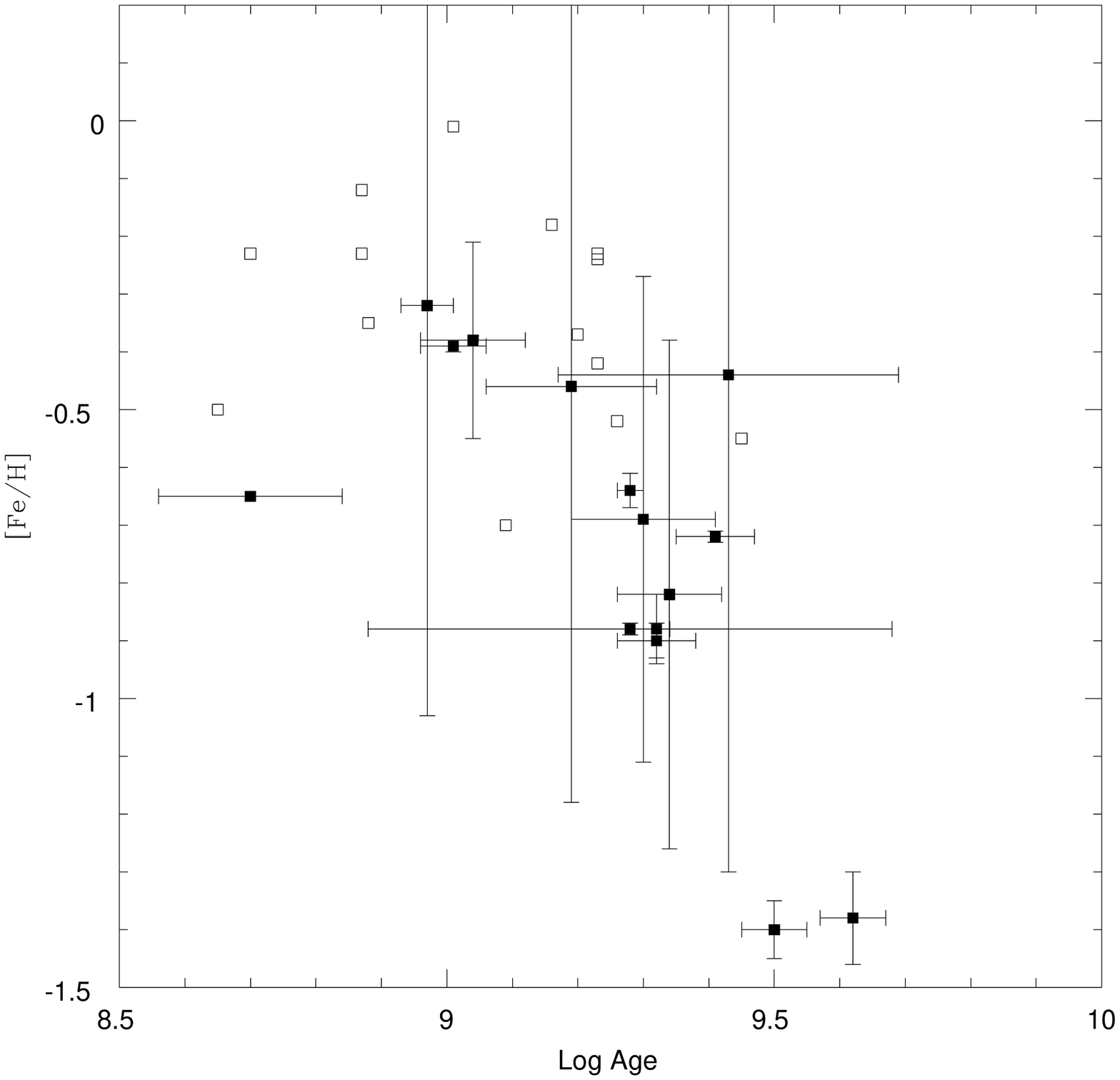}
\caption{The LMC age-metallicity relation for the derived parameters (solid
squares) and the literature values (open squares) only using clusters with
Olszewski et al. (1991) metallicity measurements.}
\label{agemeto91}
\end{figure}

\clearpage

\begin{deluxetable}{lccccc}
\tablenum{1}
\tablecaption{Observation Log \label{tab:obs}}
\tablewidth{0pt}
\tablecolumns{6}
\tablehead{
\colhead{Cluster} & \colhead{RA(2000)} & \colhead{Dec(2000)} &
\colhead{Obs. Date} & \colhead{Exp. Time} & \colhead{Trail Width} \\
\colhead{} & \colhead{(h m s)} & \colhead{($^{\circ}$ \ ' '')} &
\colhead{} & \colhead{(sec)} & \colhead{(arcsec)}}

\startdata
\cutinhead{SMC}
NGC 411 & 1:06:14.70 & -72:00:58.40 & 24/11/95 & 2700 & 40  \\
NGC 416 & 1:07:51.00 & -72:21:29.50 & 23/11/95 & 2700 & 20  \\
NGC 419 & 1:08:09.90 & -72:53:22.20 & 22/11/95 & 2700 & 60  \\
\cutinhead{LMC}
NGC 1651 & 4:37:48.00 & -70:40:34.60 & 24/11/95 & 3600 & 50\\
NGC 1751 & 4:54:27.80 & -69:52:42.20 & 22/11/95 & 2700 & 40\\
NGC 1754 & 4:54:42.10 & -70:31:05.60 & 20/11/95 & 1800 & 20\\
NGC 1777 & 4:56:54.90 & -74:21:28.30 & 22/11/95 & 2700 & 30\\
NGC 1783 & 4:58:56.90 & -66:03:40.10 & 22/11/95 & 2100 & 50\\
NGC 1795 & 5:00:01.10 & -69:52:23.30 & 24/11/95 & 3600 & 40\\
NGC 1806 & 5:02:14.40 & -68:03:33.10 & 22/11/95 & 2700 & 50\\
NGC 1818 & 5:04:04.40 & -66:29:36.50 & 23/11/95 & 2700 & 30\\
NGC 1831 & 5:05:59.80 & -64:59:23.60 & 20/11/95 & 2700 & 53\\
NGC 1846 & 5:07:33.40 & -67:30:38.10 & 21/11/95 & 2700 & 90\\
NGC 1866 & 5:13:24.30 & -65:31:35.50 & 23/11/95 & 1200 & 50\\
NGC 1868 & 5:14:12.70 & -64:00:32.00 & 21/11/95 & 2700 & 24\\
NGC 1978 & 5:28:38.30 & -66:15:47.90 & 22/11/95 & 2700 & 50\\
NGC 2010 & 5:31:03.30 & -70:50:47.20 & 24/11/95 & 2700 & 60\\
NGC 2133 & 5:52:05.50 & -71:10:45.60 & 20/11/95 & 2700 & 30\\
NGC 2134 & 5:52:32.50 & -71:06:03.90 & 21/11/95 & 2700 & 40\\
NGC 2136 & 5:53:16.70 & -69:29:37.70 & 23/11/95 & 2700 & 30\\
NGC 2155 & 5:58:20.10 & -65:28:39.40 & 22/11/95 & 2700 & 30\\
NGC 2162 & 6:00:08.90 & -63:43:09.70 & 22/11/95 & 2700 & 30\\
NGC 2164 & 5:59:06.80 & -68:31:15.30 & 23/11/95 & 2700 & 30\\
NGC 2173 & 5:58:56.60 & -72:58:23.90 & 20/11/95 & 2700 & 42\\
NGC 2190 & 6:02:24.00 & -74:42:56.20 & 21/11/95 & 2700 & 46\\
NGC 2193 & 6:06:03.30 & -65:05:18.30 & 22/11/95 & 2700 & 20\\
NGC 2203 & 6:06:12.80 & -75:25:20.00 & 24/11/95 & 2700 & 60\\
NGC 2210 & 6:11:46.90 & -69:06:14.20 & 23/11/95 & 2700 & 30\\
NGC 2213 & 6:11:21.90 & -71:30:47.70 & 20/11/95 & 2700 & 23\\
NGC 2214 & 6:13:05.40 & -68:14:44.20 & 23/11/95 & 2700 & 24\\
NGC 2249 & 6:26:00.80 & -68:53:22.70 & 23/11/95 & 2700 & 30\\

\cutinhead{Galactic Globulars}
47 Tuc & 0:23:57.90 & -72:04:36.60 & 23/11/95 & 600 & 50\\
M15 & 21:30:01.30 & 12:10:39.10 & 20/11/95 & 1200 & 60\\
M79 & 5:22:08.40 & -24:34:28.40 & 21/11/95 & 1200 & 40\\
NGC 1851 & 5:12:30.00 & -40:05:32.30 & 20/11/95 & 900 & 50\\
\enddata
\label{obs_tab}
\end{deluxetable}

\begin{deluxetable}{lccccc}
\tablenum{2}
\tablecaption{MC Cluster Adopted Ages and Metallicities \label{tab:litagemet}}
\tablewidth{0pt}
\tablecolumns{6}
\tablehead{
\colhead{Cluster} & \colhead{SWB Type\tablenotemark{a}} & \colhead{Age(Gyr)} &
\colhead{Reference} & \colhead{[Fe/H]} & \colhead{Reference}}
\startdata

\cutinhead{SMC}
NGC 411 & V-VI & 1.8 & 11 & -0.84 & 11 \\
NGC 416 & VI & 6.9 & 11 & -1.44 & 11 \\
NGC 419 & V & 1.2 & 11 & -0.70 & 11 \\
\cutinhead{LMC}
NGC 1651 & V & 1.6 & 12 & -0.37 & 1 \\
NGC 1751 & VI & 1.5 & 10 & -0.18 & 1 \\
NGC 1754 & VII & 15.5 & 11 & -1.42 & 11 \\
NGC 1777 & IVB & 0.76 & 7 & -0.35 & 1 \\
NGC 1783 & V & 0.87 & 10 & -0.75 & 3 \\
NGC 1795 & V & 1.7 & 8 & -0.23 & 1 \\
NGC 1806 & V & 0.50 & 12 & -0.23 & 1 \\
NGC 1818 & I & 0.04 & 10 & -0.9 & 4 \\
NGC 1831 & IVA & 0.37 & 9 & +0.01 & 1 \\
NGC 1846 & VI & 1.2 & 10 & -0.70 & 1 \\
NGC 1866 & III & 0.22 & 9 & -1.2 & 4 \\
NGC 1868 & IVA & 0.45 & 9 & -0.50 & 1 \\
NGC 1978 & VI & 1.7 & 10 & -0.42 & 1 \\
NGC 2010 & III & 0.16 & 9 & +0.0 & 5 \\
NGC 2133 & IVA & 0.13 & 5 & -1.0 & 5 \\
NGC 2134 & III & 0.19 & 9 & -1.0 & 5 \\
NGC 2136 & III & 0.11 & 10 & -0.4 & 4 \\
NGC 2155 & VI & 2.8 & 10 & -0.55 & 1 \\
NGC 2162 & V & 0.74 & 9 & -0.23 & 1 \\
NGC 2164 & II & 0.10 & 9 & -0.6 & 5 \\
NGC 2173 & VI & 1.7 & 10 & -0.24 & 1 \\
NGC 2190 & V & 0.74 & 9 & -0.12 & 1 \\
NGC 2193 & V & 2.0 & 7 & -0.5 & 5 \\
\tablebreak
NGC 2203 & VI & 1.8 & 8 & -0.52 & 1 \\
NGC 2210 & VII & 12.3 & 11 & -1.97 & 1 \\
NGC 2213 & V & 1.0 & 10 & -0.01 & 1 \\
NGC 2214 & II & 0.08 & 9 & -1.2 & 5 \\
NGC 2249 & IVB & 0.52 & 9 & -0.05 & 4 \\
\enddata
\tablenotetext{a} {SWB types taken from Bica et al. (1996)}
\tablerefs{
(1) Olszewski et al. 1991; (2) Olsen et al. 1998; (3) Cohen 1982;
(4) Seggewiss \& Richtler 1989; (5) Sagar \& Pandey 1989; (6) Durand, Hardy, \&
Melnick 1984; (7) Elson \& Fall 1988; (8) Geisler et al. 1997; (9) Girardi et al
.
1995; (10) Girardi \& Bertelli 1998; (11) Piatti et al. 2002; (12) Dirsch et
al. 2000
}
\label{litagemet_tab}
\end{deluxetable}

\begin{deluxetable}{lcccc}
\tablenum{3}
\tablecolumns{5}
\tablewidth{0pt}
\tablecaption{Empirical Polynomial Coefficients. \label{tab:coeffs}}
\tablehead{
\colhead{} & \colhead{Ca II K} & \colhead{Ca II H} & \colhead{Fe I $\lambda$4045
} &
\colhead{H$\delta$}
}
\startdata
1. Constant & 0.3522 & 0.3862 & 2.5286 & -3.1050\\
\nodata & \nodata & \nodata & \nodata & \nodata\\
2.  $\Theta$ & -0.2541 & -0.2606 & -2.7922 & 7.2148\\
3.  $\Theta^{2}$ & \nodata & \nodata & 1.0882 & -3.2809\\
4.  $\Theta^{3}$ & \nodata & \nodata & \nodata & \nodata\\
 & \nodata & \nodata & \nodata & \nodata\\
5.  [Fe/H] & -2.6599 & -0.3271 & \nodata & 0.1577\\
6.  [Fe/H]$^{2}$ & 0.02995 & 0.01763 & 0.01450 & \nodata\\
7.  [Fe/H]$^{3}$ & \nodata & \nodata & \nodata & 0.004024\\
 & \nodata & \nodata & \nodata & \nodata\\
8.  (log g) & 0.01866 & 0.02340 & \nodata & \nodata\\
9.  (log g)$^{2}$ & \nodata & \nodata & \nodata & \nodata\\
10. (log g)$^{3}$ & \nodata & \nodata & \nodata & \nodata\\
 & \nodata & \nodata & \nodata & \nodata\\
11. ($\Theta$)([Fe/H]) & 4.9594 & 0.2794 & -0.0862 & -0.1500\\
12. ($\Theta$)([Fe/H])$^{2}$ & \nodata & \nodata & \nodata & \nodata\\
13. ($\Theta$)(log g) & \nodata & \nodata & \nodata & \nodata\\
14. ($\Theta$)(log g)$^{2}$ & \nodata & \nodata & 0.005235 & 0.002516\\
15. ([Fe/H])(log g) & 0.01524 & 0.01166 & \nodata & \nodata\\
16. ([Fe/H])(log g)$^{2}$ & \nodata & \nodata & \nodata & \nodata\\
17. ([Fe/H])($\Theta$)$^{2}$ & -2.3205 & \nodata & \nodata & \nodata\\
18. (log g)([Fe/H])$^{2}$ & \nodata & \nodata & \nodata & \nodata\\
19. (log g)($\Theta$)$^{2}$ & \nodata & \nodata & \nodata & \nodata\\
\enddata
\label{coeffs_tab}
\end{deluxetable}

\begin{deluxetable}{cccccc}
\tablenum{4}
\tablecaption{Stellar Atmospheric Parameter Space for SYNTHE. \label{tab:syntheg
rid}}
\tablecolumns{6}
\tablewidth{0pt}
\tablehead{
\colhead{T$_{eff}$ (K)} & \colhead{$\Delta$T (K)} & \colhead{log g} &
\colhead{$\Delta$(log g)} & \colhead{[Fe/H]} & \colhead{$\Delta$([Fe/H])}
}
\startdata
4000--9750&250&2.0--5.0&1.0&-2.0 -- -0.5 & 0.5\\
 & & & & -0.3 -- 0.3 & 0.1\\
 & & & & 0.5 & \\
10000--10500&500&2.0--5.0&1.0&-2.0 -- -0.5 & 0.5\\
 & & & & -0.3 -- 0.3 & 0.1\\
 & & & & 0.5 & \\
11000--13000&500&3.0--5.0&1.0&-2.0 -- -0.5 & 0.5\\
 & & & & -0.3 -- 0.3 & 0.1\\
 & & & & 0.5 & \\
14000--26000&1000&3.0--5.0&1.0&-2.0 -- -0.5 & 0.5\\
 & & & & -0.3 -- 0.3 & 0.1\\
 & & & & 0.5 & \\
27000--35000&1000&4.0--5.0&1.0&-2.0 -- -0.5 & 0.5\\
 & & & & -0.3 -- 0.3 & 0.1\\
 & & & & 0.5 & \\
\enddata
\label{synthegrid_tab}
\end{deluxetable}

\begin{deluxetable}{ccccccc}
\tablenum{5}
\tablecolumns{7}
\tablewidth{0pt}
\tablecaption{Sample Interpolation Scheme. \label{tab:interp}}
\tablehead{
\multicolumn{7}{c}{Isochrone point - T$_{eff}$ = 16300 K, log g = 2.4, [Fe/H] =
-0.7
}\\
\multicolumn{3}{c}{Ideal} & \colhead{} & \multicolumn{3}{c}{Actual} \\
\cline{1-3} \cline{5-7} \\
\colhead{T$_{eff}$ (K)} & \colhead{log g} & \colhead{[Fe/H]} & \colhead{Present?
} &
\colhead{T$_{eff}$ (K)} & \colhead{log g} & \colhead{[Fe/H]}
}
\startdata
16000 & 2.0 & -1.0 & n & 10500 & 2.0 & -1.0\\
17000 & 2.0 & -1.0 & n & 10500 & 2.0 & -1.0\\
16000 & 3.0 & -1.0 & y & 16000 & 3.0 & -1.0\\
17000 & 3.0 & -1.0 & y & 17000 & 3.0 & -1.0\\
16000 & 2.0 & -0.5 & n & 10500 & 2.0 & -0.5\\
17000 & 2.0 & -0.5 & n & 10500 & 2.0 & -0.5\\
16000 & 3.0 & -0.5 & y & 16000 & 3.0 & -0.5\\
17000 & 3.0 & -0.5 & y & 17000 & 3.0 & -0.5\\
\enddata
\label{interp_tab}
\end{deluxetable}

\begin{deluxetable}{ccccccc}
\tablenum{6}
\tablecolumns{7}
\tablewidth{0pt}
\tablecaption{Cluster Index Values.  \label{tab:clustind}}
\tablehead{
\colhead{Cluster} & \colhead{H$\delta$/Fe I $\lambda$4045} &
\colhead{$\pm$1 $\sigma$} &
\colhead{Ca II} & \colhead{$\pm$1 $\sigma$} &
\colhead{BD} & \colhead{$\pm$1 $\sigma$}
}
\startdata
\cutinhead{SMC}
NGC 411  & 0.618 & 0.033 & 0.755 & 0.086 & 1.304 & 0.003\\
NGC 416  & 0.772 & 0.022 & 0.906 & 0.050 & 1.239 & 0.001\\
NGC 419  & 0.607 & 0.015 & 0.673 & 0.030 & 1.329 & 0.001\\
\cutinhead{LMC}
NGC 1651 & 0.674 & 0.045 & 0.762 & 0.143 & 1.188 & 0.002\\
NGC 1751 & 0.560 & 0.036 & 0.731 & 0.104 & 1.321 & 0.003\\
NGC 1754 & 0.747 & 0.020 & 0.782 & 0.037 & 1.312 & 0.001\\
NGC 1777 & 0.536 & 0.028 & 0.671 & 0.063 & 1.272 & 0.002\\
NGC 1783 & 0.647 & 0.021 & 0.787 & 0.048 & 1.335 & 0.002\\
NGC 1795 & 0.672 & 0.047 & 0.802 & 0.130 & 1.248 & 0.003\\
NGC 1806 & 0.667 & 0.028 & 0.791 & 0.056 & 1.317 & 0.002\\
NGC 1818 & 0.597 & 0.006 & 0.613 & 0.007 & 1.137 & 0.001\\
NGC 1831 & 0.462 & 0.008 & 0.493 & 0.012 & 1.435 & 0.001\\
NGC 1846 & 0.660 & 0.037 & 0.687 & 0.063 & 1.441 & 0.003\\
NGC 1866 & 0.486 & 0.010 & 0.485 & 0.012 & 1.432 & 0.001\\
NGC 1868 & 0.521 & 0.015 & 0.663 & 0.035 & 1.428 & 0.001\\
NGC 1978 & 0.727 & 0.024 & 0.881 & 0.057 & 1.246 & 0.001\\
NGC 2010 & 0.449 & 0.017 & 0.485 & 0.021 & 1.340 & 0.002\\
NGC 2133 & 0.450 & 0.011 & 0.461 & 0.014 & 1.394 & 0.001\\
NGC 2134 & 0.457 & 0.011 & 0.465 & 0.014 & 1.486 & 0.001\\
NGC 2136 & 0.499 & 0.008 & 0.539 & 0.011 & 1.352 & 0.001\\
NGC 2155 & 0.792 & 0.082 & 1.038 & 0.292 & 1.201 & 0.004\\
NGC 2162 & 0.620 & 0.038 & 0.700 & 0.073 & 1.337 & 0.003\\
NGC 2164 & 0.502 & 0.006 & 0.540 & 0.008 & 1.306 & 0.001\\
NGC 2173 & 0.691 & 0.033 & 0.749 & 0.072 & 1.326 & 0.002\\
NGC 2190 & 0.596 & 0.053 & 0.655 & 0.090 & 1.315 & 0.003\\
NGC 2193 & 0.668 & 0.040 & 0.860 & 0.097 & 1.299 & 0.003\\
NGC 2203 & 0.676 & 0.067 & 0.839 & 0.242 & 1.247 & 0.005\\
NGC 2210 & 0.707 & 0.016 & 0.799 & 0.031 & 1.310 & 0.001\\
NGC 2213 & 0.657 & 0.021 & 0.745 & 0.042 & 1.425 & 0.002\\
NGC 2214 & 0.497 & 0.012 & 0.521 & 0.014 & 1.252 & 0.001\\
NGC 2249 & 0.477 & 0.018 & 0.556 & 0.036 & 1.317 & 0.002\\

\cutinhead{Galactic Globulars}
47 Tuc   & 0.900 & 0.009 & 1.239 & 0.028 & 1.231 & 0.000\\
M15      & 0.729 & 0.004 & 0.819 & 0.008 & 1.344 & 0.000\\
M79      & 0.741 & 0.012 & 0.806 & 0.022 & 1.316 & 0.001\\
NGC 1851 & 0.788 & 0.006 & 0.968 & 0.016 & 1.310 & 0.000\\
\enddata
\label{clustind_tab}
\end{deluxetable}

\begin{deluxetable}{lcccccccccccccc}
\tablenum{7}
\tablecolumns{15}
\tablewidth{0pt}
\tablecaption{Derived MC Cluster Ages. \label{tab:agefin}}
\tablehead{
\colhead{Cluster} & \multicolumn{6}{c}{Log Age (yr)} & \multicolumn{6}{c}{[Fe/H]
} &
\colhead{Lit Age} & \colhead{Lit [Fe/H]}\\
\colhead{} & \colhead{Ca II} & \colhead{$\pm$1$\sigma$} & \colhead{BD} &
\colhead{$\pm$1$\sigma$} & \colhead{Final} & \colhead{$\pm$1$\sigma$} &
\colhead{Ca II} & \colhead{$\pm$1$\sigma$} & \colhead{BD} &
\colhead{$\pm$1$\sigma$} & \colhead{Final} & \colhead{$\pm$1$\sigma$} &
\colhead{} & \colhead{}
}
\startdata
NGC 411 & 9.09& 0.07& \nodata& \nodata& 9.09& 0.07& -0.43& 0.14& \nodata& \nodata& -0.43& 0.14& 9.25& -0.84\\
NGC 416 & 9.82& 0.18& \nodata& \nodata& 9.82& 0.18& -1.27& 0.03& \nodata& \nodata& -1.27& 0.03& 9.84& -1.44\\
NGC 419 & 9.23& 0.05& 9.35& 0.03& 9.29& 0.03& -0.78& 0.05& -1.03& 0.02& -0.90& 0.05& 9.08& -0.70\\
NGC 1651& 9.34& 0.08& \nodata& \nodata& 9.34& 0.08& -0.82& 0.44& \nodata& \nodata& -0.82& 0.44& 9.20& -0.37\\
NGC 1751& 9.04& 0.08& \nodata& \nodata& 9.04& 0.08& -0.38& 0.17& \nodata& \nodata& -0.38& 0.17& 9.16& -0.18\\
NGC 1754& 9.71& 0.05& 9.77&  0.05& 9.74& 0.04& -1.46& 0.02& -1.43& 0.02& -1.44&0.02& 10.19& -1.42\\
NGC 1777& 9.01& 0.05& \nodata& \nodata& 9.01& 0.05& -0.39& 0.01& \nodata& \nodata& -0.39& 0.01& 8.88& -0.35\\
NGC 1783& 9.15& 0.05& 9.24& 0.02& 9.19& 0.03& -0.48& 0.12& -0.59& 0.01& -0.54& 0.12& 8.94& -0.75\\
NGC 1795& 9.30& 0.11& \nodata& \nodata& 9.30& 0.11& -0.69& 0.42& \nodata& \nodata& -0.69& 0.42& 9.23& -0.23\\
NGC 1806& 9.30& 0.03& 9.26& 0.03& 9.28& 0.02& -0.70& 0.03& -0.58& 0.00& -0.64& 0.03& 8.70& -0.23\\
NGC 1818& 7.21& 0.03& 7.39& 0.01& 7.30& 0.02& \nodata& \nodata& \nodata& \nodata& \nodata& \nodata& 7.54& -0.9\\
NGC 1831& 8.95& 0.02& 8.44& 0.27& 8.70& 0.14& -0.69& 0.00& -0.61& 0.02& -0.65& 0.00& 8.57& 0.01\\
NGC 1846& 9.54& 0.08& 9.47& 0.07& 9.50& 0.05& -1.31& 0.05& -1.50& 0.00& -1.40& 0.05& 9.09& -0.70\\
NGC 1866& 7.93& 0.09& 9.02& 0.00& 8.48& 0.77& \nodata& \nodata& \nodata& \nodata& \nodata& \nodata& 8.34 & -1.2\\
NGC 1868& 8.99& 0.06& 8.95& 0.05& 8.97& 0.04& -0.37& 0.71& -0.27& 0.22& -0.32& 0.71& 8.65& -0.50\\
NGC 1978& 9.41& 0.06& \nodata& \nodata& 9.41& 0.06& -0.72& 0.01& \nodata& \nodata& -0.72& 0.01& 9.23& -0.42\\
NGC 2010& 7.90& 0.03& 7.95& 0.02& 7.92& 0.02& \nodata& \nodata& \nodata& \nodata& \nodata& \nodata& 8.20& 0.0\\
NGC 2133& 8.06& 0.04& 8.26& 0.04& 8.16& 0.03& \nodata& \nodata& \nodata& \nodata& \nodata& \nodata& 8.11& -1.0\\
NGC 2134& 8.81& 0.11& 8.79& 0.03& 8.80& 0.06&\nodata& \nodata& \nodata& \nodata&\nodata& \nodata& 8.28& -1.0\\
NGC 2136& 7.71& 0.04& 8.11& 0.00& 7.91& 0.02& \nodata& \nodata& \nodata& \nodata& \nodata& \nodata& 8.04& -0.4\\
NGC 2155& 9.43& 0.26& \nodata& \nodata& 9.43& 0.26& -0.44& 0.86& \nodata& \nodata& -0.44& 0.86& 9.45& -0.55\\
NGC 2162& 9.25& 0.07& 9.39& 0.09& 9.32& 0.06& -0.75& 0.03& -1.06& 0.07& -0.90& 0.03& 8.87& -0.23\\
NGC 2164& 7.63& 0.03& 7.88& 0.00& 7.76& 0.02& \nodata& \nodata& \nodata& \nodata& \nodata& \nodata& 7.99& -0.6\\
NGC 2173& 9.61& 0.04& 9.63& 0.09& 9.62& 0.05& -1.35& 0.08& -1.42& 0.15& -1.38& 0.08& 9.23& -0.24\\
NGC 2190& 9.18& 0.11& 9.37& 0.80& 9.28& 0.40& -0.73& 0.01& -1.02& 0.06& -0.88& 0.01& 8.87& -0.12\\
NGC 2193& 9.17& 0.06& \nodata& \nodata& 9.17& 0.06& -0.41& 0.05& \nodata& \nodata& -0.41& 0.05& 9.30& -0.5\\
NGC 2203& 9.19& 0.13& \nodata& \nodata& 9.19& 0.13& -0.46& 0.72& \nodata& \nodata& -0.46& 0.72& 9.26& -0.52\\
NGC 2210& 9.61& 0.09& 9.55& 0.04& 9.58& 0.05& -1.22& 0.20& -1.09& 0.02& -1.16& 0.20& 10.09& -1.97\\
NGC 2213& 9.31& 0.01& 9.33& 0.03& 9.32& 0.02& -0.79& 0.06& -0.98& 0.09& -0.88& 0.06& 9.01& -0.01\\
NGC 2214& 7.63& 0.09& 7.80& 0.03& 7.72& 0.05& \nodata& \nodata& \nodata& \nodata& \nodata& \nodata& 7.91& -1.2\\
NGC 2249& 8.91& 0.59& 7.97& 0.01& 8.44& 0.30& -0.40& 0.02& \nodata& \nodata& -0. 40& 0.02& 8.72& -0.05\\
\enddata
\label{agefin_tab}
\end{deluxetable}

\end{document}